\documentclass{article}

\usepackage[utf8]{inputenc}
\usepackage[T1]{fontenc}    
\usepackage{amsfonts}       
\usepackage{amsmath,amssymb}
\usepackage{xfrac}
\usepackage{mathtools}
\usepackage{graphicx}
\usepackage{array} 
\usepackage{float}
\usepackage{multirow}
\usepackage[margin=1in]{geometry}
\usepackage[thinc]{esdiff}

\usepackage[
backend=biber,
style=authoryear,
citestyle=authoryear,
sorting=nyt
]{biblatex}
\usepackage[splitrule]{footmisc}
\AtBeginDocument{
  \newgeometry{
    textheight=9in,
    textwidth=6.5in,
    top=1in,
    headheight=14pt,
    headsep=25pt,
    footskip=30pt
  }
}
\usepackage{booktabs}
\usepackage{verbatim} 
\usepackage{xltabular}

\usepackage[hidelinks]{hyperref}
\usepackage{endnotes}
\let\footnote=\endnote
\usepackage{setspace}
 \usepackage{authblk}

\title{Small-area Population Forecast in a Segregated City using Density-Functional Fluctuation Theory}

\author[1,5]{Yuchao Chen}
\author[2,5]{Yunus A. Kinkhabwala}
\author[1]{Boris Barron}
\author[3,4]{Matthew Hall}
\author[1]{Tomas A. Arias}
\author[1]{Itai Cohen}
\affil[1]{\footnotesize Department of Physics, Cornell University, Ithaca, NY 14853, USA.}
\affil[2]{\footnotesize Department of Applied and Engineering Physics, Cornell University, Ithaca, NY, 14853, USA.}
\affil[3]{\footnotesize Policy Analysis and Management, Cornell University, Ithaca, NY, 14853, USA.}
\affil[4]{\footnotesize Cornell Institute for Public Affairs, Cornell University, Ithaca, NY, 14853, USA.}
\affil[5]{\footnotesize These authors contributed equally to this work.}

\begin{document}

\maketitle % Title page without authors 
%\doublespacing
%\linenumbers
\begin{abstract}
Decisions regarding housing, transportation, and resource allocation would all benefit from accurate small-area population forecasts. While various tried-and-tested forecast methods exist at regional scales, developing an accurate neighborhood-scale forecast remains a challenge partly due to complex drivers of residential choice ranging from housing policies to social preferences and economic status that cumulatively cause drastic neighborhood-scale segregation. Here, we show how to forecast the dynamics of neighborhood-scale demographics by extending a novel statistical physics approach called Density-Functional Fluctuation Theory (DFFT) to multi-component time-dependent systems. In particular, this technique observes the fluctuations in neighborhood-scale demographics to extract effective drivers of segregation. As a demonstration, we simulate a segregated city using a Schelling-type segregation model, and found that DFFT accurately predicts how a city-scale demographic change trickles down to block scales. Should these results extend to actual human populations, DFFT could capitalize on the recent advances in demographic data collection and regional-scale forecasts to improve upon current small-area population forecasts.
\end{abstract}

% \keywords{Statistical physics \and Racial residential segregation \and Migration \and Density-Functional Fluctuation Theory}

\section{Introduction}

%%%%%%%%%%%%%%%
%We choose to use simulated data here because it provides access to limitless data and the ability to control interactions leading to segregation, which is not possible with empirical data. However, the use of simulated data is clearly problematic since it may not accurately represent the true behavior of segregation in a census population counts.

%%%%%%%%%%%%%

Forecasting the neighborhood-scale dynamics of residential populations remains an outstanding problem in demography with the potential to inform and significantly affect local planning of social and economic developments \parencite{siegel2004methods}. For example, such forecasts could be used to achieve optimal allocation of educational, health and safety resources by determining the need for new schools \parencite{swanson1998k}, hospitals \parencite{humphreys1998delimiting} and fire stations \parencite{parrott1997locating} in each neighborhood. In addition, such forecasts would be important for estimating housing demands \parencite{mason1996population}, and might help combat socioeconomic inequalities by predicting the need for low-income housing developments \parencite{anderson2003providing} and public transportation \parencite{weiner2016urban,wellman2014transportation}. Despite this wide range of important potential applications, methods for accurate neighborhood-scale population forecasts remain limited. 

At regional scales, there are already numerous methods for predicting population change and dynamics. Traditionally, the demographic equation is applied to different cohorts \parencite{rowland2003demographic,preston2000demography,siegel2004methods,smith2013practitioner,wachter2014essential,keyfitz2005applied,land2005mathematical}. In particular, estimates of birth and death rates from past data are already quite accurate. Models with different amounts of sophistication have also been developed to estimate the migration rates from available data \parencite{rogers2008demographic}. For example, the gravity model \parencite{ramos2017gravity,grogger2011income,karemera2000gravity,kim2010determinants,poot2016gravity,foot1984net} and the gravity-like Weidlich-Haag Migratory Model \parencite{weidlich2012concepts,weidlich1988interregional,weidlich2006sociodynamics,haag2017modelling} fit migration data by including relative preferences of origin and destination regions as well as preferences to make moves to closer locations. The challenge then, is how to relate the data at the regional level to forecasts at the neighborhood scale.  

%%%%%%%%Summary of small-area population forecasts paragraph%%%%%55

One of the many hurdles for making such relations is that, at the neighborhood scale, drivers of segregation can significantly affect the resulting distributions. At regional scales, most methods either ignore the drivers of segregation \parencite{ramos2017gravity} or assume some simple forms for such segregation effects \parencite{haag2017modelling}. These simplifications are most likely justified for regional or national population forecasts, because the drivers of segregation, such as economic status, social preferences and housing policy \parencite{clark1986residential,freeman2009neighbourhood}, are likely to average over on a large scale. Such simplifications, however, could potentially lead to bigger errors if we apply them to make neighborhood-scale forecasts. 
%For example, consider the recent growth of the Hispanic population in the United States. With traditional methods, demographers were able to predict this growth on the coarser county level, where drivers of segregation plausibly averaged. On the census block group level within a given county, however, this growth was concentrated in integrated and predominantly Hispanic block groups with relatively little increase in the Hispanic population occurring in predominantly non-Hispanic block groups. Thus, there is still a need for methods that can account for the complex neighborhood-scale effects of segregation in order to forecast neighborhood scale dynamics such as neighborhood migration.

While many small-area population forecast methods have been developed and tested, none to date explicitly account for the effects of residential segregation. Ref. \parencite{wilson2021methods} recently collected a thorough review of such methods including extrapolative, cohort-component, and small-area microsimulation. Some methods referenced here account for aspects that might be of special importance to small-area dynamics such as land use, roads, urban accessibility \parencite{mckee2015locally}, water body, country borders \parencite{boke2017high}, demographic and socioeconomic characteristics \parencite{chi2009can}. However, residential segregation remains a key feature of American society and a potent driver residential mobility at the neighborhood scale previously ignored. Thus, there is still a need to explore methods that can account for the complex neighborhood-scale effects of segregation in order to forecast neighborhood scale dynamics such as neighborhood migration.

Although there is a rich history of methods aimed at quantifying neighborhood-scale segregation and understanding its causes, it is unclear how to use these methods to create forecasts of future changes in neighborhood-level population distributions. One general approach to the analysis of human segregation relies on the use of numerical indices to characterize the degree of segregation of a neighborhood  \parencite{reardon2002measures,reardon2004measures,reardon2006conceptual,park2018beyond,white2005mapping,oka2014capturing,oka2016spatializing,mora2011entropy,s2016segregation,echenique2007measure}. Such indices have been essential to understanding how segregation correlates with residential outcomes as well as potential drivers of segregation. Recently, Ellis et al (2018) even used such indices to extrapolate which neighborhoods are more likely to change their degree of segregation, but stopped short of forecasting population changes, presumably because such indices are too coarse grained to make accurate predictions. Another approach utilizes agent-based models \parencite{schelling1971dynamic,zhang2004dynamic,zhang2004residential,zhang2011tipping,grauwin2012dynamic,vinkovic2006physical,clark2008understanding,bruch2006neighborhood,van2009neighborhood,bruch2009preferences,zou2012model,spaiser2018identifying}, such as the well studied Schelling model \parencite{schelling1971dynamic,zhang2004dynamic,zhang2004residential,zhang2011tipping,grauwin2012dynamic,bruch2006neighborhood}, to determine the degree to which different proposed interactions lead to segregation and to investigate their dynamics. Such studies have shown that even slight preferences towards segregated neighborhood compositions can lead to drastic city-wide segregation and dynamic phenomena such as residential tipping. Since these models require \emph{a priori} knowledge of the decision rules for migration that are challenging to determine, however, they have found limited use for predicting trends in human populations \parencite{benenson2009schelling}. Thus, despite great progress in understanding the nature of segregation, neither segregation indices nor agent-based models have led to widely adopted methods for predicting population dynamics at the neighborhood scale. 

Recently, a new statistical physics method called Density-Functional Fluctuation Theory (DFFT) was developed to make predictions of how crowds will distribute in different environments \parencite{mendez2018density}. DFFT is a top-down data-driven approach that extracts functions to separately quantify effective social and spatial preferences from observations of fluctuations in the local density. By recombining these functions, DFFT is able to forecast population distributions in new environments and with different total population numbers.

\begin{figure}
	\centering
	\includegraphics[width=0.65\linewidth]{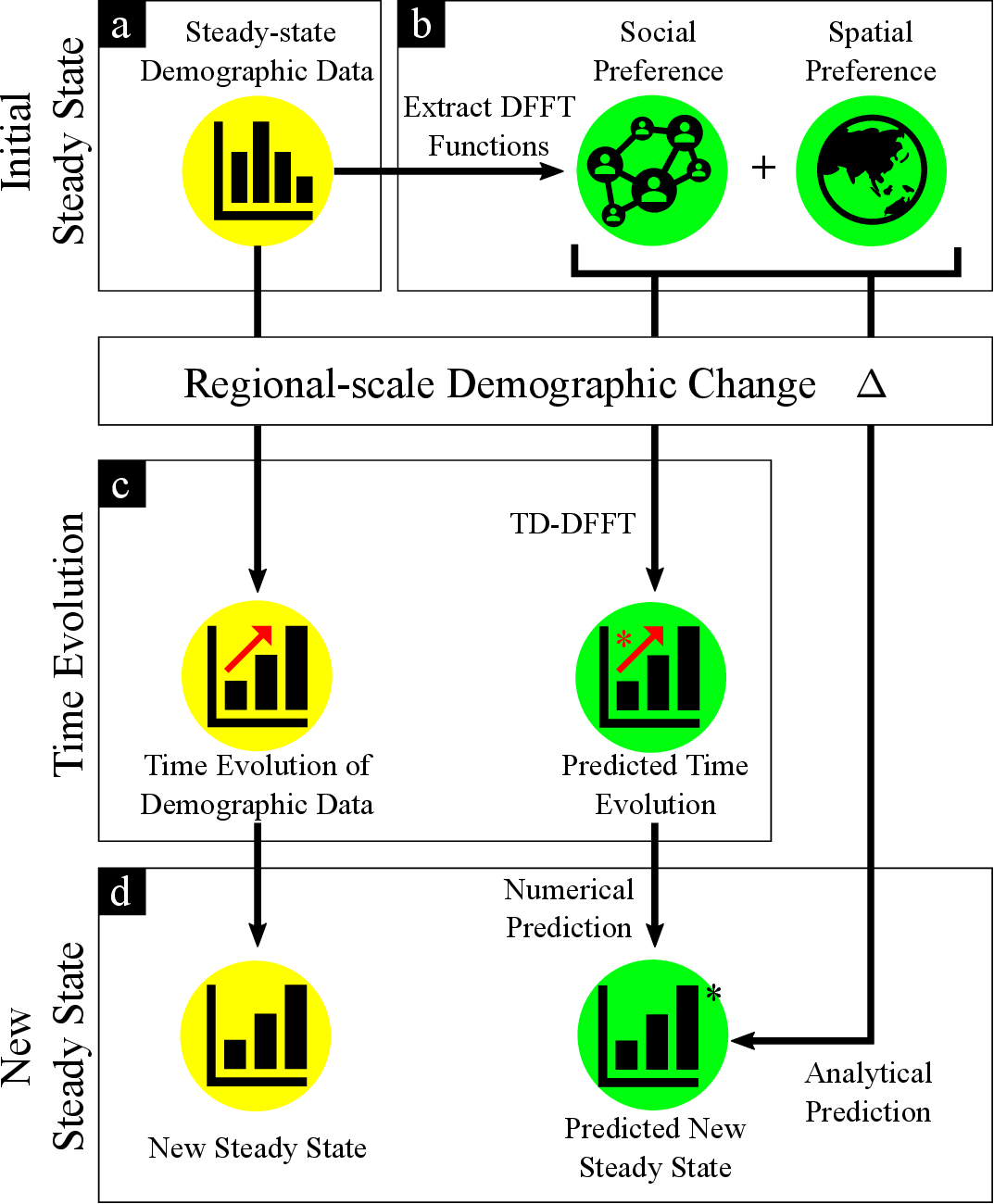}
	\singlespace \caption{\textbf{General Workflow of Applying DFFT to Demographic Data} \textbf{(a.)} Collect neighborhood-scale steady-state demographic data in the form of probability distributions of local densities. Steady state is reached when the population distribution no longer changes drastically over time. In our example, we simulate the steady-state data from a Schelling model (yellow bubble). \textbf{(b.)} Extract DFFT functions from steady-state data. The DFFT functions characterize social and spatial preferences separately.  \textbf{(c.)} After a demographic change, we predict the time evolution of neighborhood-scale demographic data with TD-DFFT using the extracted DFFT functions. We compare our prediction with the observed time evolution from the Schelling model simulations (yellow bubble). \textbf{(d.)} We predict the new neighborhood-scale steady state after the regional-scale demographic change either numerically using TD-DFFT or analytically using DFFT functions alone. We compare our prediction with the observed new steady state of the Schelling model simulations (yellow bubble).}
	\label{Flowchart}
\end{figure}

Here, we demonstrate how to extend DFFT to predict neighborhood-scale demographic data for multi-component time-dependent systems, using data generated from an extended Schelling model simulation of residential segregation. While these simulated data are not expected to accurately represent complex segregation and dynamics in real data, they provide a controlled first test of DFFT when applied to demographic systems without introducing errors typically found in small-area demographic data. We first describe application of a Schelling-type model to systems comprised of two types of agents in heterogeneous environments. We use these simulations to create the demographic steady-state data (i.e., data when population distribution no longer changes drastically over time) for our analysis (Fig.~\ref{Flowchart}a, Section~\ref{sec:Schelling}). Second, we apply DFFT to this data and extract functions describing the effective spatial and social preferences of the population (Fig.~\ref{Flowchart}b, Section~\ref{sec:DFFT}). Importantly, this quantification of preferences is generic so that it can in theory capture cumulative effects of the drivers of segregation without making specific assumptions about their properties or relative strengths. Next, we institute a sudden regional-scale demographic change to achieve a redistribution of the populations. We again use the Schelling model to generate the time evolution and new steady-state distribution of the neighborhood-scale demographic data. To predict the time evolution of the demographic data, we develop a time-dependent version of DFFT (TD-DFFT) using the extracted DFFT functions. We then compare the predictions from TD-DFFT to the data generated by the Schelling model (Fig.~\ref{Flowchart}c, Section~\ref{sec:TimeDep}). Finally, we predict the new steady-state joint densities resulting from the demographic change, either through numerical computation using TD-DFFT or analytic calculation using the DFFT functions extracted from the original steady-state data. These predictions are then compared to the Schelling model data for the new steady state (Fig.~\ref{Flowchart}d, Section~\ref{sec:NewSteadyState}). While we demonstrate this approach on data generated from a Schelling model, it should be possible to apply this method to publicly available demographic population counts for real populations. If application to real data is successful, DFFT would be the first method of its kind to extend population forecasts at the regional-scale down to the neighborhood scale while accounting for effects of segregation, and yield additional predictive power to small-area population forecasts.

%and yield valuable new insights and predictive power to the fields of segregation and migration. \par

\section{Modified Schelling Simulation}
\label{sec:Schelling}
\begin{figure}
	\centering
	\includegraphics[width=0.7\linewidth]{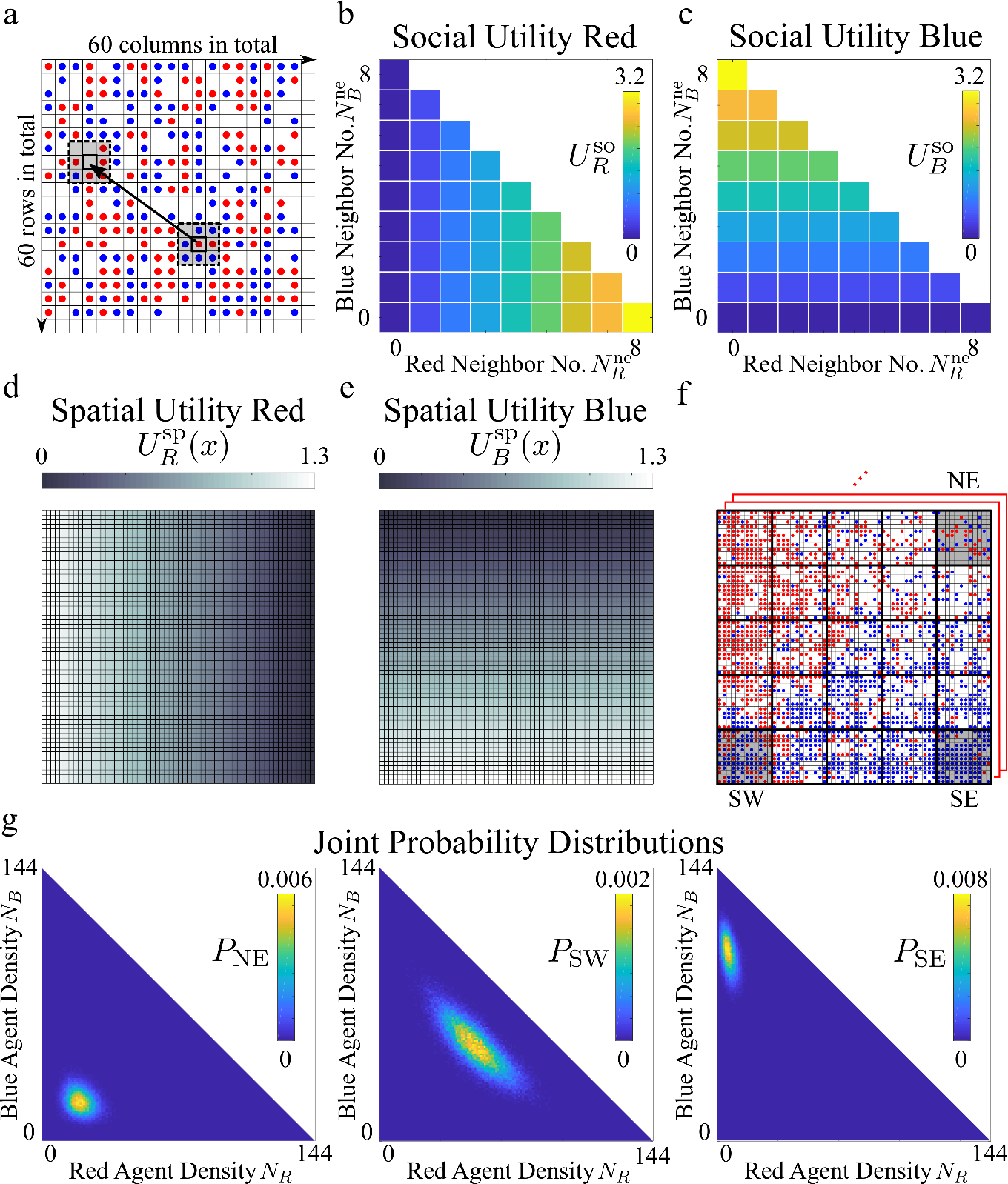}
	\singlespace \caption{\textbf{Extended Schelling-type simulation and steady-state data} \textbf{(a.)} Top-left corner of the Schelling lattice grid with 1000 red and 1000 blue agents. At each step in time, an agent and an empty cell are randomly chosen, and the agent will make probabilistic move to the empty cell. In this example, a red agent is chosen to move to a randomly chosen empty cell. The 8-connected neighborhood of the red agent and empty cell are shown as dashed boxes. \textbf{(b.)} Social Utility for red agents is defined by $U_R^\text{\text{so}}(N^\text{ne}_R,N^\text{ne}_B)=0.4\cdot N^\text{ne}_R$, where $N^\text{ne}_R$ and $N^\text{ne}_B$ are the number of 8-connected red and blue neighbors respectively. For the red agent in \textbf{a}, the change in social utility due to the proposed move is given by $U_{R}^{\text{so}}(5,1)-U_{R}^{\text{so}}(1,6)=+1.6$, making this move more socially attractive. \textbf{(c.)} Social Utility for blue agents is defined by $U_B^\text{\text{so}}(N^\text{ne}_R,N^\text{ne}_B)=0.4\cdot N^\text{ne}_B$. \textbf{(d.)} Spatial Utility for red agents $U^{\text{sp}}_{R}(x)$ is a function of location $x$, that decreases linearly in the horizontal direction. The change in spatial utility for the red agent in \textbf{a} for the proposed move is $\Delta U_{R}^{\text{sp}}\approx +0.17$, making this move more spatially attractive. So, according to Eq.~\eqref{eq:Schelling}, the red agent has a 85\% chance of moving. \textbf{(e.)} Spatial Utility for blue agents $U^{\text{sp}}_{B}(x)$ is a function of location $x$, that decreases linearly in the vertical direction.   \textbf{(f.)} A sample steady-state configuration of our simulation after reaching steady state ($>10000$ steps). We divide the Schelling lattice grid into 25 blocks with 144 sites each, three of which are shaded and labeled as 'NE', 'SW', and 'SE' for reference. To obtain a collection of steady-state configurations (red stack and red ellipses), we run an ensemble of Schelling simulations. \textbf{(g.)} From the collection of steady-state configurations, one can observe the steady-state joint probability distribution of observing a given agent densities for each block. Distributions for blocks 'NE', 'SW', 'SE' are shown.}
	\label{Schelling}
\end{figure}

To generate sample demographic data, we use a dynamic Schelling-type agent-based model \parencite{grauwin2012dynamic} modified to include spatial dependence. In this model, two type of agents, 1000 red and 1000 blue, make probabilistic moves to new empty cells on a 60-by-60 lattice grid with periodic boundary conditions (Fig.~\ref{Schelling}a). The moves are based on changes in utility functions that specify social (``Social Utility'' $U^\text{so}_R$, $U^\text{so}_B$) and spatial (``Spatial Utility'' $U^\text{sp}_R$, $U^\text{sp}_B$) preferences. In particular, at each step in time, we randomly choose an agent and an empty cell, and the agent will move to the empty cell with probability 

\begin{equation}
P_\text{Schelling}=
\begin{dcases}
\frac{1}{1+e^{-\Delta( U^\text{so}_R+U^\text{sp}_R)}} & \text{if agent is red}\\
\frac{1}{1+e^{-\Delta( U^\text{so}_B+ U^\text{sp}_B)}} & \text{if agent is blue}
\end{dcases},
\label{eq:Schelling}
\end{equation}
where $\Delta$ denotes the change in utilities due to the proposed move \parencite{mcfadden1973conditional} so that agents are more likely to move if the total utility increases. The social and spatial utility functions are defined in Figs.~\ref{Schelling}b,c and \ref{Schelling}d,e respectively. In this particular case, we define the social utility of an agent to linearly increase with the number of 8-connected neighbors (dashed box in Fig.~\ref{Schelling}a) that are of the same type. This dependence is illustrated by the homogeneous color of the columns and rows in Fig.~\ref{Schelling}b and \ref{Schelling}c respectively. We set the spatial utilities as shown in Figs.~\ref{Schelling}d,e, where red agents prefer the West side of the city and blue agents prefer the South side of the city. Additional results for more complex social utility functions are presented in SI Section S12. We use these simulations to generate the data throughout this paper. 

From these simulations we obtain coarse grained data of local agent densities and their steady-state joint probability distributions. In particular, we run an ensemble of Schelling simulations, wait until they reach a steady state, where the agent densities only fluctuate about some fixed distributions over time, and obtain the agent configuration from each simulation (See SI Section S3 for a discussion of obtaining an ensemble for real data). A sample steady-state configuration is shown in Fig.~\ref{Schelling}f. We coarse-grain the Schelling lattice grid into 25 blocks (outlined in the figure by thick lines) and record the total number of red agents $N_{R,b}$ and blue agents $N_{B,b}$ in each block $b$. Since all the blocks have the same area, $N_{R,b}$ and $N_{B,b}$ indicate local densities. By sampling the different steady-state configurations (denoted in the figure by the stack and ellipses) we measure the joint local probability distribution of agent densities for each block $P_b(N_{R,b},N_{B,b})$. For simplicity, we will abbreviate $N_{R,b}$ and $N_{B,b}$ as $N_R$ and $N_B$ when there is no ambiguity. We show the steady-state joint probability distributions for the North East (NE), South West (SW) and South East (SE) blocks in fig.~\ref{Schelling}g. We find that the SE block is likely to be occupied by a high density of blue agents, while the NE block is likely to be occupied by a low density of red and blue agents. We also find that the SW block is occupied by high densities of agents with a wide distribution of red and blue agent compositions. This wide distribution reflects the inherent biases for red and blue agents to segregate.  

\section{2-Component Density-Functional Fluctuation Theory}\label{sec:DFFT}
Single-component Density-Functional Fluctuation Theory conjectures that, by observing the steady-state probability distribution of a single type of agent in a block, one can extract information about the location-dependent preferences and social interactions of the agent \parencite{mendez2018density}. In particular, by observing the means of the distributions, it is possible to rank the agent preference for each block. Further, the shapes of the distributions provide information about social preference. For example, a Poisson-like distribution indicates no social interactions, a narrowly peaked distribution indicates strong repulsion, and a bimodal distribution indicates strong attractive interactions. As such, Méndez-Valderrama et al. (2018) write the block-dependent steady-state probability distribution as:
\begin{equation}
    P_b(N) = z_b^{-1}\frac{1}{N!}\exp[-v_{b}N-f(N)],\label{eq:DFFT_Single}
\end{equation}
where $z_b^{-1}$ is a normalization constant; $v_{b}$ is defined as the ``vexation'' and is constant for each block $b$; and $f(N)$ is defined as the ``frustration'' and is block-independent function of local densities. Since all the blocks have the same area (See SI for the general case), the number of agents $N$ indicates the density. When $f(N)$ is zero, the distribution is Poisson and the mean is proportional to $\exp[-v_b]$, indicating agents avoid blocks with high $v_b$. The deviation of the distribution from Poisson is captured by the function $f(N)$ that depends only on the density of agents. When $f(N)$ is concave up (e.g. $P_b$ is narrowly peaked), agents disperse or segregate. When $f(N)$ is concave down (e.g. $P_b$ is bimodal), agents aggregate. Thus, frustration and vexations respectively capture effective social and spatial interactions at the coarse-grained block scale. In previous work, it was shown that this functional form for the probability distribution is remarkably accurate for data on crowd distributions in not only model, but also living systems \parencite{mendez2018density}. 

To extend this theory to multiple agents, we use a multivariate distribution to describe the block dependent vexations for all types of agent. Additionally, the frustration capturing the interactions between all the agents becomes a joint function of the density of each type of agent. For the case of two types of agents (Red and Blue) discussed in the present work, DFFT conjectures that the steady-state joint probability distribution in each block $b$ is given by:
\begin{equation}
    P_b(N_R,N_B) = z_b^{-1}\frac{1}{N_R!N_B!(s-N_R-N_B)!}\exp[-v_{R,b}N_R-v_{B,b}N_B-f\left(N_R,N_B\right)],\label{eq:DFFT_form}
\end{equation}
where $z_b^{-1}$ is, again, a normalization constant; $v_{R,b}$ and $v_{B,b}$ are the block-dependent vexations for the Red and Blue agents; $s$ is the total number of sites within a block; and $f(N_R,N_B)$ is the block-independent frustration and is a function of the local densities of Red and Blue agents. As before, agents avoid blocks with high vexation. The frustration now captures the social interaction between two types of agents. Therefore, instead of a single curve that depends on the number of agents, the frustration becomes a surface that depends on the density of blue and the density of red agents. The concavities of curves on this surface indicates the social preferences for having greater or fewer agents of a particular type (See SI Section S4 for detailed interpretations of frustration and vexation). Thus, this frustration function is a functional measure of segregation and can capture a variety of segregation behaviors (SI Section S5). Finally, the term $1/(s-N_R-N_B)!$ is introduced to better account for the fact that each block in our system can fit a maximum density of $s=144$ agents (See SI Sections S1.1, S1.2 and S1.5 for a detailed derivation of Eq.~\eqref{eq:DFFT_form}). 

\begin{figure}
	\centering
	\includegraphics[width=0.9\linewidth]{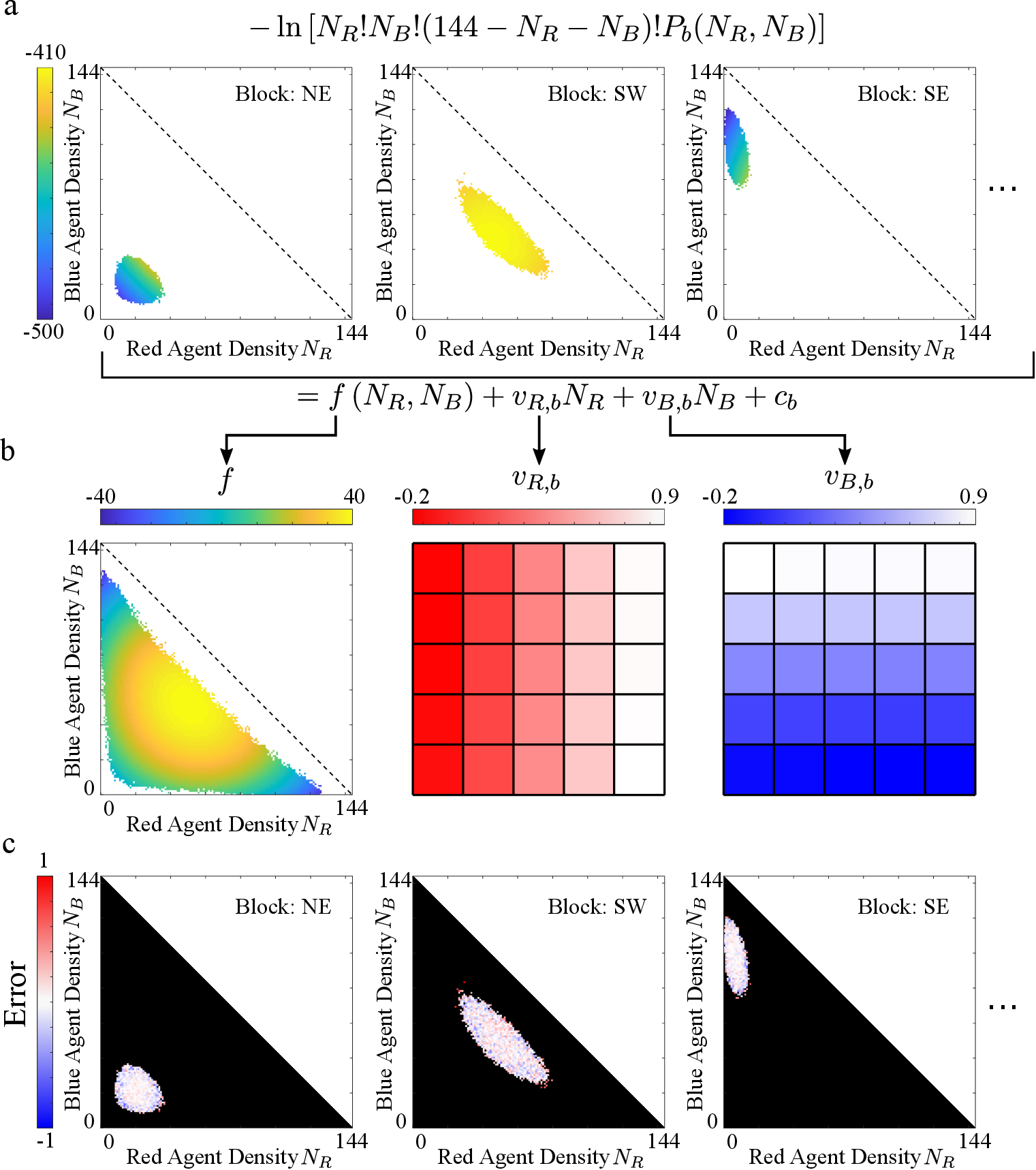}
	\singlespace \caption{\textbf{Extracting effective social and spatial preferences} \textbf{(a.)} The LHS of Eq.~\eqref{eq:H} is determined by our observed probability $P_b$ (Fig.~\ref{Schelling}g), and is plotted for blocks 'NE', 'SW', 'SE'. We only keep data for cases where more than 10 observations are recorded for a particular agent combination. \textbf{(b.)} Using Maximum Likelihood Estimation, we fit each of the 25 LHS surfaces by a block-independent surface called ``frustration'' together with a block-dependent planar shift $v_{R,b}N_R+v_{B,b}N_B+c_b$, where $v_{R,b}$ and $v_{B,b}$ are two block-dependent constants called ``vexations'', and $c_b$ is a block-dependent normalization constant. Frustration describes social preference, while vexations describe spatial preference. \textbf{(c.)} The errors of the fit in \textbf{b} are determined by the difference between the right hand and left hand sides of Eq.~\eqref{eq:H}, for the NE, SW, and SE blocks.}
	\label{Extract}
\end{figure}

To check if Eq.~\eqref{eq:DFFT_form} can indeed be used to fit the Schelling model steady-state data (Fig.~\ref{Schelling}g), we first rearrange Eq.~\eqref{eq:DFFT_form} to obtain
\begin{equation}
    -\ln[N_R!N_B!(s-N_R-N_B)!P_b(N_R,N_B)]=f\left(N_R,N_B\right)+v_{R,b}N_R+v_{B,b}N_B+c_b\label{eq:H},
\end{equation}
where $c_b=\ln(z_b)$ is a normalization constant. The left-hand side (abbr.~LHS) of Eq.~\eqref{eq:H} is determined by our observed probability $P_b$ (Fig.~\ref{Schelling}g), and is plotted for three sample blocks in Fig.~\ref{Extract}a. Next, we use a Maximum Likelihood Estimation algorithm to infer the frustration and vexations that best fit the data and plot these in Fig.~\ref{Extract}b. We find that the fits are remarkably accurate as illustrated by the small errors (differences between left-hand side and right-hand side of Eq.~\eqref{eq:H}) shown for three sample blocks in Fig.~\ref{Extract}c. When comparing the observed joint probability distributions (Fig.~\ref{Schelling}g) with the distributions modeled by Eq.~\eqref{eq:DFFT_form} using the extracted frustration and vexations (Fig.~\ref{Extract}b), we observe a mean absolute percentage error of 14\% for joint densities with at least 10 observations. The extracted frustration and vexations can then be used to predict how populations will redistribute in response to demographic changes, as we show next. 

\section{Predicting Time Evolution}\label{sec:TimeDep}
To generate a demographic change in the simulation data, we abruptly switch 350 randomly chosen red agents on the north side of the Schelling lattice into blue agents. We then record the evolution of an ensemble of agent configurations (illustrated by the stack and ellipses) as it transitions from this new altered state at $t=0$ to the new steady state at $t\rightarrow\infty$ (Fig.~\ref{Dynamic}a). As before, we coarse grain these data at the block level to extract the density of red and blue agents at each time. The challenge is to predict the evolution of these neighborhood-scale data using DFFT parameters extracted from the initial steady state data and knowledge of the regional-scale demographic change. Note that an abrupt regional-scale demographic change should result in a more extreme neighborhood-scale time evolution, and therefore be harder to predict than a more realistic continuous demographic change over time.

\subsection{Time-Dependent DFFT model (Kohn-Sham TD-DFFT)}\label{sec:TD-DFFT}
To predict the evolving joint density distributions of the ensemble described above, we construct a Time-Dependent DFFT model in which agents choose to move from block to block based on changes in the coarse-grained spatial and social preferences. The coarse-grained preferences can be combined into block-level ``Headache'' functions: 
\begin{equation}
    H_b(N_R,N_B)=v_{R,b}N_R+v_{B,b}N_B+f(N_R,N_B).\label{eq:Hdef}
\end{equation}
Specifically, at every step in time, we choose an agent randomly and choose a block with a weight proportional to the amount of empty spaces it has. The agent then moves from its current block $b$ to the chosen block $b'$ with probability:
\begin{equation}
P_{b\to b'}=\frac{1}{1+e^{\Delta H_{b}+\Delta H_{b'}}},
\label{eq:coarseDFFTagent}
\end{equation}
where $\Delta$ denotes the change due to the proposed move. By evolving the ensemble of altered states according to the above rule, we can numerically predict the evolution of the joint density distributions, up to a constant-time scale difference (See SI Section S6.3 for reasons of introducing a time scale). We can extract the time scale easily by comparing the rate of steady-state fluctuations in the Schelling simulation and the TD-DFFT model. This TD-DFFT model corresponds to the adiabatic approximation of Kohn-Sham Time-Dependent Density-functional Theory \parencite{kohn1965self,runge1984density,thiele2008adiabatic} (See SI Section S7.1), whose predictions can be determined exactly through a master equation description as well (See SI Section S6.2).\par

\subsection{Mean Value Equation (Hohenberg-Kohn TD-DFFT)}
For situations where the full analysis presented above is too computationally expensive (when the numbers of blocks and agents are high), we develop a simplified mean value approach. In particular, to obtain a good distribution of states, the above agent-based simulation approach to solving Kohn-Sham TD-DFFT model requires that we evolve an ensemble with a size much bigger than the number of frequently observed states. Since the number of states scales steeply with the number of blocks and agents, the Kohn-Sham TD-DFFT model may be computationally expensive to run for real cities with hundreds of blocks and millions of people. When the probability distributions are single-peaked, as is the case here, one may approximate the time evolution of the joint mean densities of the above model according to the following Mean Value Equation (MVE, See derivation in SI Section S6.2):
\begin{equation}
\left\{
\begin{aligned}
&\frac{\text{d}}{\text{d} t}\overline{N_{R,b}}=\sum_{b'\ne b}\nu_{R,b'\to b}-\nu_{R,b\to b'}\\
&\frac{\text{d}}{\text{d} t}\overline{N_{B,b}}=\sum_{b'\ne b}\nu_{B,b'\to b}-\nu_{B,b\to b'}
\end{aligned},
\right.
\label{eq:ODE}
\end{equation}
where $\overline{N_{R,b}}$ and $\overline{N_{R,b}}$ denotes the mean red and blue agent density for block $b$, respectively. $\nu_{R,b\to b'}$ and $\nu_{B,b\to b'}$ are the density flow rates for red and blue agents from block $b$ to block $b'$, calculated from
\begin{equation}
\left\{
\begin{aligned}
&\nu_{R,b\to b'}\approx\frac{\overline{N_{R,b}}}{N_\text{tot}}\cdot \frac{s-\overline{N_{R,b'}}-\overline{N_{B,b'}}}{s_\text{tot}-N_\text{tot}}\cdot P_{b\to b'}\\
&\nu_{B,b\to b'}\approx\frac{\overline{N_{B,b}}}{N_\text{tot}}\cdot \frac{s-\overline{N_{R,b'}}-\overline{N_{B,b'}}}{s_\text{tot}-N_\text{tot}}\cdot P_{b\to b'}
\end{aligned}
\right..
\label{eq:Nu}
\end{equation}
Eq.~\eqref{eq:ODE} says that the rate of change in the mean density of agents in block $b$ is given by the sum of inflow rates from all other blocks $b'\ne b$ into block $b$, minus the sum of outflow rates from block $b$ into all other blocks $b'\ne b$. We can also easily interpret the flow rate approximations in Eq.~\eqref{eq:Nu} following the rules of the TD-DFFT Model (Section \ref{sec:TD-DFFT}): The first term in the product represents the probability of choosing the corresponding type of agent in block $b$, where the denominator $N_\text{tot}=2000$ is the total number of agents in the city; The second term in the product represents the probability of choosing an empty cell in block $b'$, where the denominator $s_\text{tot}-N_\text{tot}=3600-2000=1600$ is the total number of empty cells in the city; The third term corresponds to the probability of transition defined in Eq.~\eqref{eq:coarseDFFTagent}. Note that this MVE might fail to capture the behavior of the TD-DFFT Model with extreme segregation, when Eq.~\eqref{eq:ODE} exhibits `bifurcation behavior' \parencite{weidlich2012concepts,haag2017modelling} (See SI Section S8). Since the MVE deals with average numbers in each block, it corresponds more closely to a Hohenberg-Kohn \parencite{hohenberg1964inhomogeneous} TD-DFT (See SI Section S7.2). \par

\subsection{Results}
\begin{figure}
	\centering
	\includegraphics[width=0.8\linewidth]{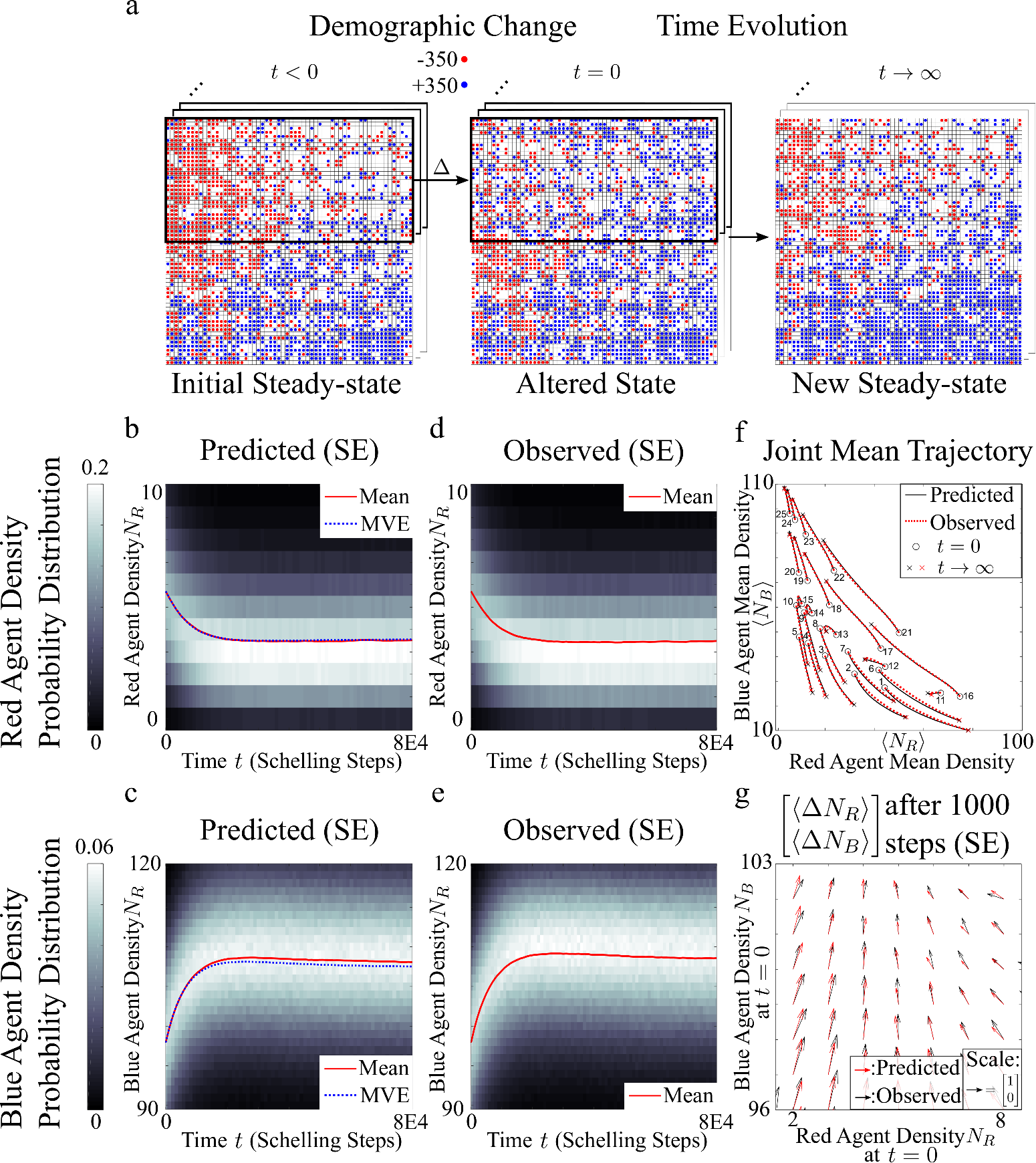}
	\caption{ \textbf{Predicting time evolution} \textbf{(a.)} Starting from a steady-state configuration, at $t=0$, we introduce our demographic change by abruptly switching 350 randomly chosen red agents on the north side of the Schelling lattice into blue agents to obtain an altered state. The system is then evolved according to the Schelling model. The above procedure is repeated over an ensemble of Schelling simulations (shown as stacks and ellipses). \textbf{(b.,c.)} Predicted time evolution of the probability distribution for observing red or blue agents for the 'SE' block using the TD-DFFT model. Note that the distribution for red agent is skewed away from the mean towards more segregated values. The MVE predictions agree well with the TD-DFFT Model. \textbf{(d.,e.)} Observed time evolution of the probability distribution for observing red or blue agents for the 'SE' block from the Schelling simulation. \textbf{(f.)} Observed versus predicted Joint-mean density trajectories for all blocks (counted left-to-right then top-to-bottom, in normal English reading order). Blocks 13-15 show interesting trajectories, which TD-DFFT model predict well. \textbf{(g.)} Observed versus predicted average changes in number of agents in block 'SE' after 1000 Schelling steps for various initial number of agents. We note that a calibration factor is necessary to match the time scales between the density based model predictions and the Schelling simulation time steps.}
	\label{Dynamic}
\end{figure}

Overall, we obtain excellent agreement between the TD-DFFT model/MVE predictions and the simulated Schelling data. As an example, we compare the predicted (Figs.~\ref{Dynamic}b,c) and observed (Figs.~\ref{Dynamic}d,e) time evolution of the probability distribution of red ($N_R$) and blue ($N_B$) agents for the South East (SE) block (and all other blocks in SI Section S9). We find that the model accurately captures the trends in the means as well as the skews in the distributions about the means. In particular, the MVE (blue dotted lines in Figs.~\ref{Dynamic}b,c) accurately tracks the mean values of the TD-DFFT model (red lines in Figs.~\ref{Dynamic}b,c). Additionally, we compare the trajectory of the joint means for all the blocks in Fig.~\ref{Dynamic}f. Once again, we find excellent agreement throughout the entire trajectory even when the evolution is non-monotonic as is the case for blocks 12-15. Finally, given an initial joint density ($N_R$,$N_B$) for a particular block we are able to predict the joint probability distribution for the block after 1000 Schelling steps. We plot the observed average change in the joint density for each initial condition for the SE block in Fig.~\ref{Dynamic}g. These predictions, indicated by the red arrows, are compared with the observed data, indicated by the black arrows. Once again, we observe excellent agreement between the predictions of the TD-DFFT model and the demographic data. The arrow directions reflect the constraints induced by local interactions over short time scales, despite the fact that the long time equilibrium of the joint probability distribution resides in the upper left region of the plot. It is this capacity to model the step-wise evolution that allows the TD-DFFT model to accurately track the time dependent trajectories of the Joint Means as shown in Fig.~\ref{Dynamic}f. Finally, we have conducted similar studies involving the same demographic change for cases where the demographic data is generated using more complicated Social and Spatial utility functions and have obtained TD-DFFT predictions of similar fidelity.\par   

In part, the reason the Time-dependent DFFT model is able to accurately describe the evolution of the Schelling data after a demographic change, is that it is itself a type of agent-based model. There are, however, a number of important distinctions. First, the TD-DFFT model relies on coarse-grained data. Details of the Schelling simulation such as the lattice grid structure, the 8-connected neighbors, and empty spots are averaged over to obtain the density in each block. As such the TD-DFFT model keeps only the essential information necessary to make predictions about the density. Second, the TD-DFFT model relies on empirically extracted parameters. As such it does not require that we impose specific rules governing the system evolution. Instead, it determines the rules from the original steady-state data in order to make predictions. These distinctions will lead to discrepancies in certain extreme conditions. For instance, when block-size is very small (e.g. 4 by 4 cells), the interaction of agents between neighboring blocks will greatly affect the dynamics. In the TD-DFFT model, however, such effects are ignored in the coarse-graining procedure, resulting in different dynamics (See SI, Section S10).    \par     

\section{Predicting the New Steady-state}\label{sec:NewSteadyState}
\begin{figure}
	\centering
	\includegraphics[width=0.9\linewidth]{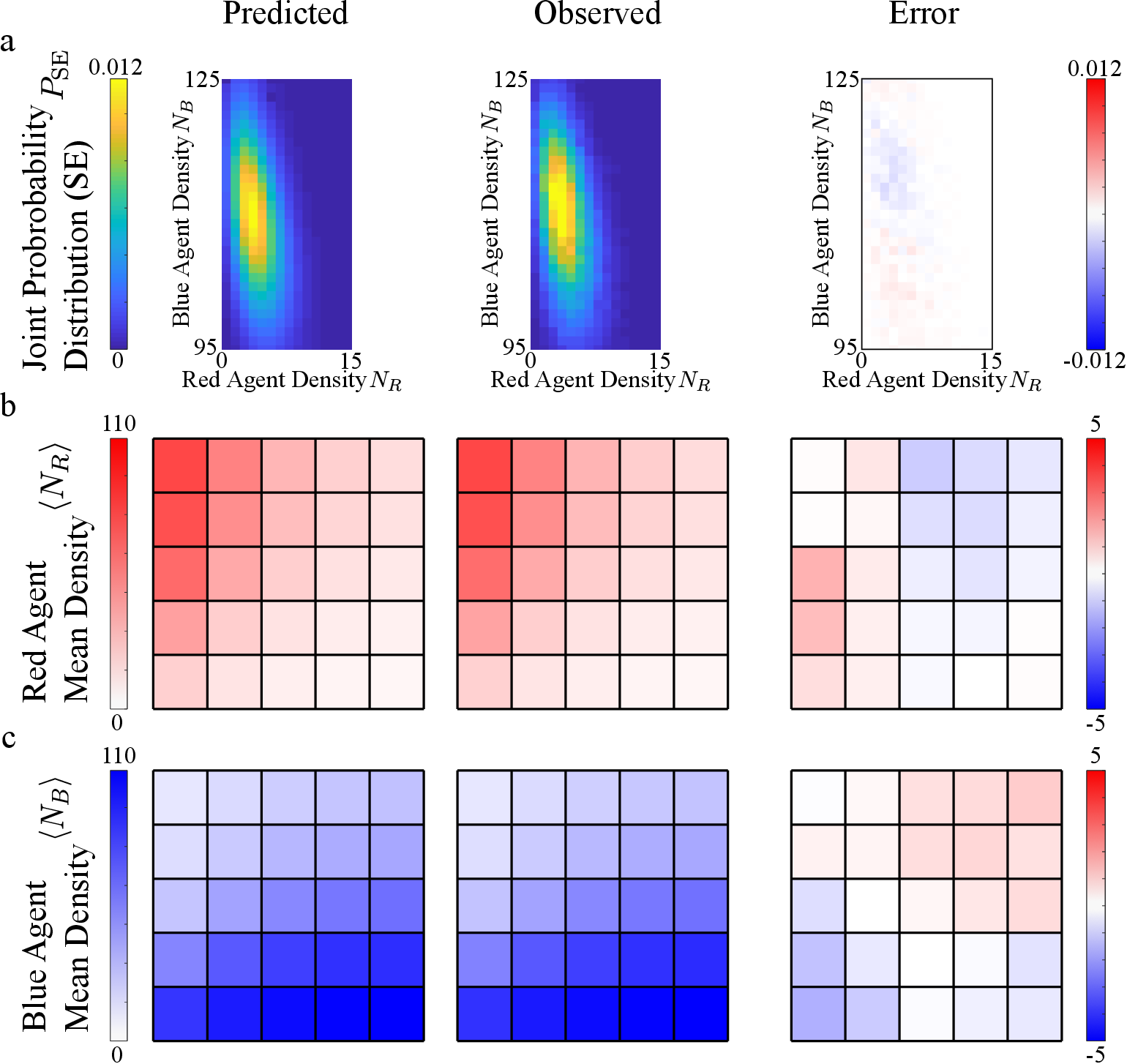}
	\caption{\textbf{Analytically predict new steady state} \textbf{(a.)} Predicted versus observed new steady-state joint probability distribution for block 'SE'. \textbf{(b.)} Predicted versus observed mean densities of red agents for all blocks. \textbf{(c.)} Predicted versus observed mean densities of blue agents for all blocks.}
	\label{Steadystate}
\end{figure}

To predict the new steady-state joint probability distributions, $P_b(N_R,N_B,t\rightarrow\infty)$ we can either run the TD-DFFT model until it reaches a steady state, or calculate the new distribution analytically. To analytically predict the new distribution, we take advantage of the fact that the social and spatial preferences of the individuals remain the same throughout this demographic change in the total number of each agent. Since these social and spatial preferences manifest themselves at the coarse grained level as differences in the headache functions of two blocks, $\Delta H_b + \Delta H_b'$, we can rewrite the headache function as:
\begin{equation}
        H_b \longrightarrow H_b-\mu_RN_R-\mu_BN_B
        \label{eq:AgentPot}
\end{equation}
without affecting these preferences. Here, $\mu_R$ and $\mu_B$ are block-independent constants called `agent potentials' (analogous to chemical potentials in statistical physics) that tune the expected total number of each type of agent over all the blocks (see SI, Sections S1.2 and S1.4). To determine these constants, we modify the exponent in Eq.~\ref{eq:DFFT_form} and use Newton's method to converge on values for $\mu_R$ and $\mu_B$ so that the means of the probability distribution, when averaged over all the blocks, equals the new mean for the entire system resulting from the demographic change. We compare the predictions for the joint probability distributions, as well as the mean of the agent densities for each block with the observed data from the Schelling model in Fig.~\ref{Steadystate} and SI Section S9. We find excellent agreement between the predictions and the simulated demographic data. Importantly, this analytic approach arrives at the same new steady state distribution that the previously described time dependent DFFT model predicts, but with orders of magnitude increase in computational speed and without the need for modelling dynamic behaviors.

\section{Implications and Future Directions}
The ability of DFFT to accurately predict demographic changes in the Schelling model suggests that it may improve upon current methods for small-area demographic forecasting. Unlike bottom-up agent-based approaches that postulate specific rules, DFFT empirically extracts these rules from observations. Additionally, in contrast to top-down data-driven approaches that only extract descriptive measures of segregation, DFFT uses the more detailed extracted rules to forecast population dynamics.

Importantly, this framework could easily be extended to include an even greater number of agent types. As long as residential choice is driven separately by the composition of local neighborhoods and the coarse agent spatial preferences, it is straightforward to include vexations for each agent type and a multidimensional frustration that accounts for the multi-agent social interactions.

We also anticipate a pathway towards modeling other demographic changes such as changes in social or spatial preferences. In these particular cases it is necessary to know how to map such changes to the DFFT functions extracted from the initial steady state data. Such a mapping can be achieved in a number of ways. First, it may be possible to directly guess the change in the DFFT functions. For example, if we knew that the spatial utility functions for the agents were exchanged, we could use knowledge from previous simulations to determine the new vexations: $v_B \rightarrow v_R$ and $v_R \rightarrow v_B$. Second, even if we had no \emph{a priori} knowledge of this change, we could still use intermediate time points in the evolution of the demographic data to adjust the vexations and frustration. For example, after a certain number of Schelling time steps we would modify our DFFT functions so that we get the best agreement between the dynamics of the predicted and observed density changes for the agents (Fig.~\ref{Dynamic}g). These examples illustrate the flexibility that DFFT provides for making predictions for a broad range of demographic changes within the context of the Schelling model.     

% Should be keep this paragraph?
%The ability to recast the Schelling model into the DFFT framework suggests that it should be possible to extend this approach to other agent based models. In the field of ecological demography, DFFT could provide a new description for predator-prey dynamics in the stochastic spatial Lotka-Volterra model \parencite{dobramysl2018stochastic}. In the fields of evolutionary biology or economics, DFFT could give new insights into the success of different strategies such as cooperation by coarse graining stochastic game models \parencite{wu2018environment,nowak1992evolutionary}. As long as the agents in a given system are driven separately by local interactions with their neighbors and spatial interactions, we expect the DFFT analysis presented here will make accurate predictions of the evolving system.

Finally, to the extent that agent-based approaches like the Schelling model inform trends in demographic patterns, it may be possible to apply DFFT to real data. For example, decennial American census data provides block level counts of the number of people by race and ethnicity. Given these data, DFFT could be used to measure the frustration between different races or ethnicities and their vexations throughout the country. Then, likely in conjunction with current methods of small-area population forecasts \parencite{wilson2021methods} using model averaging \parencite{rayer2010factors}, DFFT could potentially generate improved forecasts of demographic changes at the neighborhood-scale. 
%As such DFFT could serve as a detailed lens into the social and spatial nature of racial residential segregation and be a powerful tool for forecasting the composition of American neighborhoods.   \par 

\section{Code Availability}
Code for performing simulations and performing statistical analysis are available at \url{https://github.com/yunuskink/DFFT-Schelling-model}.

\section{Authors' Contributions}

T.A.A. developed initial theoretical extensions to DFFT work including multi-component and time-dependent systems. Y.A.K. and I.C. proposed initial application of DFFT onto demographic systems. M.H. provided context of proposed methods within the broader field of demography. Y.C., Y.A.K., B.B., T.A.A. and I.C. developed and refined multi-component and time-dependent applications of DFFT onto simulated data. Y.C. implemented all simulations, performed all analyses of data, and created all figures. Y.A.K. wrote code for statistical extraction of parameters from data. Y.C., Y.A.K., T.A.A. and I.C. wrote the manuscript with all authors contributing. Y.C., T.A.A and B.B. wrote the SI with all authors contributing. I.C. mentored Y.C. and Y.A.K.. T.A.A. mentored Y.C. and B.B..

\section{Acknowledgements}
The authors thank The Cohen and Arias groups for helpful discussions throughout this work. The work was funded by NSF BRAIN EAGER 1546710, ARO W911NF-18-1-0032, and NIH R01NS116595-01 grants. Y.K. was also supported in part by funding from the National Science Foundation Graduate Research Fellowship Award No. DGE-1650441. B.B. was supported in part by funding from the Postgraduate Scholarship-Doctoral (PGS D) from the Natural Sciences and Engineering Research Council of Canada (NSERC).

\pagebreak

\pagebreak

\pagebreak

\pagebreak

\pagebreak

\pagebreak

%\doublespace
\pagebreak
\printbibliography[title={References}]

\pagebreak

%\end{document}

\begin{table}[ht!]
\singlespace \caption{List of Important Variables Used} 
\centering 
\begin{tabular}{l l l} 
\hline\hline 
Variable & Meaning & Usage \\ [0.5ex]
\hline 
$P_\text{Schelling}$ & Probability of moving for each Schelling step&\\ [0.5ex]
$U^\text{so}_R$ & Social Utility for a red agent& \multirow{8}{4em}{Schelling}  \\  [0.5ex]

$U^\text{so}_B$ & Social Utility for a blue agent&\\ [0.5ex]
$U^\text{sp}_R$ & Spatial Utility for a red agent&\\ [0.5ex]
$U^\text{sp}_B$ & Spatial Utility for a blue agent&\\ [0.5ex]
$x$ & Position in Schelling city &\\ [0.5ex]
$N_R^\text{ne}$ & Number of red agents in the 8-connected neighborhood &\\ [0.5ex]
$N_B^\text{ne}$ & Number of blue agents in the 8-connected neighborhood &\\ [0.5ex]
\hline 

$b$ & Block index & \multirow{12}{4em}{Schelling \& DFFT}\\ [0.5ex]
$t$ & Time &\\ [0.5ex]
$s$ & Number of cells (spaces) in a block& \\ [0.5ex]
$s_\text{tot}$ & Total number of cells (spaces) in a city &\\ [0.5ex]
$N_\text{tot}$ & Total number of agents in a city &\\ [0.5ex]
$P_b$ & Probability Distribution of agents in block $b$      &\\ [0.5ex]
$N_{R,b}$ & Number (Density) of red agents in block $b$    & \\ [0.5ex]
$N_{R}$ &  Abbreviated $N_{R,b}$ when there is no ambiguity   &\\ [0.5ex]
$N_{B,b}$ & Number (Density) of blue agents in block $b$     &\\ [0.5ex]
$N_{B}$ & Abbreviated $N_{B,b}$ when there is no ambiguity &\\ [0.5ex]
\hline 

$z_b$ & Normalization constant for $P_b$     &\multirow{12}{4em}{DFFT}\\ [0.5ex]
$v_{R,b}$ & Vexation for red agents in block $b$    &\\ [0.5ex]
$v_{B,b}$ &  Vexation for blue agents in block $b$    & \\ [0.5ex]
$f$ & Frustration& \\ [0.5ex]
$H_b$ & Headache function for block $b$& \\ [0.5ex]
$P_{b\to b'} $ & Probability of transition of an agent from block $b$ to $b'$& \\ [0.5ex]

$\nu_{R,b\to b'}$ & Number (Density) flow rate for red agent from block $b$ to $b'$ &\\ [0.5ex]
$\nu_{B,b\to b'}$ & Number (Density) flow rate for blue agent from block $b$ to $b'$ &\\ [0.5ex]
$\mu_R$ & Red agent potential in a city &\\ [0.5ex]
$\mu_B$ & Blue agent potential in a city &\\ [0.5ex]
\hline
\end{tabular}
\end{table}

\renewcommand{\theequation}{S\arabic{equation}}
\renewcommand{\thefigure}{S\arabic{figure}}
\renewcommand{\thetable}{S\arabic{table}}
\renewcommand{\thesection}{S\arabic{section}}

\setcounter{equation}{0}
\setcounter{figure}{0}
\setcounter{table}{0}
\setcounter{section}{0}

\newpage

\begin{titlepage}
\vspace*{\fill}
\begin{center}
  \Huge{Supplemental Information}
\end{center}
\vspace*{\fill}
\end{titlepage}

% \title{Supplemental Information}
% \maketitle
\newpage

\section{Multi-component DFFT Framework}
\label{sec:underlyingDFFT}

In this section, we describe the underlying migration model central to the DFFT framework, mostly extending the model presented by Mendez et. al (2018) to the context of two types of agents. In particular, we will derive Eq. (3) of the main text from the underlying migration model, extend the DFFT function extraction method discussed in the main text, provide more rigorous reasoning behind the ``New Steady-state'' analytical prediction, and lay the conceptual groundwork and notation for future sections within this Supplemental Information. Extension beyond two agent types is straightforward.

\subsection{Underlying Migration Model and its Steady-State Distribution}\label{sec:Under}
One path towards deriving our DFFT equations is through the coarse-graining of a generic agent-based model, referred to as the underlying migration model. 

\subsubsection{Underlying Migration Model}
Consider a dynamically changing city where two types of agents, red and blue, frequently propose to move to new locations. Specifically, at each step in time, an agent at a particular location $x'$ proposes to move to a new location $x''$. The agent will accept the move from $x'$ to $x''$ with probability
\begin{equation}
\frac{1}{1+e^{h(x'')-h(x')}},\label{eq:DensityP}
\end{equation}
where $h(x)$ is the agent's dissatisfaction at location $x$. In other words, agents are unlikely to move to locations with a significantly higher dissatisfaction value. We further split the dissatisfaction, $h$, into social and spatial contributions, defined separately for red and blue agents:  
\begin{equation}
    h(x)\equiv\begin{cases}
    f_R(n_R(x),n_B(x))+V_R(x) & \text{if agent is red}\\
    f_B(n_R(x),n_B(x))+V_B(x) & \text{if agent is blue}
\end{cases}.
\end{equation}
We shall refer to the social contributions, $f_R$ or $f_B$, as the ``frustration'' functions. These functions characterize the social dissatisfaction felt by the agent due to red agent density $n_R(x)$ and blue agent density $n_B(x)$ at location $x$.  We shall refer to the spatial contributions, $V_R$ or $V_B$, ``vexation'' constants. These constants characterize the spatial dissatisfaction felt by an agent due solely to being at location $x$, independent of the densities of agents. The above considerations give a complete agent-based model of a city once we specify the functions $f_R$, $V_R$, $f_B$, $V_B$, the total number of red $N_R^\text{tot}$ and blue agents $N_B^{\text{tot}}$, the geometry of the city (which we take to have total area $A$), and the process by which $x'$ and $x''$ are chosen.

It is important to note that the process by which $x'$ and $x''$ are selected, namely the rules by which agents propose moves (abbr. \emph{agent rule}), affects the overall behavior of the system and can be modified to better fit different scenarios. For instance, instead of randomly chosen agents proposing to move to uniformly randomly chosen locations, agents might propose to move to locations with more empty housings at a higher rate. Such a version better matches the Schelling model and is used in the main text. Alternatively, agents in a certain neighborhood might propose to move to nearby neighborhoods more frequently than more distant neighborhoods, giving rise to a gravity-like model. Such a choice leads to a model that is nearly identical to the Weidlich-Haag Migratory Model (see chapter 8 of Ref.~\parencite{haag2017modelling} for a detailed description). We will consider such agent rules and explore how they change the DFFT framework in Section \ref{sec:version}.

Application of the above models to predict migration requires reasonable estimates of the frustrations and vexations. 
We prefer to extract these functions from empirical steady-state data, without making detailed assumptions on the functional form of the frustration. Therefore, we shall first calculate a generic expression for the steady-state probability distribution of the above model. 

Observe that the functional form of the probability in Eq.~\eqref{eq:DensityP} corresponds exactly to the Metropolis-Barker algorithm \parencite{metropolis1953equation,barker1965monte} to draw samples from a Boltzmann distribution, provided that we can re-express the change in dissatisfaction for a single agent $\Delta h=h(x'')-h(x')$ as a change in `energy' between two `states' of the system. We identify states by the pair of densities $(n_R(x),n_B(x))$. The energy of each state can thus be represented as a density-functional $H[n_R(x),n_B(x)]$. To find the functional $H$ such that $\Delta H\equiv\Delta h$, let us first look at the state of lowest dissatisfaction (the `ground state') for which the variation $\delta H$ (the differential of the functional $H$) vanishes for infinitesimal changes in densities $[\delta n_R,\delta n_B]$ that keep the total numbers of agents fixed. At ground-state densities, we therefore expect $\Delta h$ to vanish for any proposed move and $h$ to be uniform for red and blue agents so that no move can lower the dissatisfaction in the system. This gives the condition:
\begin{equation}
    \left\{\begin{aligned}
    f_R(n_R(x),n_B(x))+V_R(x)=\mu_R \\
    f_B(n_R(x),n_B(x))+V_B(x)=\mu_B 
    \end{aligned}\right.,\label{eq:LME}
\end{equation}
where $\mu_R$ and $\mu_B$ are constants. These equations correspond to the Lagrange-multiplier equations for minimization of a candidate density-functional:
\begin{equation}
    H\left[n_R(x),n_B(x)\right]\equiv\int_A f\left(n_R(x),n_B(x)\right) \text{d} x+\int_A V_R(x)\cdot n_R(x) \text{d} x+\int_A V_B(x)\cdot n_B(x) \text{d} x,\label{eq:defH}
\end{equation}
subject to the constraints that the total numbers of agents are fixed:
\begin{equation}
    \left\{\begin{aligned}
    \int_An_R(x)\text{d}x=N_R^\text{tot} \\
    \int_An_B(x)\text{d}x=N_B^\text{tot} 
    \end{aligned}\right..\label{eq:constraints}
\end{equation}
Specifically, the parameters $\mu_R$ and $\mu_B$ in Eq.~\eqref{eq:LME} correspond to the Lagrange-multipliers for each constraint in Eq.~\eqref{eq:constraints}. Note that $f(n_R,n_B)$ in Eq.~\eqref{eq:defH} is called the ``global frustration'' which is a function of joint densities $(n_R,n_B)$ and satisfies both $\sfrac{\partial f}{\partial n_R}=f_R$ and $\sfrac{\partial f}{\partial n_B}=f_B$.\footnote{In general such a function $f$ may not exist. In the case where $f_R$ and $f_B$ are differentiable, $f$ exists if and only if $\sfrac{\partial f_R}{\partial n_B}=\sfrac{\partial f_B}{\partial n_R}$. So the global description is not as general as the underlying migration model. This additional requirement on the functional form of $f_R$ and $f_B$ ensures that change in dissatisfaction only depends on the initial and final states, and independent of the path.}  We can further show\footnote{Suppose a red agent considers a move from location $x'$ to $x''$, which corresponds to a change in the final density $n''_R(x)$ from the initial density $n'_R(x)$, while $n''_B(x)=n'_B(x)$ remains fixed. We show that
$H[n''_R(x),n''_B(x)]-H[n'_R(x),n'_B(x)]\approx h_R(x'')-h_R(x')$. We first approximate the change in $H$ with the variation $\delta H$:
\begin{multline}
    H[n''_R(x),n''_B(x)]-H[n'_R(x),n'_B(x)]\approx \delta H\\
    \equiv \int_A \delta n_R(x)\cdot \left[\frac{\partial f}{\partial n_R}(n'_R(x),n'_B(x))+V_R(x)\right] \text{d} x+\int_A \delta n_B(x)\cdot\left[\frac{\partial f}{\partial n_B}(n'_R(x),n'_B(x))+V_R(x)\right] \text{d} x.
    \label{eq:deltaH}
\end{multline}
where the changes in densities are defined and approximated by 
\begin{equation}
\left\{
    \begin{aligned}
        &\delta n_R(x)=n''_R(x)-n'_R(x)\approx \delta(x-x'')-\delta(x-x')\\
        &\delta n_B(x)=n''_B(x)-n'_B(x)=0
    \end{aligned}
\right..\label{eq:deltan}
\end{equation}
In Eq.~\eqref{eq:deltan}, the change in red agent density $\delta n_R(x)$ is approximated with Dirac delta functions centered at $x''$ and $x'$. Substituting Eq.~\eqref{eq:deltan} into Eq.~\eqref{eq:deltaH}, and identifying $\sfrac{\partial f}{\partial n_R}=f_R$ and $\sfrac{\partial f}{\partial n_B}=f_B$ gives the desired result:
\begin{equation}
\begin{aligned}
    H[n''_R(x),n''_B(x)]-H[n'_R(x),n'_B(x)]&\approx \left[f_R(n'_R(x''),n'_B(x''))+V_R(x'')\right]-\left[f_R(n'_R(x'),n'_B(x'))+V_R(x')\right]\\
    &=h(x'')-h(x'). 
\end{aligned}
    \label{eq:deltaH2}
\end{equation}

Switching labels 'R' and 'B' shows the same result for a blue agent moving.} that the candidate density-functional defined in \eqref{eq:defH} indeed satisfies $\Delta H\approx \Delta h$. Now, we can describe the underlying migration model with $\Delta H$ instead of $\Delta h$, and express the steady-state joint-density distribution as a Boltzmann distribution \parencite{barker1965monte}: 
\begin{equation}
    P[n_R(x),n_B(x)]=Z^{-1}\Omega[n_R(x),n_B(x)]\exp (-H[n_R(x),n_B(x)]),\label{eq:Bdist}
\end{equation}
where $Z$ is a normalization constant (known in statistical physics as the partition function), and ``multiplicity'' $\Omega$ counts the number of microscopic states that correspond to the given densities $[n_R(x),n_B(x)]$. Note that $\Omega$ depends upon the particular agent rule used, and we shall derive appropriate forms for $\Omega$ in Section \ref{sec:version} (for the coarse-grained underlying migration model).

\subsubsection{Coarse-grained Underlying Migration Model}\label{sec:coarsegrain}
So far we have been working with densities of agents and heuristically treating each agent as a `chunk of density' to inspire the link between the underlying migration model and Density-Functional Theory (see Section \ref{sec:DFFTDFT} for this link). In real demographic systems, data is readily available in terms of population counts across sample areas (e.g. blocks) as opposed to continuous densities. To match the nature of block-level census data, we need to coarse-grain the underlying migration model and obtain its steady-state distribution through analogous arguments.   

We start by defining average densities of agents for each block $n_{R,b}\equiv N_{R,b}/A_b$ and $n_{B,b}\equiv N_{B,b}/A_b$, where $N_{R,b}$ and $N_{B,b}$ represent the number of red and blue agents in block $b$, and $A_b$ is the area of block $b$. Similarly, we define ``average vexations'' over each block $b$ as $v_{R,b}\equiv[\int_{A_b}V_R(x)\text{d}x]/A_b$ and $v_{B,b}\equiv[\int_{A_b}V_B(x)\text{d}x]/A_b$. Now we can define the coarse-grained underlying migration model: At each step in time, an agent proposes to move from block $b'$ to $b''$ according to some agent rule. The agent will accept the move with probability $$P_{b'\to b''}=\frac{1}{1+e^{h(b'')-h(b')}},$$ where $h(b)$ is the agent's dissatisfaction in block $b$. We still have 
\begin{equation*}
    h(b)\equiv\begin{cases}
    f_R(n_{R,b},n_{B,b})+v_{R,b} & \text{if agent is red}\\
    f_B(n_{R,b},n_{B,b})+v_{B,b} & \text{if agent is blue}
\end{cases}.
\end{equation*}

Next, we define the coarse-grained global headache $H$:
\begin{equation} 
    H(\{N_{R},N_{B}\})\equiv\sum_b\left[f\left(\frac{N_{R,b}}{A_b},\frac{N_{B,b}}{A_b}\right)A_b+v_{R,b}N_{R,b}+v_{B,b}N_{B,b}\right]\equiv \sum_b H_b(N_{R,b},N_{B,b}),\label{eq:CoarseH}
\end{equation}
where we have further defined the ``block-level headache functions'' for convenience:
\begin{equation}
     H_b(N_{R,b},N_{B,b})\equiv f\left(\frac{N_{R,b}}{A_b},\frac{N_{B,b}}{A_b}\right)A_b+v_{R,b}N_{R,b}+v_{B,b}N_{B,b}.\label{eq:DefBlockH}
\end{equation}
In Eq.~\eqref{eq:CoarseH}, $\{N_{R},N_{B}\}$ is a short-hand notation for a particular state of the coarse-grained model. More explicitly, such a state is determined by a particular list of block occupations: $$ \{N_{R},N_{B}\}\equiv[ N_{R, 1}, N_{B, 1}, \cdots, N_{R, b_\text{tot}}, N_{B, b_\text{tot}}].$$ $\Delta H\approx\Delta h$ can be derived again if the global frustration $f$ satisfies $\sfrac{\partial f}{\partial n_R}\approx f_R$ and $\sfrac{\partial f}{\partial n_B}\approx f_B$, where the partial derivatives are now interpreted as the appropriate finite-difference approximations. For instance, if a red agent moves from block $b'$ to block $b''$, we have
\begin{equation*}
\begin{aligned}
    \Delta H=&\Delta H_{b''}+\Delta H_{b'}\\
    =& \left\{\left[f\left(\frac{N_{R,b''}+1}{A_{b''}},\frac{N_{B,b''}}{A_{b''}}\right)-f\left(\frac{N_{R,b''}}{A_{b''}},\frac{N_{B,b''}}{A_{b''}}\right)\right]A_{b''}+v_{R,b''}\right\}\\
    & \qquad \qquad\qquad -\left\{\left[f\left(\frac{N_{R,b'}}{A_{b'}},\frac{N_{B,b'}}{A_{b'}}\right)-f\left(\frac{N_{R,b'}-1}{A_{b'}},\frac{N_{B,b'}}{A_{b'}}\right)\right]A_{b'}+v_{R,b'}\right\}\\
    \approx& h(b'')-h(b')=\Delta h.
\end{aligned}
\end{equation*}

Finally, we consider the coarse-grained underlying migration model where $\Delta h$ is replaced by $\Delta H$, so that $$P_{b'\to b''}=\frac{1}{1+e^{\Delta H_{b'}+\Delta H_{b''}}}.$$
Then, an argument similar to those presented in Section \ref{sec:version} and Ref.~\parencite{barker1965monte} will lead to coarse-grained version of Eq.~\eqref{eq:Bdist}
\begin{equation}
    P(\{N_{R},N_{B}\})=Z^{-1}\Omega(\{N_{R},N_{B}\})\prod_b\exp\left[-f\left(\frac{N_{R,b}}{A_b},\frac{N_{B,b}}{A_b}\right)A_b-v_{R,b}N_{R,b}-v_{B,b}N_{B,b}\right].\label{eq:GeneralP}
\end{equation}
Again, the functional form of multiplicity $\Omega(\{N_{R},N_{B}\})$ depends on the agent rule used, and we shall derive appropriate forms for $\Omega$ in Section \ref{sec:version}.

\subsection{Marginal Block Distributions and Agent Potentials} \label{sec:mbd}
Eq.~\eqref{eq:GeneralP} is not easy to use when extracting the frustration and vexations because the number of possible states $\{N_{R},N_{B}\}$ grows quickly with the number of blocks. This complexity makes it difficult to obtain sufficient statistics for $P(\{N_{R},N_{B}\})$. Therefore, we evaluate the approximate joint probability distribution $P_b(N_{R,b},N_{B,b})$ for a particular block $b$, independent of the states of the other blocks, which is known as the marginal probability distribution for block $b$.

To do so, we begin by noting that the multiplicity factor $\Omega(\{N_{R},N_{B}\})$ in Eq.~\eqref{eq:GeneralP} represents the number of ways agents can be rearranged while maintaining the counts $N_{R,b},N_{B,b}$ in each block $b$.  Because agents can be rearranged \emph{independently} within each block, under quite general conditions as discussed in Section~\ref{sec:version}, generally $\Omega(\{N_{R},N_{B}\})$ is proportional to the product of a series of independent factors $\omega_b(N_{R,b},N_{B,b})$ for each block $b$, rendering the probability distribution as a product of seemingly independent factors,
\begin{equation}
    P(\{N_{R},N_{B}\})={Z'}^{-1}\prod_b \omega_b(N_{R,b},N_{B,b}) \exp\left[-f\left(\frac{N_{R,b}}{A_b},\frac{N_{B,b}}{A_b}\right)A_b-v_{R,b}N_{R,b}-v_{B,b}N_{B,b}\right],\label{eq:GlobalP}
\end{equation}
where $Z'$ is the new normalization factor.

Despite the surface appearance of Eq.~\eqref{eq:GeneralP}, the blocks are not fully independently distributed, but are in fact correlated through the constraint that the total population of red and blue agents does not change. Consequently, the corresponding factors in Eq.~\eqref{eq:GlobalP} do not give directly probability distributions for each block, but rather should be considered independent {\em likelihood factors},
\begin{equation} \label{eq:LHF}
L_b(N_{R,b},N_{B,b})\equiv \omega_b(N_{R,b},N_{B,b}) \exp\left[-f\left(\frac{N_{R,b}}{A_b},\frac{N_{B,b}}{A_b}\right)A_b-v_{R,b}N_{R,b}-v_{B,b}N_{B,b}\right].
\end{equation}
The conversion of these likelihood factors to block probability distributions from the constraint of fixed numbers of agents is a well-known issue from statistical physics and ultimately requires multiplication by a correction factor to produce the marginal block distribution functions. In statistical physics, this factor becomes $e^{\mu_R N_R+\mu_R N_B}$, where $N_R,N_B$ are the number of red and blue agents in the corresponding block and $\mu_R$ and $\mu_B$ are ``chemical potentials'' corresponding to the headache associated with removing agents from the given block, thereby corresponding to the definition of these quantities in Eq.~\eqref{eq:LME}.

To see how this factor arises, we consider the probability distribution for the first block (labeling of blocks is arbitrary, so there is no loss of generality). Because this probability considers only the numbers of agents in a single block, and does not consider how agents are distributed in the remaining blocks, we must sum the joint probability distribution over all the possible ways these `external' agents may distribute themselves in the rest of the system, resulting in  
\begin{eqnarray}
P_1(N_{R,1},N_{B,1})&=&Z^{-1} L_1(N_{R,1},N_{B,1}) \left[\sum_{N_{R,2}=0}^{N_R^\text{tot}-N_{R,1}}\right.\ \  \sum_{N_{B,2=0}}^{N_B^\text{tot}-N_{B,1}} L_2(N_{R,2},N_{B,2}) 
\times \nonumber\\ 
&& \sum_{N_{R,3}=0}^{N_R^\text{tot}-N_{R,1}-N_{R,2}}\ \  \left.\sum_{N_{B,3=0}}^{N_B^\text{tot}-N_{B,1}-N_{B,2}} L_3(N_{R,3},N_{B,3}) \times\cdots L_{b_\text{tot}}(N_{R,b_\text{tot}},N_{B,b_\text{tot}})\right] \nonumber \\
& \equiv & Z^{-1} L_1(N_{R,1},N_{B,1}) {\mathcal L}(N_R^\text{tot}-N_{R,1},N_B^\text{tot}-N_{B,1}), \label{eq:conv}
\end{eqnarray}
where $N_R^\text{tot}$ and $N_B^\text{tot}$ represent the fixed total number of red and blue agents in the system, respectively, and $b_\text{tot}$ denotes the total number of blocks. Here, we have obtained the marginal joint probability for block 1 from Eq.~\eqref{eq:GlobalP} by summing over all possible arrangements in the other blocks and using our definition of the likelihood factors in Eq.~\eqref{eq:LHF}. Note that the upper limits of each sum are adjusted properly to reflect the number of agents already accounted in previous terms.  From Eq.~\eqref{eq:conv}, it is apparent that a correction factor ${\mathcal L}$ is needed to convert the likelihood factor $L_1$ into the true probability distribution $P_1$.  It is also apparent that the needed factor ${\mathcal L} \equiv L_2 \circ L_3 \circ \ldots \circ L_{b_\text{tot}}$ fits precisely the mathematical definition of the discrete convolution of the likelihood factors for all bins other than $b=1$. 

The convolution of the $L_b$'s ($\mathcal{L}$) may be evaluated through the (discrete) central-limit theorem.  We first note that, because the $L_b$'s are likelihood factors, the presence of the normalization factor $Z^{-1}$ in Eq.~\eqref{eq:conv} allows, without loss of generality, us to rescale the $L_b$'s so that they are normalized ($\sum_{N_{R},N_{B}} L_b(N_{R},N_{B})=1$) and may then be interpreted mathematically as probability distributions. (We again stress that, due to the fixed agent-number constraints, the $L_b$'s do not represent probability distributions for the blocks, we merely make this identification to allow us to exploit the mathematical machinery of the central-limit theorem.) Under this interpretation, the convolutions give the probability distribution of the sum of a collection of random variables $\left\{N_{R,b>1},N_{B,b>1}\right\}$ {\em independently} distributed according to the distributions $L_b(N_{R,b},N_{B,b})$.  The central-limit theorem then ensures that, for systems with large numbers of blocks, the resulting distribution ${\mathcal L}(N_R,N_B)$ will approach a bivariate Gaussian distribution $\mathcal{N}_2$ with mean values
$\langle N_R\rangle = \sum_{b>1} \langle N_{R,b} \rangle_b \equiv (b_\text{tot}-1) \bar{N}_R$, $\langle N_B\rangle = \sum_{b>1} \langle N_{B,b} \rangle_b \equiv (b_\text{tot}-1) \bar{N}_R$, where averages $\langle \cdots \rangle_b$ are evaluated according to the effective distributions $L_b$, and $\bar{N}_R$ and $\bar{N}_B$ are effective average bin populations that do not scale with the number of bins $b_\text{tot}$ in the system.  Likewise, the covariance matrix associated with ${\mathcal L}$ is then ensured to be 
%$$
%\Sigma^2 = \left(\begin{array}{cc}
%\sum_{b>1} \sigma^2_{R,b} & \sum_{b>1} %\mbox{cov}\,(N_{R,B},N_{N,b})\\
%\sum_{b>1}\mbox{cov}\,(N_{R,B},N_{N,b}) & %\sum_{b>1} \sigma^2_{B,b}
%\end{array}\right) \equiv (b_\text{tot}-1) %\bar{\Sigma}^2,
%$$
$\mathbf{\Sigma} = \sum_{b>1}\mathbf{\Sigma}_b \equiv (b_\text{tot}-1) \bar{\mathbf{\Sigma}}$, where $\mathbf{\Sigma}_b$ are the covariance matrices of the effective distributions $L_b$ and, again, $\bar{\mathbf{\Sigma}}$ is an effective average block covariance matrix that does not scale with the number of bins $b_\text{tot}$ in the system.  As a result, if we gather $(N_R,N_B)$ and $(\bar{N}_R,\bar{N}_B)$ into a column vectors $\mathbf{N}$ and $\bar{\boldsymbol{\mu}}$, respectively, we have
\begin{align*}
    \mathcal{L}(\mathbf{N})&\approx\mathcal{N}_2((b_\text{tot}-1)\bar{\boldsymbol{\mu}},(b_\text{tot}-1)\bar{\mathbf{\Sigma}})\\
    &\equiv\frac{1}{2\pi \sqrt{(b_\text{tot}-1)\det\bar{\mathbf{\Sigma}}}} e^{-\frac{1}{2(b_\text{tot}-1)}[\mathbf{N}- (b_\text{tot}-1)\bar{\boldsymbol{\mu}}]^\intercal \bar{\mathbf{\Sigma}}^{-1} [\mathbf{N}- (b_\text{tot}-1)\bar{\boldsymbol{\mu}}]}.
\end{align*}

Finally, we can compute the needed correction factor by defining $(N_{R}^\text{tot},N_{B}^\text{tot})$ and $(N_{R,1},N_{B,1})$ as the column vectors $\mathbf{N}^\text{tot}$ and $\mathbf{N}_1$, respectively, making the needed correction factor in Equation~\eqref{eq:conv} equal to
\begin{multline}
\mathcal{L}(\mathbf{N}^\text{tot}-\mathbf{N}_1)\approx\left(\frac{1}{2\pi \sqrt{(b_\text{tot}-1)\det\bar{\mathbf{\Sigma}}}} e^{-\frac{1}{2(b_\text{tot}-1)}[\mathbf{N}^\text{tot}-(b_\text{tot}-1)\bar{\boldsymbol{\mu}}]^\intercal \bar{\mathbf{\Sigma}}^{-1} [\mathbf{N}^\text{tot}-(b_\text{tot}-1)\bar{\boldsymbol{\mu}}]}\right)\\
\left(
e^{-\mathbf{N}_1^\intercal \bar{\mathbf{\Sigma}}^{-1} \left[\frac{\mathbf{N}^\text{tot}}{b_\text{tot}-1}-\bar{\boldsymbol{\mu}}\right]}\right) \left( e^{-\frac{1}{2(b_\text{tot}-1)} \mathbf{N}_1^\intercal \bar{\mathbf{\Sigma}}^{-1} \mathbf{N}_1}\right).\label{eq:large_L}
\end{multline} 
\\
Each of the three terms above deserves comment.  The first term is independent of the occupancy of the first bin and thus ultimately absorbed in the normalization factor $Z^{-1}$ in Eq.~\eqref{eq:conv}.  The exponent of the final term has the form of a quadratic function of $N_{R,1}$ and $N_{B,1}$.  Because this term has very nearly the same functional form regardless of the bin chosen and does not depend on the total number of agents $\mathbf{N}^\text{tot}$ or blocks $b_\text{tot}$, it will ultimately be absorbed into the frustration, where it represents a frustration arising from the reduction in entropy in the rest of the system as agents gather into one particular bin.  Moreover, for systems with large numbers of bins compared to the square of the typical occupancy of a bin, this term is negligible.\footnote{In cases where this term is not negligible, future work should explore the possibility of correcting for this term when extracting frustrations in order to improve description of the system dynamics through TD-DFFT.}  The one remaining term, the second term, cannot be absorbed into other factors in Equation~\eqref{eq:conv}, and must be included explicitly.  The term has precisely the form expected from statistical physics, $e^{\mu_R N_{R,1}+\mu_B N_{B,1}}$, provided we identify the needed ``agent potentials'' (referred to as chemical potentials in the Physics literature) as $(\mu_R,\mu_B)=-\bar{\mathbf{\Sigma}}^{-1}[\mathbf{N}^\text{tot}/(b_\text{tot}-1)-\bar{\boldsymbol{\mu}}]$, which we expect to remain essentially constant for all blocks so long as there are a sufficient number of blocks of sufficiently little variation among blocks that the averages over other blocks $\bar{\mathbf{\Sigma}}$ and $\bar{\boldsymbol{\mu}}$ are all essentially the same. Note, however, that we do expect these agent potentials to depend on the values of $\mathbf{N}^\text{tot}$, so that we may account, within this framework, for changes in the total numbers of agents of the two types simply by adjusting the values of our agent potentials.

We therefore obtain the approximate joint probability distribution for each block $b$:
\begin{equation} 
    P_b(N_{R},N_{B})\approx z_b^{-1}\omega_b(N_R,N_B)\exp\left[-f\left(\frac{N_{R}}{A_b},\frac{N_{B}}{A_b}\right)A_b-(v_{R,b}-\mu_R)N_{R}-(v_{B,b}-\mu_B)N_{B}\right].\label{eq:DFFT_full}
\end{equation}
To simplify Eq.~\eqref{eq:DFFT_full} further for DFFT function extraction, we redefine the vexations up to constant shifts so that the chemical potentials vanish, namely $v_{R,b}\leftarrow v_{R,b}-\mu_R$ and $v_{B,b}\leftarrow v_{B,b}-\mu_B$. With these shifts, we can then extract (See Section \ref{sec:ExtractNUA}) the frustration $f(n_R,n_B)$ and shifted vexations $v_{R,b},v_{B,b}$ by fitting observational data to 
\begin{equation} 
    P_b(N_{R},N_{B})\approx z_b^{-1}\omega_b(N_R,N_b)\exp\left[-f\left(\frac{N_{R}}{A_b},\frac{N_{B}}{A_b}\right)A_b-v_{R,b}N_{R}-v_{B,b}N_{B}\right].\label{eq:DFFT_shift}
\end{equation}
Once the extraction is performed, the distributions for systems with different total numbers of agents are then given directly by replacing $v_{R,b}-\mu_R\leftarrow v_{R,b}$ and $v_{B,b}-\mu_B\leftarrow v_{B,b}$ in \eqref{eq:DFFT_shift} (i.e., using \eqref{eq:DFFT_full})
for appropriate values of $(\mu_R,\mu_R)$ with values set to ensure the correct total number of agents of each type (See Section \ref{sec:PredSteadyState}).

Two important simplifications to \eqref{eq:DFFT_shift} are possible when, as in the text, the blocks, apart from social preferences, are geometrically identical. We may measure area in units of blocks so that $A_b=1$, and exploit the fact that the functional form for the block multiplicity factors must be identical $\omega_b(N_R,N_B)=\omega(N_R,N_B)$, resulting in 
\begin{equation}
    P_b(N_{R},N_{B})\approx z_b^{-1}\omega(N_R,N_B)\exp\left[-f\left(N_R,N_B\right)-v_{R,b}N_{R}-v_{B,b}N_{B}\right],\label{eq:DFFT_full2}
\end{equation}
which corresponds directly to the form used in the main text for the particular case when $\omega(N_R,N_B)=1/[N_R!N_B!(s-N_R-N_B)!]$, which we shall derive in Section \ref{sec:uniform}. Second, in the absence of sufficient knowledge of the underlying agent behavior to determine $\omega(N_R,N_B)$, one may define $\tilde{f}\left(N_R,N_B\right)\equiv f\left(N_R,N_B\right) - \ln\omega(N_R,N_B)$, resulting in
\begin{equation}
    P_b(N_{R},N_{B})\approx {z_b'}^{-1}\exp\left[-\tilde{f}\left(N_R,N_B\right)-v_{R,b}N_{R}-v_{B,b}N_{B}\right].\label{eq:DFFT_full3}
\end{equation}
\eqref{eq:DFFT_full3} may then be fit to observational data to extract $\tilde{f}(n_R,n_B),v_{R,b},v_{B,b}$ and used to predict system behavior for different numbers of agents by introduction of agent potentials $(\mu_R,\mu_R)$ in the same way as described above.

The formulations of \eqref{eq:DFFT_full2} and \eqref{eq:DFFT_full3} are mathematically equivalent and thus, apart from numerical issues, lead to identical predictions for the steady-state distributions $P_b(N_{R},N_{B})$. Differences appear, however, when making dynamical predictions of time evolving systems. Section~\ref{sec:version} describes the relationship between agent behaviors and the multiplicity factors $\omega_b(N_R,N_B)$, and Section~\ref{sec:altR} explores the sensitivity of dynamical predictions to the selection of multiplicity factors. 

Finally, to connect the agent potentials $(\mu_R,\mu_B)$ appearing in Equation~\eqref{eq:DFFT_full3} to the Lagrange multipliers appearing in Eq.~\eqref{eq:LME}, we note that the formulation leading to Eq.~\eqref{eq:LME} corresponds to a particular limit. Specifically, Eq.~\eqref{eq:LME} arose from the assumption that the system reaches a state of least possible dissatisfaction, at which point all change in the system ceases. This corresponds to an agent rule where moves are always accepted (rejected) if the move decreases (increases) the dissatisfaction, corresponding to probabilities in Equation~\eqref{eq:DensityP} always being 0 or 1. For \eqref{eq:DensityP} to yield such values, the system must be in a limit where the changes in frustrations and vexations as agents are all much larger than unity.  Moreover, the steady state of the system in this limit is the state of lowest possible dissatisfaction and thus shows no variations or fluctuations, and each block assumes a single, final set of occuptation numbers. Without fluctuations, the mean and mode of the distribution Eq.~(\ref{eq:DFFT_full3}) thus must correspond, and we can determine the numbers of agents by maximizing Eq.~(\ref{eq:DFFT_full3}) or, equivalently, minimizing the negative of the exponent with respect to $N_R$ and $N_B$, leading directly to the following equation for each bin $b$:
\begin{equation}
    \left\{\begin{aligned}
        &\frac{\partial}{\partial n_R} \tilde{f}\left(N_R,N_B\right) + (v_{R,b} - \mu_R)  =  0 \\ 
        &\frac{\partial}{\partial n_B} \tilde{f}\left({N_R},{N_B}\right) + (v_{B,b} - \mu_R)  =  0.
    \end{aligned}\right..\label{eq:mumeaning}
\end{equation}

Finally, because this limit requires that the changes in frustration values  ${f}\left(N_R,N_B\right)$ be large compared to unity, we can replace $\tilde{f}$ with $f$ in \eqref{eq:mumeaning}, yielding exactly \eqref{eq:LME}, thereby directly connecting the meaning of the quantities $(\mu_R,\mu_B)$ in both equations.

\subsection{Extracting Frustration and Vexations when Blocks have Non-uniform Areas}\label{sec:ExtractNUA}
To extract the frustration and vexations when area of the blocks are different, we follow a similar procedure as described in the main text, starting from Eq.~\eqref{eq:DFFT_full} (with agent potentials set to zeros). First, we rearrange Eq.~\eqref{eq:DFFT_full} to obtain 
\begin{equation}
    -A_b^{-1}\ln \left[P_b(n_{R}A_b,n_{B}A_b)/\omega_b(n_{R}A_b,n_{B}A_b)\right]\approx f\left(n_{R},n_{B}\right)+v_{R,b}n_{R}+v_{B,b}n_{B}+c_b,\label{eq:H_full}
\end{equation}
where we have defined the block-level average density for red agents $n_{R}\equiv N_R/A_b$ and blue agents $n_{B}\equiv N_B/A_b$, and block-dependent constants $c_b\equiv A_b^{-1}\ln(z_b)$. The LHS of Eq.~\eqref{eq:H_full} is determined from observed $P_b$ and is defined only for discrete values of $n_R$ and $n_B$ such that $n_RA_b$ and $n_BA_b$ are integers. If the observed $P_b$ can be described by our DFFT framework, the LHS of Eq.~\eqref{eq:H_full} for each block would be fitted well by a block-independent global frustration $f$ plus a block-dependent linear shift $v_{R,b}n_{R}+v_{B,b}n_{B}+c_b$, where the average vexations of the block $v_{R,b}$ and $v_{B,b}$ are the slopes of this shift along the axes. We could then simply find the global frustration $f$ by interpolating the discrete data points from the LHS of Eq.~\eqref{eq:H_full} for a block with a given size, $A_b$. Then, we can determine the vexations from the planar shifts required to fit the LHS of Eq.~\eqref{eq:H_full} for the remaining blocks. To most accurately extract these DFFT functions, however, we use a maximum likelihood estimation (MLE) algorithm, as done in the main text. The Code Availability section describes where to find the corresponding MATLAB code.

It is important to keep in mind that the extraction of DFFT frustrations and vexations is not unique. In fact, the following transformation to the functions (analogous to a gauge transformation in physics) preserves the fit to the probability distributions:
\begin{equation}\label{eq:gauge}
\left\{
\begin{aligned}
    f&\longrightarrow f+a_1n_R+a_2n_B+a_3\\
v_{R,b}&\longrightarrow v_{R,b}-a_1\\
v_{B,b}&\longrightarrow v_{B,b}-a_2\\
c_b&\longrightarrow  c_b-a_3
\end{aligned}\right.,
\end{equation}
where $a_1$, $a_2$ and $a_3$ are any arbitrary constants. Luckily, this transformation affects neither the steady-state prediction (Eq.~\eqref{eq:DFFT_full} is only affected by $a_3$, whose effect is removed by the normalization constant $z_b$) nor the time-evolution prediction. (Changes in $H$, as shown in Eq.~\eqref{eq:DefBlockH}, are not affected by any of these constants.) Finally, the interpretations of the DFFT functions are also preserved: the differences, $\sfrac{\partial f}{\partial n_R}(n_R'',n_B'')-\sfrac{\partial f}{\partial n_R}(n_R',n_B')$, $\sfrac{\partial f}{\partial n_B}(n_R'',n_B'')-\sfrac{\partial f}{\partial n_B}(n_R',n_B')$, $v_{R,b''}-v_{R,b'}$ and $v_{B,b''}-v_{B,b'}$ are not affected by the above transformation. Therefore, one may equally well use any particular set of extracted DFFT functions related by the transformation of Equation~\eqref{eq:gauge}.

\subsection{Analytically Predicting New Steady State or Quasi-static Time Evolution}\label{sec:PredSteadyState}

As we mentioned in Section \ref{sec:mbd}, Eq.~\eqref{eq:DFFT_full} gives the steady-state marginal block distribution for any set of total population numbers $N^\text{tot}_R,N^\text{tot}_B$. So long as agents in the new system shares the same social and spatial preferences as agents from whom the frustration and vexations function is extracted, we can easily calculate the new marginal block distributions, by adjusting agent potentials $\mu_R$ and $\mu_B$. To find the values of $\mu_R$ and $\mu_B$, we use Newton's Method (See Code Availability) to solve for the values of these agent potentials such that `the means of the probability distribution, when averaged over all the blocks, equals the new mean for the entire system resulting from the demographic change'. More specifically, we solve for $\mu_R$ and $\mu_B$ such that
\begin{equation}
\left\{
    \begin{aligned}
        &N^\text{tot}_R=\sum_{b,N_R,N_B}N_R\cdot P_b(N_R,N_B)\\
        &N^\text{tot}_B=\sum_{b,N_R,N_B}N_B\cdot P_b(N_R,N_B)
    \end{aligned}
\right..
\end{equation}

We can extend this method to analytically predict quasi-static changes in block populations $N_{R,b}(t),N_{B,b}(t)$ as the population redistributes in response to demographic changes in the population totals $N^\text{tot}_R(t),N^\text{tot}_B(t)$.  In particular, we can use the population totals to solve for the agent potentials $\mu_{R}(t),\mu_{B}(t)$ as above, and then use Eq.~\eqref{eq:DFFT_full} to predict the population content of each of the blocks as time evolves.  

\subsection{Multiplicity Factors and Agent Rules}\label{sec:version}

We now turn to the question of the appropriate form for the multiplicity factor $\Omega(\{N_{R},N_{B}\})$. As the discussion surrounding \eqref{eq:DFFT_full3} demonstrates, it is often possible to make predictions of steady-state probability distributions without knowledge of the mathematical form for $\Omega(\{N_{R},N_{B}\})$. However, the extracted functions $\tilde{f}\left(N_R,N_B\right)\equiv f\left(N_R,N_B\right) - \ln\omega(N_R,N_B)$ combine information regarding social interactions with multiplicity effects, thereby limiting the ability to extract meaningful information regarding social interactions from the data and potentially introducing numerical difficulties in extracting such information, particularly if the multiplicity term $- \ln\omega(N_R,N_B)$ dominates the frustration $f\left(N_R,N_B\right)$ and reduces the signal to noise ratio in the statistical data. Moreover, although the extracted $\tilde{f}\left(N_R,N_B\right)$ contains, in principle, exactly the information required to predict steady-state distributions, this function does not contain the exact social-preference information needed to predict dynamical evolution in time, but rather confounds underlying social interactions with combinatorial effects arising from the agent rules. Thus, in general, the more precisely the correct form for $\Omega(\{N_{R},N_{B}\})$ is captured, the more accurate will the dynamical predictions be. 

It is perhaps surprising that the multiplicity of the number of ways a given state $\{N_{R},N_{B}\}$ can be realized as its ``microstates'', a purely combinatorial question, is related to the agent rules (akin to ``kinematic'' effects in physics systems. This effect is independent from the social interactions, akin to ``dynamical'' effects in physical systems, because it originates from specifically how proposed moves are generated as opposed to whether proposed moves are accepted.) As we shall see below, the connection between multiplicity and agent rules arises because, ultimately, it is the agent rules that determine what literally counts as microstates of the system. We note that microstates of our system are only theoretical constructs to interpret $\Omega$, rather than direct implications of the coarse-grained underlying migration model. In fact, there can be multiple microstate-interpretations of the same agent rule.

We first discuss how agent rules are defined. Recall that potential agent transitions are proposed according to the agent rule and then accepted or rejected according to the probability from $P_{b'\to b''}$ (the coarse-grained version of Eq.~\eqref{eq:DensityP}, as discussed in Section \ref{sec:coarsegrain}).
%The behaviors we consider differ based on how, for a given proposed move, the moving agent and empty destination location are selected. We consider two specific methods for choosing moving agents and destination locations. 
Thus, to specify an agent rule, one needs to specify how ``originating block'' $b'$ and ``destination block'' $b''$ are chosen. The methods we use to choose these blocks can be more intuitively described if we make the following interpretations: ``choosing block $b'$'' corresponds to ``choosing a moving agent (in block $b'$)'', and ``choosing $b''$'' corresponds to ``choosing a destination location (in block $b''$)''. We note that in the coarse-grained underlying migration model, agent location and destination location are only specified at the block level, unlike the Schelling model where agents and empty cells have specific arrangements within blocks. We consider two methods of choosing blocks. The first method, which we shall call ``direct'', involves selecting the moving agent or destination location at random from among all $N_\text{tot}$ available agents or $s_\text{tot}-N_\text{tot}$ empty locations, respectively. With this approach, the probability that a particular block is the origin of a moving agent or a particular block contains the destination location is directly proportional to the number of agents $N_{R,b}+N_{B,b}$ or empty locations $s-N_{R,b}-N_{B,b}$ within that block, respectively. Variable $s$ is defined as the maximum agent occupancy of a block, representing the number of housing units in a real system, or the number of cells in the Schelling Model. $s$ can be block-dependent in general, but for our discussion we have assumed geometrically identical blocks. $s_\text{tot}$ is defined as the maximum agent occupancy of the entire city ($s_\text{tot}\equiv s\cdot b_\text{tot}$).  The second method, which we shall call ``block based'', involves first selecting the block that contains the moving agent or destination location at random from among the available \emph{blocks}, and then selecting an agent or empty location from within the chosen block. (In this description, selection of agent and empty location is mentioned only to match the description of the first method, and does not carry much information. Agent rule in the coarse-grained model only concerns how $b'$ and $b''$ are chosen.) With this approach, the probability of choosing a particular block for the origin or destination of the move is independent of the number of agents or number of empty locations in that block. 

Below we consider the multiplicity factors associated with three different types of agent rules, obtained from different combinations of the two methods of choosing moving agents and destination locations we just discussed.
\begin{itemize}
    \item \textbf{Agent Rule (1), ``uniform consideration of moves'', main-text behavior:} agent and destination location selected with the direct method.
    \item \textbf{Agent Rule (2), ``block-focused locations'', Ref.~\parencite{mendez2018density} behavior:} agent selected with direct method and destination location selected with the block based method. 
    \item \textbf{Agent Rule (3), ``block-focused agents and locations'', simplest multiplicity factor:} agent and destination location selected with the block based method.
\end{itemize}

Agent Rule (1) corresponds to a situation where all agents propose moves at the same rate, selecting among all possible moves equally. (In all cases, the decision whether the move actually takes place is decided later based upon change in dissatisfaction.) This corresponds to a scenario where, for example, the possibility of moving occurs to all people in a given city at the same rate, and they then consider all housing listings equally. Agent Rule (2) corresponds to a situation, where again, all agents propose moves at the same rate, but then explore potential moves based on the various blocks. This corresponds to a scenario where again the possibility of moving occurs to all people at the same rate, but they do not consider all possible moves equally. Rather, they first select among the blocks and then focus their search on housing listings within the given block. In addition to describing a particular type of human behavior, a second reason to consider Agent Rule (2) is that, when modeling real-world conditions, the total population capacity of a block may not be known, thus making it impossible to scale the appropriate rates by the number of available empty units. We note that (the single-agent-type version of) Agent Rule (2) is used in Ref.~\parencite{mendez2018density}.   
Finally, Agent Rule (3) corresponds to a more abstract case where moves originate from all blocks with equal rates regardless of the population of the block. We include this case primarily out of mathematical interest and because of the extremely simple multiplicity factor with which it is associated.

To simplify the analyses of multiplicity factors below, we consider the special case of systems with agent rules but no social interactions or preferences ($f(N_R,N_B)=v_{R,b}=v_{B,b}=0$). The same arguments, with somewhat more complex algebra, will hold for the general case and lead directly to Equation~\eqref{eq:GeneralP} as an exact result; however, by considering only the effects of agent rules, we are able to focus the presentation below on the direct connections between agent rules and multiplicity factors. Under the conditions of no interactions, the Markov chains corresponding to our scenarios are ergodic\footnote{all states can be reached from any other state through a series of allowed transitions} and symmetric\footnote{the transition probability between any two states is the same regardless of the direction of the transition}, so that all possible arrangements of agents in the system become equal, so that $P(\{N_{R},N_{B}\}) \propto \Omega(\{N_{R},N_{B}\})$, exactly as \eqref{eq:GeneralP} would predict if $\Omega(\{N_{R},N_{B}\})$ counts the number of of distinct states corresponding to the block counts $\{N_{R},N_{B}\})$.  Finally, to carry out the analyses, we will employ the principle of detailed balance, which states that, for many systems, in steady-state the statistical rate of any transition in the system matches exactly the rate of the reverse transition. Not all systems satisfy detailed balance in their steady state, but as we shall see from explicit construction, the systems that we consider below do.

\subsubsection{Agent Rule (1), Uniform consideration of moves, main-text behavior } \label{sec:uniform}

We begin with the case corresponding to the main text, where the agent rule is for proposed moves, an agent and an empty location are selected at random, both according to a uniform distribution.
The selection of empty locations in this way implies that the probability of an agent considering a move to a particular block is proportional to the number of empty locations in that block, corresponding to behavior where agents consider all open real-estate listings equally, rather than, for example, focusing on particular blocks and then evaluating the listings within those blocks.

To see how the multiplicity factor arises directly from the agent rule, we employ the principle of detailed balance. We consider any possible proposed move and let $N_{R},N_B$ be the number of agents of each type in the originating block, $M_R,M_B$ be the number of agents in the destination block, and $\Omega(N_R,N_B;M_R,M_B;\ldots)$ be the multiplicity factor in the original state.  Detailed balance requires that the probability for this transition must match the probability of the reverse.  If the moving agent happens to be red, this gives:
\begin{multline} \label{eq:kin1}
Z^{-1} \Omega(N_R,N_B;M_R,M_B;\ldots)\cdot  \frac{N_R}{N_\mathrm{tot}} \frac{s-M_R-M_B}{s_\mathrm{tot}-N_\mathrm{tot}} \cdot \frac{1}{2} =  \\ 
 Z^{-1} \Omega(N_R-1,N_B;M_R+1,M_B;\ldots) \cdot \frac{M_R+1}{N_\mathrm{tot}} \frac{s-(N_R-1)-N_B}{s_\mathrm{tot}-N_\mathrm{tot}} \cdot \frac{1}{2}, 
\end{multline}
where, on the left-hand side, $Z^{-1} \Omega(N_R,N_B;M_R,M_B;\ldots)$ gives the probability of being in the originating state, ${N_R}/{N_\mathrm{tot}}$ is the probability of selecting a red agent from the originating block, $\sfrac{s-M_R-M_B}{s_\mathrm{tot}-N_\mathrm{tot}}$ is the probability of selecting one of the empty locations in the destination block, and the final factor $\frac{1}{2}$ is the acceptance probability for when $\Delta H=0$ in the absence of social interactions or location preferences. Finally, the right-hand side represents the reverse transition with exactly the same logic, and an exactly corresponding equation follows if the moving agent happens to be blue.

Rearranging the above equation and its blue counterpart gives two relationships
\begin{equation}
\left\{
    \begin{aligned}
        \frac{\Omega(N_R,N_B;M_R,M_B;\ldots)}{\Omega(N_R-1,N_B;M_R+1,M_B;\ldots)} & = \frac{s-(N_R-1)-N_B}{N_R} \cdot\frac{M_R+1}{s-M_R-M_B}  \\
        \frac{\Omega(N_R,N_B;M_R,M_B;\ldots)}{\Omega(N_R,N_B-1;M_R,M_B+1;\ldots)} & = \frac{s-N_R-(N_B-1)}{N_B} \cdot\frac{M_B+1}{s-M_R-M_B}
    \end{aligned}\right.,\label{eq:recurr1}
\end{equation}
that we can apply recursively to obtain $\Omega(\{N_R,N_B\})$ for any state $\{N_R,N_B\}$ with the same $N_\text{tot}$ in relation to some fixed reference state $\{N_R^\text{ref},N_B^\text{ref}\}\equiv [ N_{R, 1}^\text{ref}, N_{B, 1}^\text{ref}, \cdots, N_{R, b_\text{tot}}^\text{ref}, N_{B, b_\text{tot}}^\text{ref}]$. The $N,M$-separable form of the right-hand sides of \eqref{eq:recurr1} indicates that\footnote{It is not immediately clear that $\Omega(\{N_R,N_B\})/\Omega(\{N_R^\text{ref},N_B^\text{ref}\}$ or $R_b$ are well defined, as they can be calculated from different sequences of state transitions. In case they are not well defined, the system will not satisfy detailed balance. We will see that for our agent rule, $R_b$, and thus $\Omega(\{N_R,N_B\})/\Omega(\{N_R^\text{ref},N_B^\text{ref}\}$, are in fact independent of the sequence and are functions of only the initial and final states of the sequence.}
\begin{equation}
    \frac{\Omega(\{N_R,N_B\})}{\Omega(\{N_R^\text{ref},N_B^\text{ref}\})}=\prod_b R_b,\label{eq:RecurRatio}
\end{equation}
where $R_b$ consists of ratios concerning the number of agents in block $b$ as we recursively apply Eq.~\eqref{eq:recurr1}. 
If there is a function $\omega$ satisfying the following recursive relation 
\begin{equation}
    \left\{
    \begin{aligned}
        \frac{\omega(N_R,N_B)}{\omega(N_R-1,N_B)} & = \frac{s-(N_R-1)-N_B}{N_R}\\
        \frac{\omega(N_R,N_B)}{\omega(N_R,N_B-1)} & = \frac{s-N_R-(N_B-1)}{N_B}
    \end{aligned}
    \right.,\label{eq:RecurBlock}
\end{equation}
one can check that 
$$R_b=\frac{\omega(N_{R,b},N_{B,b})}{\omega(N_{R,b}^\text{ref},N_{B,b}^\text{ref})}.$$
(In the simplest case, Eq.~\eqref{eq:recurr1} can be shown to satisfy Eq.~\eqref{eq:RecurRatio} with the above form of $R_b$.\footnote{For instance, the first line of Eq.~\eqref{eq:recurr1} can be written as $$\frac{\Omega(N_R,N_B;M_R,M_B;\ldots)}{\Omega(N_R-1,N_B;M_R+1,M_B;\ldots)} = \frac{\omega(N_R,N_B)}{\omega(N_R-1,N_B)} \cdot\frac{\omega(M_R,M_B)}{\omega(M_R+1,N_B)}=\prod_b R_b.$$})  Eq.~\eqref{eq:RecurRatio} can then be written more simply as : 
\begin{equation}
    \Omega(\{N_R,N_B\})=C \prod_b \omega(N_{R,b},N_{B,b}), \label{eq:sepfull}
\end{equation}
where $C$ is an undetermined and ultimately irrelevant normalization factor. Note that Eq.~\eqref{eq:sepfull} (with a more general block-dependent $\omega_b$, which we would have obtained if our agent rule and thus Eq.~\eqref{eq:recurr1} were block-dependent) is used to transition from \eqref{eq:GeneralP} to \eqref{eq:GlobalP} in Section~\ref{sec:mbd}.
Finally, we can tentatively find $\omega$ by applying Eq.~\eqref{eq:RecurBlock} recursively:
\begin{align}
    \omega(N_R,N_B) & = \frac{s-(N_R-1)-N_B}{N_R} \cdot\omega(N_R-1,N_B) \nonumber\\
    & = \frac{\left[s-(N_R-1)-N_B\right] \cdots \left[s-N_B\right]}{N_R \cdots 1} \cdot\omega(0,N_B) \nonumber\\
    &=\frac{\left[s-(N_R-1)-N_B\right] \cdots \left[s-N_B\right]}{N_R \cdots 1} \cdot 
    \frac{\left[s-(N_B-1)\right]\cdots s}{N_B\cdots 1} \cdot\omega(0,0) \nonumber \\
    & =  \frac{s!}{N_R!N_B!(s-N_R-N_B)!} \equiv \left(\begin{array}{c}s\\N_R\ N_B\ (s-N_R-N_B)\end{array}\right)\label{eq:omega1},
\end{align}
where we set the undetermined and ultimately irrelevant normalization constant by taking $\omega(0,0)\equiv 1$ to reflect the fact that there is only one way to arrange zero agents. We can further check that Eq.~\eqref{eq:omega1} does satisfy Eq.~\eqref{eq:RecurBlock} for all possible $N_R$ and $N_B$ pairs.

From the above result come two important conclusions.  First, substituting \eqref{eq:omega1} into \eqref{eq:sepfull} and setting the ultimately irrelevant normalization constant $C=N^\mathrm{tot}_R! N^\mathrm{tot}_B!$, we obtain 
\begin{align}
\Omega(\{N_{R},N_{B}\}) & = N^{\mathrm{tot}}_R!\, N^{\mathrm{tot}}_B!\,
\prod_b{\left( 
\begin{array}{c}s\\ N_{R,b}\ N_{B,b}\ (s-N_{R,b}-N_{B,b})\end{array}
\right)} \label{eq:comb1} \\
& \equiv  N^\mathrm{tot}_R!\, N^\mathrm{tot}_B!\, \prod_b{\frac{s!}{N_{R,b}!\,N_{B,b}!\,(s-N_{R,b}-N_{B,b})!}}. \nonumber
\end{align} 
Second, we find that \eqref{eq:omega1}, apart from the ultimately irrelevant normalization constant $s!$ in the numerator, corresponds precisely to the combinatorial prefactors used in the main text.

To alternatively derive and interpret the multiplicity factor (Eq.~\eqref{eq:comb1}) as the number of microstates corresponding to state $\{N_R,N_B\}$, we note that the proportionality of the rate of transition into a block to the number of empty sites in that block implies that each possible placement represents a distinct microstate of the system.  As the population of each block is built up from individual moves, this implies that each possible arrangement of agents is to be regarded as distinct. The multiplicity factor thus counts the total number of distinct ways to arrange $N^\mathrm{tot}_R,N^\mathrm{tot}_B$ agents among $b_\mathrm{tot}$ blocks of size $s$, again giving rise to Eq.~\eqref{eq:comb1}, where the multinomial factors $\Large(\,\Large)$ count all possible ways of selecting which sites within each block are occupied by red or blue agents, and the factorial prefactors count all possible ways of arranging the red and blue agents, respectively, among the selected sites.\footnote{The argument presented in this paragraph is very subtle. In this footnote we present more details for interested readers.

To define microstates of a state $\{N_R,N_B\}$, one needs to specify additional structures that distinguish the microstates apart. For instance, we can define a microstate as an ordered list of individual red and blue agents for each block. Then, the multiplicity of state $\{N_R,N_B\}$, $\Omega(\{N_R,N_B\})$, is defined as the number of microstates that corresponds to the same state $\{N_R,N_B\}$ (known as a ``macrostate'' in physics). 

By interpreting $\Omega$ as the number of microstates, one can obtain an alternative method to find solutions of Eq.~\eqref{eq:recurr1}. Consider the macrostate transition from $[N_R,N_B;M_R,M_B;\ldots]$ to $[N_R-1,N_B;M_R+1,M_B;\ldots]$. For some definition of allowed microstate transitions (satisfying reversibility, where each transition has a reverse transition), suppose each microstate of $[N_R,N_B;M_R,M_B;\ldots]$ can transition to $E$ microstates of $[N_R-1,N_B;M_R+1,M_B;\ldots]$, and each microstate of $[N_R-1,N_B;M_R+1,M_B;\ldots]$ can transition to $F$ microstates of $[N_R,N_B;M_R,M_B;\ldots]$. Since the total number of microstate transitions between two macrostates is equal for opposite directions (as each microstate transition is reversible), we must have:
$$\frac{\Omega(N_R,N_B;M_R,M_B;\ldots)}{\Omega(N_R-1,N_B;M_R+1,M_B;\ldots)}=\frac{F}{E}.$$
Thus, if we can define microstates and allowed transitions such that the ratio $F/E$ satisfies
\begin{equation}
    \frac{F}{E}=\frac{s-(N_R-1)-N_B}{N_R} \cdot\frac{M_R+1}{s-M_R-M_B},\label{eq:microForm}
\end{equation}
and the corresponding form for blue agent transitions, for all possible macrostate transitions, then the multiplicity of macrostates $\Omega$ for the given definition of microstates will satisfy Eq.~\eqref{eq:recurr1}. 

It may seem hard to find the desired definition for microstates and allowed transitions, but typically the agent rule is suggestive of the desired definition. For agent rule (1), define microstates as ordered lists of red, blue, and empty locations for each block, where all agents are distinguishable; and allowed transitions as switching list position of an agent and an empty location. We find that for the previously mentioned macrostate transition, $E=N_R\cdot (s-M_R-M_B)$, since there are $N_R$ ways select the agent and $(s-M_R-M_B)$ ways to select the empty location. Similarly, $F=[s-(N_R-1)-N_B]\cdot (M_R+1)$. Thus Eq.~\eqref{eq:microForm} is satisfied (we can similarly check the corresponding form for blue agent transitions). Therefore, according to our microstate definition, the number of microstates, given by Eq.~\eqref{eq:comb1}, satisfies Eq.~\eqref{eq:recurr1}.

We note that there might be different definitions of microstates and allowed transitions that satisfy the desired relations (Eq.~\eqref{eq:microForm}). For instance, in the above definition, we may treat all agents of the same color as indistinguishable from one another. Then we obtain $\Omega$ without the prefactor $N_R^\text{tot}!N_B^\text{tot}!$.\label{ft:micro}} 

the $\omega'(M_R,M_B)'$ factors, which implies that, apart from a normalization constant, $\omega(N_R,N_B)$ and $\omega'(M_R,M_B)$ must have the same functional form when referring to the same block.

\subsubsection{Agent Rule (2), Block-focused locations} \label{sec:grav}

The main text focuses on rules where agents consider transitions to all possible empty sites directly, and thus equally, rather than focusing first on a block and then seeking empty sites within the considered block, thereby treating blocks equally. The overall system will behave differently if the agents exhibit the latter behavior instead of the former, even if agents share precisely the same dissatisfaction functions. To reflect this, there will be a corresponding change in the multiplicity factors in \eqref{eq:GeneralP} and \eqref{eq:GlobalP}.

The detailed-balance equation corresponding to \eqref{eq:kin1} now becomes,
\begin{equation}
Z^{-1} \Omega(N_R,N_B;M_R,M_B;\ldots)\cdot  \frac{N_R}{N_\mathrm{tot}} \frac{1}{b_\mathrm{tot}} \cdot \frac{1}{2} = 
 Z^{-1} \Omega(N_R-1,N_B;M_R+1,M_B;\ldots) \cdot \frac{M_R+1}{N_\mathrm{tot}} \frac{1}{b_\mathrm{tot}} \cdot \frac{1}{2}, \label{eq:kin2} 
\end{equation}
where, on the left-hand side, $Z^{-1} \Omega(N_R,N_B;M_R,M_B;\ldots)$ gives the probability of being in the originating state, ${N_R}/{N_\mathrm{tot}}$ is the probability of selecting a red agent from the originating block, $1/b_\mathrm{tot}$ is the probability of selecting the destination block with $b_\mathrm{tot}$ being the total number of blocks in the system, and the final factor $1/2$ is the acceptance probability for when $\Delta H=0$ in the absence of social interactions or location preferences. As before, the right-hand side represents the reverse transition with exactly the same logic, and an exactly corresponding equation follows if the moving agent happens to be blue. 
Rearranging the above equation and its blue counterpart gives
\begin{equation}
    \left\{
    \begin{aligned}
        \frac{\Omega(N_R,N_B;M_R,M_B;\ldots)}{\Omega(N_R-1,N_B;M_R+1,M_B;\ldots)} & =  \frac{1}{N_R} \cdot (M_R+1)\\
        \frac{\Omega(N_R,N_B;M_R,M_B;\ldots)}{\Omega(N_R,N_B-1;M_R,M_B+1;\ldots)} & =  \frac{1}{N_B} \cdot (M_B+1) 
    \end{aligned}
    \right., \label{eq:recurr21} 
\end{equation}
which again are separable. Following the argument presented in Section \ref{sec:uniform}, we look for a function $\omega$ satisfying
\begin{equation}
    \left\{
    \begin{aligned}
        \frac{\omega(N_R,N_B)}{\omega(N_R-1,N_B)} & = \frac{1}{N_R}\\
        \frac{\omega(N_R,N_B)}{\omega(N_R,N_B-1)} & = \frac{1}{N_B}
    \end{aligned}
    \right..\label{eq:RecurBlock2}
\end{equation}
Applying Eq.~\eqref{eq:RecurBlock2} recursively, we obtain
\begin{equation} \label{eq:omega21}
\omega(N_R,N_B) = \frac{1}{N_R\cdots 1}\cdot \omega(0,N_B) = \frac{1}{N_R \cdots 1}\cdot \frac{1}{N_B \cdots 1}\cdot \omega(0,0) \equiv \frac{1}{N_R!\,N_B!},
\end{equation}
where again we set the undetermined, ultimately irrelevant normalization constant by setting $\omega(0,0)\equiv 1$, reflecting that a block containing zero agents is a unique state.
Finally, inserting the above result into the separated form for the full multiplicity factor \eqref{eq:sepfull} gives
\begin{equation} \label{eq:mult2}
    \Omega(\{N_{R},N_{B}\}) = C \, \prod_b \frac{1}{N_{R,b}!\,N_{B,b}!} = \left(\begin{array}{c} N^\mathrm{tot}_R\\N_{R,1} \ \ldots \ N_{R,b_\mathrm{tot}}\end{array}\right) \left(\begin{array}{c} N^\mathrm{tot}_B\\N_{B,1} \ \ldots \ N_{B,b_\mathrm{tot}}\end{array}\right),
\end{equation}
where, with the choice $C=N^\mathrm{tot}_R!\,N^\mathrm{tot}_B!$ for the undetermined and irrelevant normalization factor, we find that the multiplicity factor corresponds to counting all possible ways of selecting which of the $N^\mathrm{tot}_R$ red agents and 
$N^\mathrm{tot}_B$ blue agents fall within each of the blocks of the system.

The key to understanding how the above microstate counting corresponds to the block-focused agent rule is to recognize the fact that destination blocks are now chosen at rates independent of the number of empty sites within them.  By not considering the empty locations as distinct locations that generate greater possibility for moving to a block with many empty locations, the block-focused rule treats all empty locations essentially as identical options, thereby making all spatial arrangements of agents within the block as identical microstates as well.  The only relevant characteristic left distinguishing microstates is then the identities of which agents are present in which block, corresponding to Eq.~\eqref{eq:mult2}.  To confirm this counting, one can generate exactly the same result by taking the original total number of possible states (Eq.~\eqref{eq:comb1}), which includes the spatial arrangement of agents within each block, and then divide by the total number of possible spatial arrangements of the agents within the blocks, $$\prod_b N_{R,b}! N_{B,b}! \left(\begin{array}{c}s\\N_{R,b}\,N_{B,b}\,(s-N_{R,b}-N_{B,b})\end{array}\right),$$ where the multinomial factors count the number of possible choices of red and blue sites within each block and the factorial factors count the number of ways to arrange the agents within the block among the chosen sites.\footnote{Following our detailed discussion in footnote\footref{ft:micro}, for agent rule (2), we can define microstates as unordered lists of distinguishable red and blue agents for each block; and allowed transitions as moving one agent from one block to another. One can check that the desired condition (corresponding to Eq.~\eqref{eq:microForm}) $F/E=(M_R+1)/N_R$ and its blue counterpart are satisfied. Thus the number of microstates Eq.~\eqref{eq:mult2} satisfies Eq.~\eqref{eq:recurr21}.}

As a final set of considerations, we explore what happens when a system actually controlled by the direct agent rules from Section~\ref{sec:uniform} is analyzed with the block-focused rules of this section. From the steady-state distribution \eqref{eq:DFFT_full2}, we would expect according to \eqref{eq:omega1} and \eqref{eq:mult2}, respectively, 
\begin{eqnarray}
P_b(N_R,N_B) & = & z_b^{-1}\frac{1}{N_R!\ N_B!\ (s-N_R-N_B)!} \exp\left[-f\left(N_R,N_B\right)-v_{R,b}N_{R}-v_{B,b}N_{B}\right]\label{eq:equivFs}\\
& = &  z_b^{-1} \frac{1}{N_R!\ N_B!} \exp\left[-\tilde{f}\left(N_R,N_B\right)-v_{R,b}N_{R}-v_{B,b}N_{B}\right], \nonumber
\end{eqnarray}
which will both therefore describe the steady-state equally well provided
\begin{equation} \label{eq:effF}
\tilde{f}\left(N_R,N_B\right) \equiv f\left(N_R,N_B\right) + \ln\left[\left( s-N_R-N_B\right)!\right].
\end{equation}
This means that the present, block-focused description will describe the higher incoming rate for low-population blocks resulting from the direct agent rule (which ultimately lead to the $\left( s-N_R-N_B\right)!$ terms) as an \emph{effective} additional social interaction term $\ln\left[\left( s-N_R-N_B\right)!\right]$. This effective social interaction term corresponds to higher frustrations at lower occupations, thereby appropriately reducing the probability of finding low occupation blocks. It is thus unnecessary to understand the precise form of the agent rules to describe fluctuations in the steady-state or to predict the response of the system to changes in total populations $N^\mathrm{tot}_R,N^\mathrm{tot}_B$ or changes in spatial preferences $v_{R,b},v_{B,b}$, using the considerations in Section~\ref{sec:mbd}.

Although equilibrium behaviors can be understood regardless of the model used for the multiplicity factors, care is needed to disentangle true preferences from effective preferences due to
the underlying rules when using frustrations to interpret social preferences, an issue best mitigated by employing multiplicity factors appropriate to the kinematics of the system under study. Also, misunderstanding of the underlying agent rules for a system can also distort predictions of time evolution. However, Eq.~\eqref{eq:equivFs} ensures that time evolution under both descriptions converges to the same steady state. Thus, so long as the changes in the system as it evolves are not sufficiently extreme to cause large changes in the effective interaction $\ln\left[\left( s-N_R-N_B\right)!\right]$, we may expect the time evolution to be well represented using either representation. Section~\ref{sec:altR} explores in detail the issue of the accuracy of dynamical predictions when using incorrect multiplicity factors.

We end our discussion of Agent Rule (2) by noting that certain gravity-type migration models also leads to the same multiplicity factor (Eq.~\eqref{eq:mult2}).  Consider an agent rule where agent is selected with direct method (in block $i$) and destination location (block $j$) is selected with a probability proportional to $v^\text{M}_{i,j}$. $v^\text{M}_{i,j}$ is called the `strength of migratory interaction' or `mobility factor' from block $i$ to block $j$. We further assume that $v^\text{M}_{i,j}=v^\text{M}_{j,i}$, following the treatment in the Weidlich-Haag Migratory Model \parencite{haag2017modelling} (where $v_{i,j}$ is used instead of $v^\text{M}_{i,j}$.\footnote{We add the additional superscript `M' to avoid confusion with our definition of vexations.})  and many other gravity-type migration models. The strength of migratory interaction may depend on the distance between the blocks (as people are less likely to move long-distance), or other socioeconomic factors. (See Ref. \parencite{haag2017modelling,weidlich1988interregional} for a more detailed discussion.) Under such conditions, the detailed balance condition \eqref{eq:kin2} will appear much the same, but with the factors of $1/b_\mathrm{tot}$ replaced with $v^\text{M}_{i,j}$ and $v^\text{M}_{j,i}$ on the left and right, respectively.
Due to the condition that $v^\text{M}_{i,j}=v^\text{M}_{j,i}$, ultimately, even though the time evolution of the system will be different, the detailed-balance condition (Eq.~\eqref{eq:kin2}) is equivalent, and we obtain the same results as above. To estimate $v^\text{M}_{i,j}$ from migration data for the purposes of time-evolution studies, one may use a least square procedure or Maximum likelihood estimation, where frustration and vexations are first determined from the steady-state distributions, a key benefit of the DFFT framework.
\subsubsection{Agent Rule (3), Block-focused agents and locations, simplest multiplicity factor}

As a final example, we consider the case where selection of both the moving agent and empty location are block-based. The detailed-balance equation is then

$$
Z^{-1} \Omega(N_R,N_B;M_R,M_B;\ldots)\cdot  \frac{1}{b_\mathrm{tot}} \frac{1}{b_\mathrm{tot}} \cdot \frac{1}{2} = 
 Z^{-1} \Omega(N_R-1,N_B;M_R+1,M_B;\ldots) \cdot \frac{1}{b_\mathrm{tot}} \frac{1}{b_\mathrm{tot}} \cdot \frac{1}{2}, %\label{eq:kin3} 
$$

which simplifies directly to
\begin{equation}
\Omega(N_R,N_B;M_R,M_B;\ldots) = \Omega(N_R-1,N_B;M_R+1,M_B;\ldots), \label{eq:recurr3}
\end{equation}
so that the multiplicity factors associated with any set of occupation numbers $\{N_{R},N_{B}\}$ connected by a transition are equal. Because we expect, for all but pathological examples, that any two sets of occupation numbers can be reached through some set of transitions, we conclude that $\Omega(\{N_{R},N_{B}\})$ must be a constant. Alternately, one can reach the same conclusion by proceeding with the analysis of the previous two examples.

To understand the combinatorial reason for this result, first note that, as in Section~\ref{sec:grav},
due to the selection of destination blocks regardless of the number of empty sites that they contain, the spatial arrangement of agents within blocks is irrelevant. Now, however, the selection of moving agents also is independent of the number within each block, so that movement of all agents within a block is equivalent and all importance of the identity of moving agents is removed. There is no longer need to to track even the identity of the agents in each block, as led to \eqref{eq:mult2}. As a result, the occupation numbers themselves are the only distinguishing feature between different states, and so\footnote{Following our detailed discussion in footnote\footref{ft:micro}, for agent rule (3), we can define microstates as unordered lists of indistinguishable red and blue agents for each block (which does not specify any additional structure to macrostates, so there is one microstate for each macrostate); and allowed transitions as moving one agent from one block to another. One can check that the desired condition (corresponding to Eq.~\eqref{eq:microForm}) $F/E=1$ and its blue counterpart are satisfied. Thus the number of microstates Eq.~\eqref{eq:mult3} satisfies Eq.~\eqref{eq:recurr3}. }
\begin{equation} \label{eq:mult3}
\Omega(\{N_{R},N_{B}\})=1.
\end{equation}
Finally, we note that \eqref{eq:mult3} corresponds precisely to the form \eqref{eq:DFFT_full3}, which is thus not a mere mathematical abstraction, but does indeed correspond to a system with a particular agent rule.

\section{Comparing DFFT and DFT}\label{sec:DFFTDFT}
In the introductory description of the underlying migration model (Section~\ref{sec:Under}), the global headache functional (Eq.~\eqref{eq:defH}) takes on the same general form $F[n]+\int_A v(x)n(x)\text{d}x$ used in density-functional theory (DFT), where $F[n]$ is a universal functional independent of the number of particles or potential $v(x)$ \parencite{hohenberg1964inhomogeneous,kohn1965self}. The specific form we consider in Section~\ref{sec:Under} corresponds to the well-known local density approximation (LDA) \parencite{kohn1965self} from the density-functional theory literature, \begin{equation}
    F[n_R,n_B]\equiv\int_A f(n_R(x),n_B(x))\text{d}x,\label{eq:LDA}
\end{equation} where the integrand at position $x$ only depends on densities at $x$.\footnote{LDA in the original form deals only with the exchange-correlation energy \parencite{kohn1965self}, which is only a part of $F$. The LDA states that $$E_\text{xc}[n]\approx\int n(x)\epsilon_\text{xc}(n(x))\text{d}x.$$ In DFFT, $F$ itself is analogous to $E_\text{xc}$. Instead of using the natural extension
\begin{equation}
    F[n_R,n_B]\equiv\int_A \left[n_R(x)f_1(n_R(x),n_B(x))+n_B(x)f_2(n_R(x),n_B(x))\right] \text{d}x,\label{eq:LDA2} 
\end{equation}
we use the simpler form in Eq.~\eqref{eq:LDA}, without loss of generality.} It is also possible to envision a form of $F[n_R,n_B]$ corresponding to the weighted density approximation (WDA) where the integrand at $x$ does not just depend on the densities at $x$, but some weighted density around $x$ \parencite{gunnarsson1976exchange}. Different forms for $F[n_R,n_B]$ can in principle describe different underlying migration models to be explored in future work. It is worth noting that in traditional Density-functional Theory (DFT), the rules for interactions between the individual entities are known exactly (e.g. the interactions between electrons) and approximate functionals, $F$, derived, in part, from these interactions are used to model their density distributions. In contrast, DFFT addresses the inverse problem by extracting a functional, $F$, from observations without knowledge of the details of the underlying interactions.

A key result of DFT is that the ground-state density of a system can be found by minimizing the energy density-functional \parencite{hohenberg1964inhomogeneous}. We applied this concept when finding the candidate headache functional (Eq.~\eqref{eq:defH}). 

\section{Ensemble of States}
An ensemble of states is required in order to obtain marginal block distributions $P_b$ and extract the DFFT functions. In the main text, we introduced an ensemble of Schelling simulations to demonstrate this concept. For real systems, there are two natural ways to obtain the ensemble. 

First, one may treat fluctuations of states over time in a chosen time window as an ensemble. Mendez et al (2018) used this approach when applying DFFT to a system of walking fruit flies. For human residential systems, however, the fluctuations of states might be too slow to obtain sufficient statistics for DFFT function extraction. In addition, a relatively large time window will interfere with time evolution prediction, because we need to treat all states in a given time window as an ensemble of samples for a fixed set of conditions, whereas the conditions may already change significantly over the time window.

Alternatively, one may build a collection of block compositions for a selected set of similar blocks at a given time as a way to extract joint probability distributions. For example, we may investigate the distribution of census-block compositions within a given county for a given census year. In this case, we can extract the DFFT functions at the county scale and then use those functions to make probabilistic predictions of how any given block within that county is likely to evolve into the future. 

\section{Interpretation of Frustration and Vexations}\label{sec:interp}
The main text indicates that ``The concavities of curves on this [global frustration $f$] surface indicates social preferences for having greater or fewer agents of a particular type'' and that ``agents avoid blocks with high vexation''. Here, we give a more precise description of these statements in the context of the underlying migration model.

Recall from Section \ref{sec:Under} that the partial derivatives of $f$ satisfy
\begin{equation}
    \frac{\partial f}{\partial n_R}=f_R,\text{ and } \frac{\partial f}{\partial n_B}=f_B.
\end{equation}
The relative value of the first partial derivatives between two locations, $$\frac{\partial f}{\partial n_R}(n_R'',n_B'')-\frac{\partial f}{\partial n_R}(n_R',n_B')\ \text{, and  }\frac{\partial f}{\partial n_B}(n_R'',n_B'')-\frac{\partial f}{\partial n_B}(n_R',n_B'),$$ 
then compare the level of dissatisfaction of a red agent or blue agent, respectively, for being at a location with one set of desities $(n_R'',n_B'')$ over another $(n_R',n_B')$. The ``concavities of curves'', by which we mean $\sfrac{\partial^2f}{{\partial n_R}^2}$ and $\sfrac{\partial^2f}{{\partial n_B}^2}$, inform us whether $\sfrac{\partial f}{\partial n_R}$ and $\sfrac{\partial f}{\partial n_B}$ increase as the density of the respective agent type increases while the other is fixed. Positive concavity thus indicates an increasing level of dissatisfaction as the density of the respective agent type increases, i.e., a preference for lower density of the respective agent type, and vice versa. 

\begin{figure}[H]
	\centering	\includegraphics[width=1\linewidth]{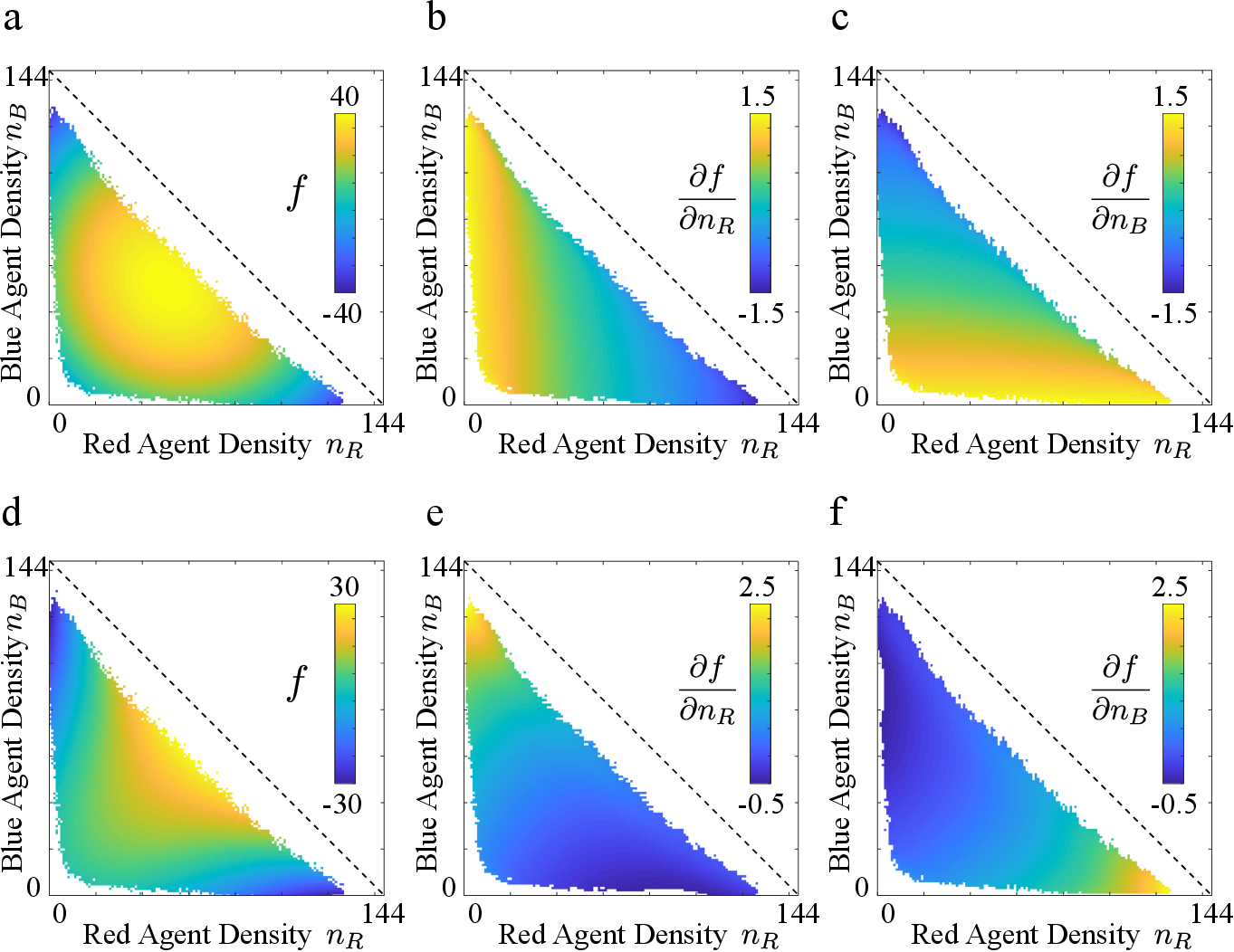}
	\caption{\textbf{Frustrations and their partial derivatives using analysis from main text (a-c) and when using block-focused multiplicity factors (d-f)} \textbf{(a.)} Frustration function presented in Fig.~3b, main text (calculated with agent rule from Section~\ref{sec:uniform}). The negative concavity along horizontal and vertical lines show that agents prefer blocks with high number of agents of their own type. \textbf{(b.)} The partial derivative of frustration in \textbf{a} with respect to red-agent density $n_R$, which shows how the level of dissatisfaction of a block for red agents decreases as the number of red agents increases, but remains roughly constant with the number of blue agents. Note that, throughout this section, we use the finite difference approximation for partial derivatives. \textbf{(c.)} The partial derivative of frustration in \textbf{a} with respect to blue-agent density $n_B$, which shows how the level of dissatisfaction of a block for blue agents decreases as number of blue agents increases, but remains roughly constant with the number of red agents. \textbf{(d.)} Frustration extracted from DFFT analysis of the same steady-state data as \textbf{a}, but using the agent rule discussed in Section \ref{sec:grav}.  We still observe negative concavities along horizontal and vertical lines, but less pronounced compared to \textbf{a}. A lower dissatisfaction at low total agent density is expected because thus frustration here additionally accounts for the tendency of agents to move to blocks with lower total density of agents.  \textbf{(e.)} The partial derivative of frustration in \textbf{d} with respect to red-agent density $n_R$, showing that the level of dissatisfaction of red agents is low both for blocks with low total density of agents and blocks with high red agent density, with the level of dissatisfaction high for blocks with high blue agent density. \textbf{(f.)} The partial derivative of frustration in \textbf{d} with respect to blue-agents density $n_B$, showing that the level of dissatisfaction of blue agents is low for both blocks with low total density of agents and with high blue agent density, with the level of dissatisfaction high for blocks with high red agent density.}\label{Der}
\end{figure}

For example, the downward concavity of the extracted frustration in the main text (repeated here in Fig.~\ref{Der}a) indicates that both types of agents prefer blocks with more agents of the same type. More specifically, the partial derivative with respect to the density of red agents $n_R$ (Fig.~\ref{Der} ) shows that the level of dissatisfaction of a block for red agents decreases with the number of red agents, and remains roughly constant with the number of blue agents. One might notice the similarity between $-\sfrac{\partial f}{\partial n_R}$ and the definition of the social utility for red agents $U_R^\text{so}$ (Fig.~2b). This is because $-\sfrac{\partial f}{\partial n_R}$ for the version of DFFT described in the main text plays a role simular to $U^\text{so}_R$ in the Schelling model, albeit using different variables as input (densities in a block versus number of 8-connected neighbors). If we instead describe this system by implementing DFFT using multiplicity factors from Section~\ref{sec:grav}, where we assume that agents propose moves to all blocks with equal probability, we would extract a different frustration (Fig.~\ref{Der}d), which captures the extra tendency that agents prefer to move to blocks with lower number of agents as an additional effective frustration as described in Equation~\eqref{eq:effF}. This effect is clear if we compare $\sfrac{\partial f}{\partial n_R}$ in both cases (Figs.~\ref{Der}b,e). We see that in Fig.~\ref{Der}e, red agents now have an almost equal preference for blocks with low total-agent density and high red-agent density. Similar observations for $\sfrac{\partial f}{\partial n_B}$ also hold. Section \ref{sec:Segregation} explores the frustration behaviors associated with different underlying social utility functions $U^\text{so}_R,U^\text{so}_B$.

Finally, the extracted block-level vexations can similarly be mapped to the underlying model. The relative values of the block-average average vexations, $v_{R,b''}-v_{R,b'}$ or $v_{B,b''}-v_{B,b'}$, compare the level of dissatisfaction of a red agent or blue agent, respectively, for one block $b''$ over another $b'$. The similarity between the extracted vexations $v_{R,b}$ and $v_{B,b}$ (Fig.~3b) and the definitions of spatial utilities $U_R^\text{sp}$ and $U_B^\text{sp}$ (Figs.~2d,e) is then a direct result of the similar roles played by the vexations and the spatial utilities in their repsective models.

\section{Frustration as a Detailed Measure of Segregation}\label{sec:Segregation}

In the main text we suggested that `DFFT could serve as a detailed lens into the social and spatial nature of racial residential segregation'. As a first attempt, we demonstrate with simulated data how frustration can serve as a more detailed measure of segregation than the Multi-Group Entropy Index \parencite{iceland2004multigroup} used by the U.S. Census Bureau \parencite{website}.

To generate data for comparison, we use the Schelling simulation described in the main text, but set
\begin{equation}
\left\{
\begin{aligned}
    &U_R^\text{so}(N^\text{ne}_R,N^\text{ne}_B)=\beta_R\cdot N^\text{ne}_R\\
    &U_B^\text{so}(N^\text{ne}_R,N^\text{ne}_B)=\beta_B\cdot N^\text{ne}_B\\
    &U_R^\text{sp}(x)=0\\
    &U_B^\text{sp}(x)=0
\end{aligned}
\right.,
\end{equation}
where we change $\beta_R$ and $\beta_B$ to test the behavior of frustration and Multi-Group Entropy Index. After running these simulations sufficiently long to obtain an ensemble of states with good statistics, we then extract the DFFT frustration as described in the main text.

We first demonstrate that the level of concavities of frustration reflects the level of segregation in much the same way as does the Multi-Group Entropy Index. Due to the similarity between the partial derivatives of frustration and the social utilities (See SI Section \ref{sec:interp}), we expect that the concavities of the extracted frustration will become more negative as we increase the level of segregation in the simulated city by increasing $\beta_R$ and $\beta_B$. Indeed, this is what we observe in Fig.~\ref{Entropy}. The frustration is a flat plane with statistical noise when $\beta_R=0$ and $\beta_B=0$ suggesting no social interactions between red and blue agents and the multi-group segregation index is a low value of 0.01. As $\beta_R$ and $\beta_B$ increase, the frustration develops more negative concavities (note the increasing color bar scales) and the multi-group entropy index likewise increases. 

We next demonstrate that frustration gives sufficiently detailed information to distinguish between different types of segregation for which the Multi-group Entropy Index gives identical values. We can easily generate a family of segregation patterns with the same Multi-group Entropy Index by altering the ratio and size of $\beta_R$ and $\beta_B$. For example, in Fig.~\ref{Entropy2}a, c, and e, we show segregation where red and blue agents show different degrees of clustering that result in the same Multi-Group Entropy Index of 0.03. Comparing the concavities of the frustrations for these three scenarios (Figs.~\ref{Entropy2}b, d, and f), on the other hand, quickly reveals the differences in the underlying behavior. The DFFT frustration function therefore captures additional details regarding segregation, including differences in red and blue agent behavior, that the multi-group entropy index does not measure directly.

The reason that the frustration in Figs.~\ref{Entropy}b,d,f are defined over slightly different domains is that frustration values can be extract only from density combinations actually observed in the data, making it sometimes difficult to compare multiple frustrations. In such situations, one may construct as a segregation index the average concavities of the frustration over their respective domains, $\overline{\sfrac{\partial^2f}{{\partial n_R}^2}}$ and $\overline{\sfrac{\partial^2f}{{\partial n_B}^2}}$. In fact, with detailed information captured by frustration, one may devise a wide variety of segregation indices. For heterogeneous environments, one may also include the spatial contribution to segregation by manipulating the extracted vexations.

\section{TD-DFFT Framework}
DFFT framework provides analytical predictions of new steady state or quasi-static time evolution (Section \ref{sec:PredSteadyState}). TD-DFFT further predicts the time evolution of an out-of-equilibrium city toward its steady-state distribution. In this section we derive Kohn-Sham TD-DFFT and Hohenberg-Kohn TD-DFFT used in the main text.

\subsection{Time-Dependent DFFT model (Kohn-Sham TD-DFFT)}\label{sec:SI-TD-DFFT}
To predict the time evolution of an out-of-equilibrium city toward its steady-state distribution, one can simply evolve the ensemble of altered states according to the coarse-grained underlying migration model (Section \ref{sec:coarsegrain}), where agents propose moves from block $b$ to $b'$ according to the proposed agent rules, and accepts moves with probability $P_{b\to b'}=1/(1+e^{\Delta H_b+\Delta H_{b'}})$, where $H_b$ is defined in Eq.~\eqref{eq:DefBlockH} that can be extracted directly from coarse-grained observations as described above. We call this model the Time-Dependent DFFT (TD-DFFT) model and this approach the Kohn-Sham TD-DFFT (as discussed in Section 4.1 of the main text). TD-DFFT model, by construction, yields equilibrium distributions that follow \eqref{eq:GeneralP}. Moreover, because the underlying agent dynamics is maintained, but now with the dissatisfaction function replaced with the block-level headaches which represent an effective average utility over the block, the TD-DFFT model is expected to approximate the time evolution of the coarse-grained underlying migration model.

\subsection{Master and Mean Value Equation (Hohenberg-Kohn TD-DFFT)}
While Kohn-Sham TD-DFFT possesses the ability to model dynamics in detail, it requires the knowledge of the full probability of the system $P(\{N_R,N_B\})$ and can become computationally costly to run. In this section, we derive the (approximate) mean-value equation (Eq.~(7) in the main text) from the
Kohn-Sham TD-DFFT model. For the sake of clarity, we begin with a detailed derivation for systems consisting of a single agent type, and then generalize the result to the case for multiple agents in Section \ref{sec:MVE}.

\subsubsection{Single agent type}
For a single-agent-type system, each state is specified by the particular number of agents in each block $\{N\}\equiv[N_1,\cdots,N_{b_\text{tot}}]$. Suppose at time $t=0$, the ensemble of states has a distribution $P(\{N\},t=0)$. Denote by $\nu_{b'\to b''}(\{N\})$ the rate of transition from a state $\{N\}\equiv[N_1,\cdots,N_{b_\text{tot}}]$ to another state $\{N\}^{b'\to b''}\equiv[N_1,\cdots,N_{b'}-1,\cdots,N_{b''}+1,\cdots,N_{b_\text{tot}}]$ (the rate at which an agent from block $b'$ moves to block $b''$, starting from state $\{N\}$). The following master equation describes the time evolution of $P(\{N\},t)$:
\begin{equation}
    \frac{\partial}{\partial t}P(\{N\},t)= \left[\sum_{b'} \sum_{b''\ne b'} \nu_{b'\to b''}(\{N\}^{b'' \to b'})\cdot P(\{N\}^{b'' \to b'},t)\right]-\left[\sum_{b'} \sum_{b''\ne b'}\nu_{b'\to b''}(\{N\})\cdot P(\{N\},t)\right],\label{eq:master_simplified}
\end{equation}
where the first bracket gives the total probability flow from other states, $\{N\}^{b'' \to b'}$, to state $\{N\}$ due to transitions of agents from block $b'$ to block $b''$ (as $b'$ and $b''$ ranges over all pairs of different blocks), and the second bracket gives the total probability flow from state $\{N\}$ to other states, $\{N\}^{b' \to b''}$, due to transitions of agents from block $b'$ to block $b''$ (as $b'$ and $b''$ ranges over all pairs of different blocks). 

Astute readers may have noticed that some transitions will lead to physically meaningless states, such as states with negative number of agents in certain blocks. Eq.~\eqref{eq:master_simplified} and subsequent derivations hold if we set the transition rates to these states $\nu$ and the probability of these states $P(\{N\},t)$ to be zeros, such that summations over $\{N\}$ can be considered to sum over all states with an integer number of agents and a fixed $N_\text{tot}=\sum_bN_b$ .

The central object of concern in the mean value equation is the average number of agents in a particular block $b$, $\overline{N_b}$, given by
\begin{equation}
\overline{N_{b}} = \sum_{\{N\}} N_{b}(\{N\}) \cdot P(\{N\} ,t),\label{eq:mean_value}
\end{equation}
where $N_b(\{N\})$ is the corresponding number of agents in block $b$ for state $\{N\}$. Combining Eq.~\eqref{eq:master_simplified} and Eq.~\eqref{eq:mean_value}, the time evolution of the average can be determined:
\begin{equation}
\begin{aligned}
\diff{}{t} \overline{N_{b}} = & \sum_{\{N\}} N_{b}(\{N\}) \cdot \frac{\partial}{\partial t}P(\{N\} ,t) \\
= &  \left[\sum_{b'} \sum_{b''\ne b'} \sum_{\{N\}} N_{b}(\{N\}) \cdot \nu_{b'\to b''}(\{N\}^{b'' \to b'})\cdot P(\{N\}^{b'' \to b'},t)\right]\\
& \qquad \qquad \qquad \qquad \qquad  -\left[\sum_{b'} \sum_{b''\ne b'} \sum_{\{N\}} N_{b}(\{N\}) \cdot\nu_{b'\to b''}(\{N\})\cdot P(\{N\},t)\right]\\
=& \left[\sum_{b'} \sum_{b''\ne b'} \sum_{\{N\}} N_{b}(\{N\}^{b'\to b''}) \cdot \nu_{b'\to b''}(\{N\})\cdot P(\{N\},t)\right]\\
& \qquad \qquad \qquad \qquad \qquad -\left[\sum_{b'} \sum_{b''\ne b'} \sum_{\{N\}} N_{b}(\{N\}) \cdot\nu_{b'\to b''}(\{N\})\cdot P(\{N\},t)\right]\\
=& \sum_{b'} \sum_{b''\ne b'} \sum_{\{N\}} \left[N_{b}(\{N\}^{b'\to b''})-N_{b}(\{N\})\right]\cdot \nu_{b'\to b''}(\{N\})\cdot P(\{N\},t).
\end{aligned}\label{eq:mean_value_evolution0}
\end{equation}
The third equality in Eq.~\eqref{eq:mean_value_evolution0} is obtained from changes of variables $\{N\}\to\{N\}^{b'\to b''}$ for the first bracket. The sum over $\{N\}$ is unchanged since it is the sum over all states (with constant $N_\text{tot}$). Observe that, for $b'\ne b''$,
\begin{equation}
N_{b}(\{N\}^{b'\to b''})-N_{b}(\{N\})=
\begin{cases}
1 & \text{if $b''=b$}\\
-1 & \text{if $b'=b$}\\
0 & \text{otherwise}
\end{cases}.\label{eq:cases}
\end{equation}
Substituting Eq.~\eqref{eq:cases} into Eq.~\eqref{eq:mean_value_evolution0} gives
\begin{equation}
\begin{aligned}
\diff{}{t} \overline{N_{b}}
=& \left[\sum_{b' \neq b} \sum_{\{N\}} \nu_{b'\to b}(\{N\})\cdot P(\{N\},t)\right]-\left[\sum_{b'' \neq b} \sum_{\{N\}}\nu_{b\to b''}(\{N\})\cdot P(\{N\},t)\right]\\
=& \sum_{b'\ne b} \overline{\nu_{b'\to b}(\{N\})} - \sum_{b''\ne b} \overline{\nu_{b\to b''}(\{N\})}\\
= & \sum_{b' \neq b} \left[\overline{\nu_{b'\to b}(\{N\})}-\overline{\nu_{b\to b'}(\{N\})}\right],
\end{aligned}\label{eq:mean_value_evolution2}
\end{equation}
where we have changed the variable name $b''$ to $b'$, and defined the average rate of transition 
\begin{equation*}
    \overline{\nu_{b\to b'}(\{N\})}\equiv \sum_{\{N\}}\nu_{b\to b'}(\{N\})\cdot P(\{N\},t).
\end{equation*}

Up to this point, the derivation has been exact. However, if the probability distribution is sharply peaked, then any likely block occupation is well represented by its average. Explicitly, this corresponds to the following rate approximation,
\begin{equation}
   \overline{\nu_{b\to b'}(\{N\})}\approx\nu_{b\to b'}(\{\overline{N}\}),\label{eq:over_nu}
\end{equation}
where the function $\nu_{b\rightarrow b'}$ is interpolated to allow for non-integer inputs $\{\overline{N}\}\equiv[\overline{N_1},\cdots,\overline{N_{b_\text{tot}}}]$. Employing the approximation Equation~\eqref{eq:over_nu} in Equation~\eqref{eq:mean_value_evolution2} results in the mean-value equation for the case of a single-agent type.

\subsubsection{Two agent types}\label{sec:MVE}
For the case of two agent types, the master equation governing the time evolution of the probability distribution $P(\{N_{R},N_{B}\},t)$ becomes
\begin{equation}
\begin{aligned}
\frac{\partial}{\partial t}P(\{N_{R},N_{B}\},t)=&-\sum_{b'}\sum_{b''\ne b'}\nu_{R,b'\to b''}(\{N_{R},N_{B}\})\cdot P(\{N_{R},N_{B}\},t)\\
&-\sum_{b'}\sum_{b''\ne b'}\nu_{B, b'\to b''}(\{N_{R},N_{B}\})\cdot P(\{N_{R},N_{B}\},t)\\
&+\sum_{b'}\sum_{b''\ne b'}\nu_{R, b'\to b''}(\{N_{R},N_{B}\}^{R,b'' \rightarrow b'})\cdot P(\{N_{R},N_{B}\}^{R,b'' \rightarrow b'},t)\\
&+\sum_{b'}\sum_{b''\ne b'}\nu_{B, b'\to b''}(\{N_{R},N_{B}\}^{B,b'' \rightarrow b'})\cdot P(\{N_{R},N_{B}\}^{B,b'' \rightarrow b'},t),
\end{aligned}\label{eq:master}
\end{equation}
where $\{N_{R},N_{B}\}^{R,b'' \rightarrow b'}$ denotes a state that converts to state $\{N_{R},N_{B}\}$ when a red agent moves from block $b'$ to block $b''$. The transition rates $\nu_{R,b'\to b''}$ and $\nu_{B,b'\to b''}$ are defined as the probabilities that a red or blue agent moves from block $b'$ to block $b''$ in one time step, respectively.

For the version of DFFT described in the main text (Agent Rule (1) in section \ref{sec:version}), we find that
\begin{equation}
\left\{
\begin{aligned}
&\nu_{R,b'\to b''}(\{N_{R},N_{B}\})=\frac{N_{R,b'}}{N_\text{tot}}\cdot \frac{s-N_{R,b''}-N_{B,b''}}{s_\text{tot}-N_\text{tot}}\cdot \frac{1}{1+e^{\Delta H_{b''}+\Delta H_{b'}}}\\
&\nu_{B,b'\to b''}(\{N_{R},N_{B}\})=\frac{N_{B,b'}}{N_\text{tot}}\cdot \frac{s-N_{R,b''}-N_{B,b''}}{s_\text{tot}-N_\text{tot}}\cdot \frac{1}{1+e^{\Delta H_{b''}+\Delta H_{b'}}}
\end{aligned},\label{eq:nu_def}
\right.
\end{equation}
where the first multiplicand on each line is the probability of choosing the potentially moving an agent from block $b'$, the second is the probability of choosing an empty location for the transition from block $b''$, and the third is the probability of accepting the proposed move. Note that $\Delta H_{b''}$ and $\Delta H_{b'}$ are changes in the respective block headaches $H_b$ that would occur were the transition to happen.

To derive the mean-value equation (MVE) for the multiple agent-type case, we start again with the definition of mean value and take the derivative, employing the master equation. For example, the average number of red agents in block $b$ is
$$\overline{N_{R,b}}\equiv \sum_{\{N_{R},N_{B}\}}N_{R,b}\cdot P(\{N_{R},N_{B}\},t),$$
which gives
\begin{equation}\label{eq:Mean}
    \diff{}{t}\overline{N_{R,b}}=\sum_{\{N_{R},N_{B}\}}N_{R,b}\cdot \frac{\partial}{\partial t}P(\{N_{R},N_{B}\},t),
\end{equation}
with corresponding equations holding for the blue agents.
After substituting Eq.~\eqref{eq:master} into Eq.~\eqref{eq:Mean} and manipulate as in Eq.~\eqref{eq:mean_value_evolution0}, we obtain the direct analogue of Equation~\eqref{eq:mean_value_evolution2},
\begin{equation} 
\begin{aligned}
    \diff{}{t}\overline{N_{R,b}}&=\sum_{b'\ne b}\Bigg[\sum_{\{N_{R},N_{B}\}}(+1)\cdot\nu_{R,b'\to b}(\{N_{R},N_{B}\})\cdot P(\{N_{R},N_{B}\},t)\\
    &\qquad\quad+\sum_{\{N_{R},N_{B}\}}(-1)\cdot\nu_{R,b\to b'}(\{N_{R},N_{B}\})\cdot P(\{N_{R},N_{B}\},t)\Bigg]\\
    &=\sum_{b'\ne b}\Big[\overline{\nu_{R,b'\to b}}-\overline{\nu_{R,b\to b'}}\Big],
\end{aligned} \label{eq:mamev}
\end{equation}
with a corresponding result for the blue agents.

Finally, as above for the single agent case, if the distribution $P(\{N_{R},N_{B}\},t)$ is sharply peaked, then we can make the approximation
\begin{equation}
\left\{
\begin{aligned}
    &\overline{\nu_{R,b'\to b}}\approx\nu_{R,b'\to b}(\{\overline{N_{R}},\overline{N_{B}}\})\\
    &\overline{\nu_{R,b\to b'}}\approx\nu_{R,b\to b'}(\{\overline{N_{R}},\overline{N_{B}}\})
\end{aligned}\right.,
\end{equation}
which, when combined with Eq.~\eqref{eq:mamev}, gives exactly the MVE that appears in the main text. Note that expressions $\nu_{R,b\to b'}(\{\overline{N_{R}},\overline{N_{B}}\}$, $\nu_{B,b\to b'}(\{\overline{N_{R}},\overline{N_{B}}\}$ can be evaluated with Eq.~\eqref{eq:nu_def} as a reasonable interpolation of non-integer values, as presented in the main text.

\subsection{Time Scale Adjustment} \label{sec:timescale}

When working with observational data, there is no direct access to the underlying social utility function or agent rules.  As described above, TD-DFFT headache functions can be extracted from the observed probability distributions of fluctuations in the steady state. However, these distributions contain no direct information about the underlying agent rules and, in particular, no information about the time scales associated with the system dynamics. Therefore, when using TD-DFFT to predict dynamics directly from observational data, it is important to choose agent rules based upon the best available understanding of rules governing the underlying system (for example, whether the rate at which an agent moves is independent of or proportional to the number of available options), so that the \emph{relative} rates of moves within the system are preserved under as great a range of situations as possible. When dealing with data from a known model, the fundamental \emph{overall} time-scale is known (for example, a single, discrete time-step in a Schelling simulation). However, when working particularly with real-world data, the overall time scale for moves must be determined through empirical observation (for example, whether it is 4 weeks, 18 months, 5 years, etc.). For example, if we find $M^\text{obs}$ moves per unit time in the steady state of the empirical system under observation, and and average of $M^\text{TDDFFT}$ moves per iteration in the TD-DFFT steady states, we can multiply the number of iterations in the TD-DFFT system by $M^\text{TDDFFT}/M^\text{obs}$ to scale to real times in the observed system. This normalization has been done for all the TD-DFFT predictions we show in this paper.

This scaling can be useful even when there is no unknown underlying empirical time scale and the agent rules and time scale are known. For example, consider the use of TD-DFFT or its computationally simpler versions, such as the Master Equation or MVE approaches, to predict the outcome of an agent-based simulation whose underlying agent rule is known.  Under these conditions, the TD-DFFT calculations can use precisely the same agent rule. However, although the block headache functions represent an effective average dissatisfaction at the block level, individual agents may also tend statistically to occupy more favorable locations within the block, thus changing the perceived dissatisfaction of the same block as the agent is considering joining or leaving it\footnote{Note that we are not referring to the fact that the block has one more agent during the reverse leaving process, which is taken care of by the different choice of the finite-difference approximation (Section \ref{sec:coarsegrain})}, and eventually making the actual rate of moves different from that of the corresponding TD-DFFT model.  For example, under the agent movement rule employed in this work, the sum of the probabilities for a move and its reverse is always
$$\frac{1}{1+e^{\Delta H}}+\frac{1}{1+e^{-\Delta H}} = 1,$$
%whereas in practice we find that the underlying utility-based sum of forward and probabilities averages over time to
%$$\langle\frac{1}{1+e^{-\Delta U}}+\frac{1}{1+e^{\Delta U}}\rangle \approx 0.85.$$
whereas in the Schelling simulation the average probabilities of the corresponding moves as they happen over time give
$$\langle P_\text{Schelling}\rangle_{\text{move}}+\langle P_\text{Schelling}\rangle_{\text{reverse}} \approx 0.85.$$
(Recall that $P$ is the probability defined in Eq. (1) of the main text. Note that for a particular move identified by the number of agents in the corresponding blocks alone, there can be a wide range of $P_\text{Schelling}$ values depending on the spatial distribution of the agents within the blocks, hence the average.) As a result, we find for the simulations in the text that $M^\text{pred}/M^\text{obs}\approx1.16$, confirming that the TD-DFFT indeed approximates the dynamics and time-scales of the underlying model quite well.  Moreover, applying this factor to normalize the time constant to match the observed system gives the simplest correction for the discrepancy.  The time scales for all of the dynamic predictions presented in the text have been scaled in this way by the factor 1.16.

\section{Comparing TD-DFFT and TD-DFT}
\subsection{TD-DFFT Model and Kohn-Sham TD-DFT}
This section considers the relationship between the TD-DFFT model and the Kohn-Sham formulation of Time-dependent Density-Functional Theory \parencite{kohn1965self,runge1984density} within the adiabatic local density approximation (ALDA)\parencite{thiele2008adiabatic}.  The key to the latter formulation is the definition of an effective potential $v^\text{eff}$ in which non-interacting particles evolve so as to reproduce the time-evolution of the original interacting system. In our case\footnote{One may recognize the form of the effective potential better if we look at the continuum case. Recall in section \ref{sec:Under} that we defined  $$F[n_R,n_B]=\int_Af(n_R,n_B)\text{d}A,$$
then
\begin{equation}
    \left\{
        \begin{aligned}
            V^{\text{eff}}_{R}(x)=\frac{\delta H[n_R,n_B]}{\delta n_R}(x)=\frac{\delta F[n_R,n_B]}{\delta  n_R}(x)+V_R(x)=\frac{\partial f}{\partial n_R}\left(n_R(x),n_B(x)\right)+V_R(x)\\
            V^{\text{eff}}_{B}(x)=\frac{\delta H[n_R,n_B]}{\delta n_B}(x)=\frac{\delta F[n_R,n_B]}{\delta  n_B}(x)+V_B(x)=\frac{\partial f}{\partial n_B}\left(n_R(x),n_B(x)\right)+V_B(x)
        \end{aligned}
    \right..
\end{equation}
As we mentioned in Section \ref{sec:DFFTDFT}, $F$ is analogous to the exchange and correlation energy $E_\text{xc}$ in classical DFT \parencite{kohn1965self}},
the corresponding effective potential (vexation) of an agent in block $b$ can be found by 
\begin{equation}
    \left\{
        \begin{aligned}
            v^{\text{eff}}_{R,b}=\frac{\partial H}{\partial N_{R,b}}=\frac{\partial H_b}{\partial N_{R,b}}=\frac{\partial f}{\partial  n_R}\left(\frac{N_{R,b}}{A_{b}},\frac{N_{B,b}}{A_{b}}\right)+v_{R,b}\\
            v^{\text{eff}}_{B,b}=\frac{\partial H}{\partial N_{B,b}}=\frac{\partial H_b}{\partial N_{B,b}}=\frac{\partial f}{\partial  n_B}\left(\frac{N_{R,b}}{A_{b}},\frac{N_{B,b}}{A_{b}}\right)+v_{B,b}
        \end{aligned}
    \right.,
\end{equation}
so that the change in dissatisfaction as an agent moves from block $b'$ to block $b''$ is
\begin{equation}
\Delta h=
    \begin{dcases}
    v^\text{eff}_{R,b''}-v^\text{eff}_{R,b'} & \text{if agent is red}\\
    v^\text{eff}_{B,b''}-v^\text{eff}_{B,b'} & \text{if agent is blue}
    \end{dcases},
\end{equation}
which is precisely the approach we take in this work.

\subsection{MVE and Hohenberg-Kohn TD-DFT}
The time-dependent version of Hohenberg-Kohn DFT states that the density evolution of a system is given by the stationary point of an action functional \parencite{runge1984density}. Although we have not identified such an action, the MVE solves for the mean density evolution of the underlying migration model, and therefore corresponds to a Hohenberg-Kohn TD-DFFT.

\section{MVE and Bifurcation}
Section \ref{sec:MVE} demonstrates that MVE approximates the mean value well when the probability distributions are narrowly-peaked. When the distributions become wide or bimodal, typical with high levels of segregation in the system, MVE may yield multiple stable states. (Chapter 1 of Ref. \parencite{weidlich2012concepts} contains a more detailed discussion of such behavior.)

\begin{figure}[H]
	\centering
	\includegraphics[width=1\linewidth]{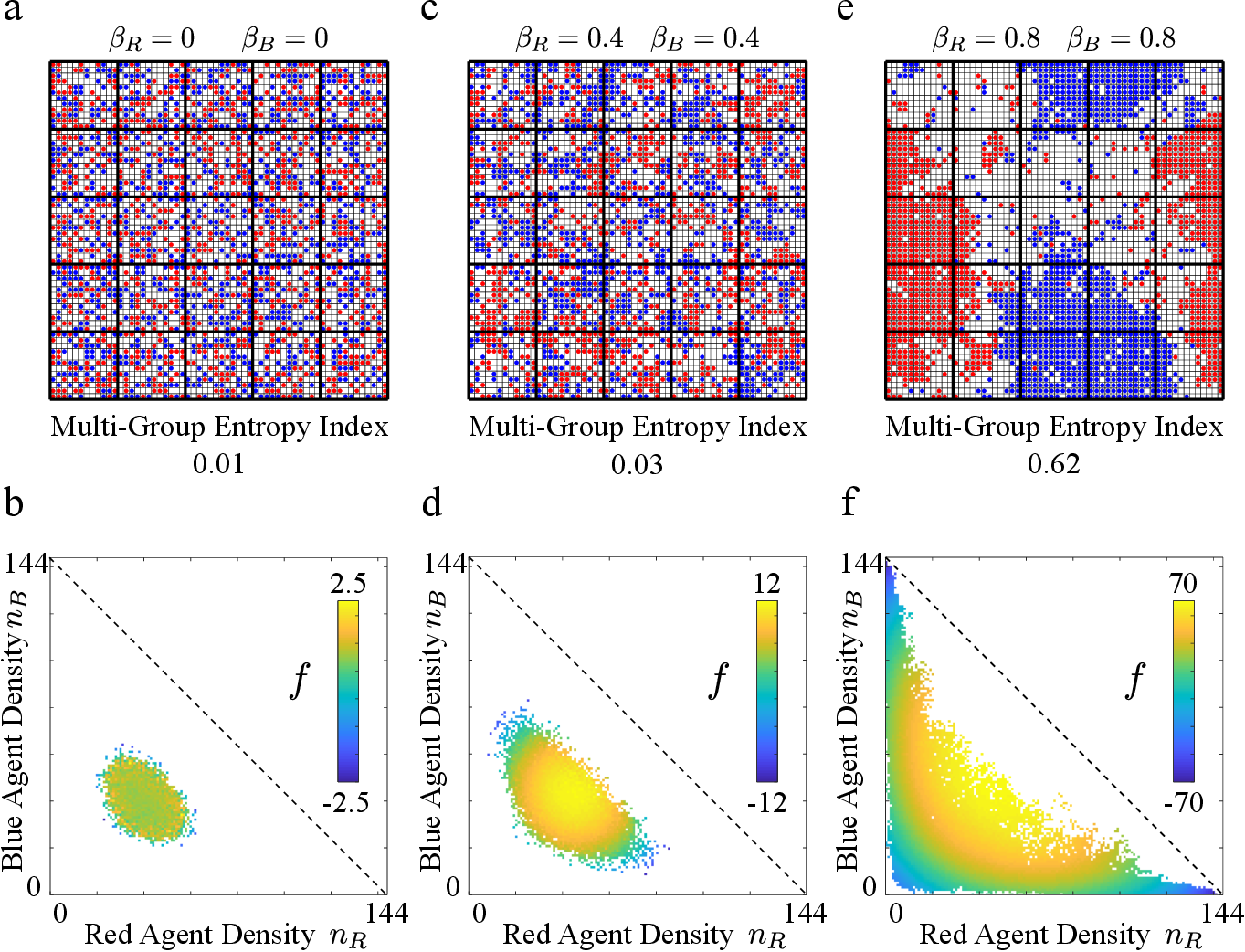}
	\caption{\textbf{Frustration as a measure of segregation} \textbf{(a.)} A snapshot of the steady-state distribution for $\beta_R=0$ and $\beta_B=0$. A low entropy index of 0.01 indicates low segregation. \textbf{(b.)} The extracted frustration for the distribution in \textbf{a} is a flat plane (zero concavities along curves in both directions), which also indicates low segregation. \textbf{(c.)} A snapshot of the steady-state distribution for $\beta_R=0.4$ and $\beta_B=0.4$. An intermediate entropy index of 0.03 indicates intermediate segregation. \textbf{(d.)} The extracted frustration for the distribution in \textbf{c} exhibits slightly negative concavities along curves in both directions, which also indicates intermediate segregation. \textbf{(e.)} A snapshot of the steady-state distribution for $\beta_R=0.8$ and $\beta_B=0.8$. A high entropy index of 0.62 indicates high segregation. \textbf{(f.)} The extracted frustration for the distribution in \textbf{e} exhibits very negative concavities along curves in both directions, also indicating high segregation.}\label{Entropy}
\end{figure}

\begin{figure}[H]
	\centering
	\includegraphics[width=1\linewidth]{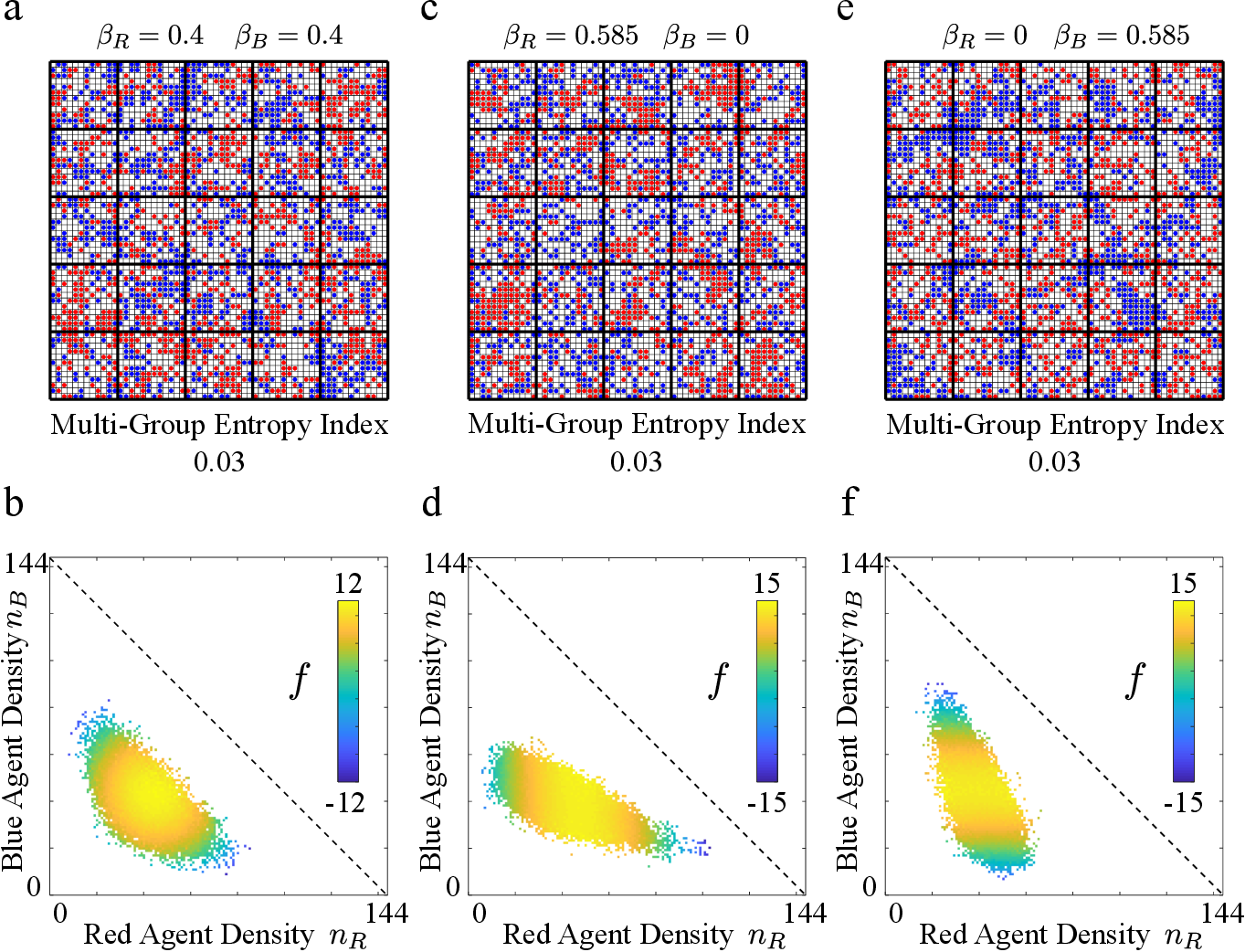}
	\caption{\textbf{Frustration functions capture more information than traditional segregation indices} \textbf{(a.)} Snapshot of the steady-state distribution for $\beta_R=0.4$ and $\beta_B=0.4$. We observe clustering of both red and blue agents to agents of the same type. An intermediate entropy index of 0.03 indicates intermediate segregation. \textbf{(b.)} The extracted frustration for the distribution in \textbf{a} exhibits slightly negative concavities along curves in both directions, indicating that segregation is effectively caused by social preferences from both types of agents. \textbf{(c.)} A snapshot of the steady-state distribution for $\beta_R=0.585$ and $\beta_B=0$. We observe clustering of red agents but not blue agents. An intermediate entropy index of 0.03 indicates the same intermediate segregation as \textbf{a}: the Entropy Index fails to capture the different nature of the segregation in this case. \textbf{(d.)} In contrast to \textbf{b}, the extracted frustration for the distribution in \textbf{c} exhibits negative concavities only along horizontal curves, which indicates that the observed segregation is due primarily to effective social preferences from the red agents. \textbf{(e.)} A snapshot of the steady-state distribution for $\beta_R=0$ and $\beta_B=0.585$. We observe clustering of blue agents but not red agents. An intermediate entropy index of 0.03 indicates the same intermediate segregation as \textbf{a} and \textbf{c}: the Entropy Index fails to capture the different nature of the segregation in this case. \textbf{(f.)} In contrast to \textbf{b} and \textbf{d}, the extracted frustration for the distribution in \textbf{e} exhibits negative concavities only along vertical curves, which indicates that the observed segregation is is due primarily to effective social preferences from the blue agents.}\label{Entropy2}
\end{figure}

\begin{figure}[H]
	\centering
	\includegraphics[width=0.8\linewidth]{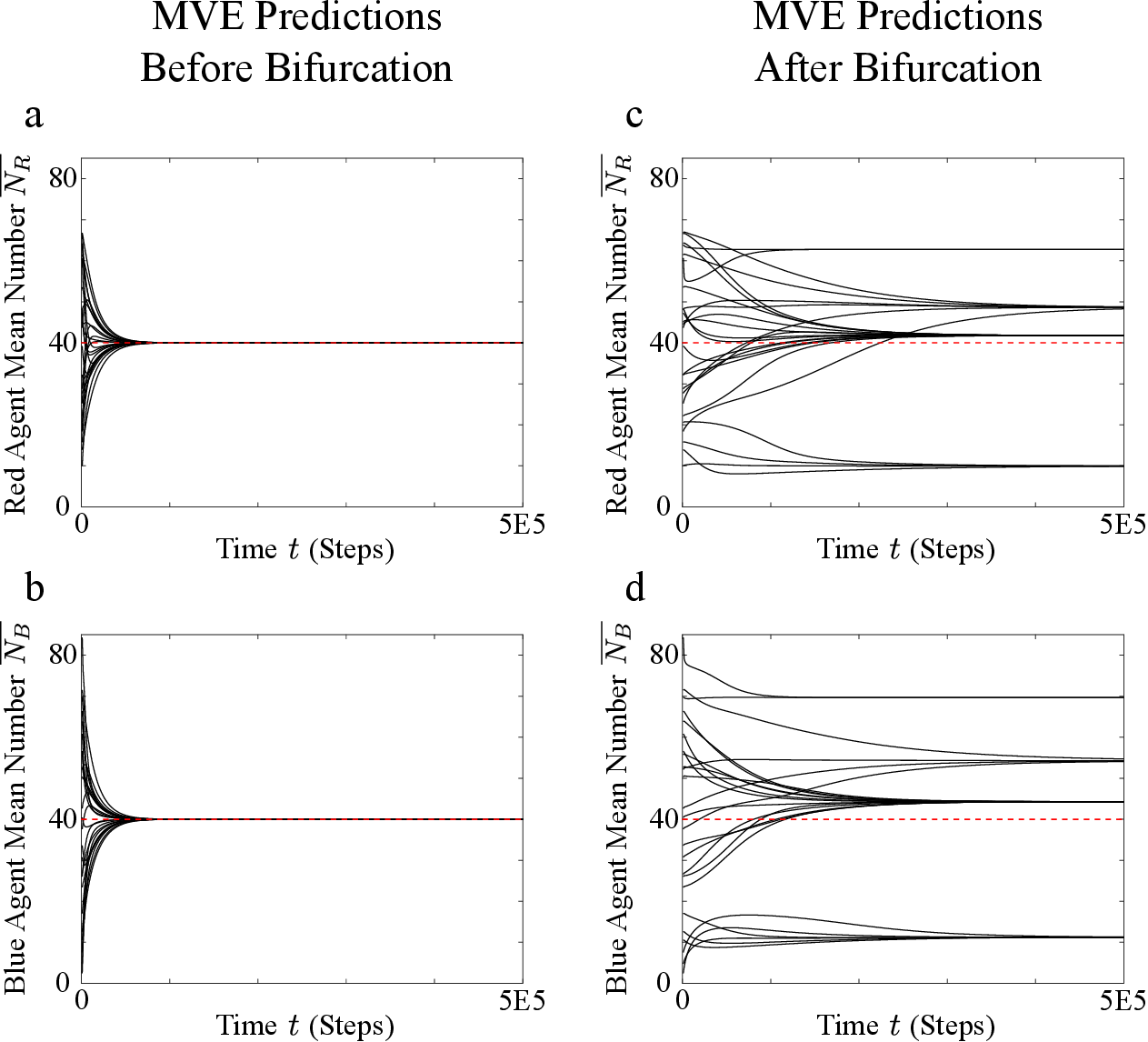}
	\caption{\textbf{Bifurcation Behavior of MVE} \textbf{(a,b.)} Before bifurcation, the MVE converges quickly to the expected state of $[40,...,40]$, the only stable node (state) in the system. \textbf{(c,d.)} After bifurcation, MVE no longer converges to the expected state (the true mean values for the TD-DFFT Model). Instead, it converges slowly to one of the many stable nodes that has been created in the bifurcation.}\label{Bifurcation}
\end{figure}

Fig.~\ref{Bifurcation} illustrates this behavior for simulations containing 1000 red agents and 1000 blue agents initially distributed randomly among 25 blocks, all with equal vexations. The figure compares the behavior of the MVE as the neighbor interaction strength increases from below to above the bifrucation threshold.   Because all blocks are identical, we would expect the steady-state mean number for both types of agent in each block to be $1000/25=40$, regardless of the frustration value. Using the frustration function from the main text (Fig. 3b), this is indeed the case (Fig.~\ref{Bifurcation}a,b). However, using the frustration extracted from a Schelling simulation with a high level of segregation (Fig.~\ref{Entropy}f), we find that all blocks converge slowly to different mean numbers of agents (Figs.~\ref{Bifurcation}c,d).

One way to interpret the post-bifrucation behavior of the MVE is to note that, at this strong level of segregation, the underlying model segregates into large, very stable islands (Fig.~\ref{Entropy}f), so that each block eventually converges to a different nearly-stable occupancy that depends randomly upon how the island boundaries happen to land with regard to the block boundaries.  The extent to which such system behavior is described by the corresponding tendency of the MVE occupancies to converge to different stable random values remains to be explored, and great caution should be exercised when interpreting MVE predictions in this regime. As a practical matter, to ensure that one is not in the bifrucation regime, we recommend running multiple MVE simulations from different random initial conditions to ascertain the stability of the results.

\section{Main Text Predictions for All Blocks}\label{sec:addC}

\begin{figure}[H]
	\centering
	\includegraphics[width=0.79\linewidth]{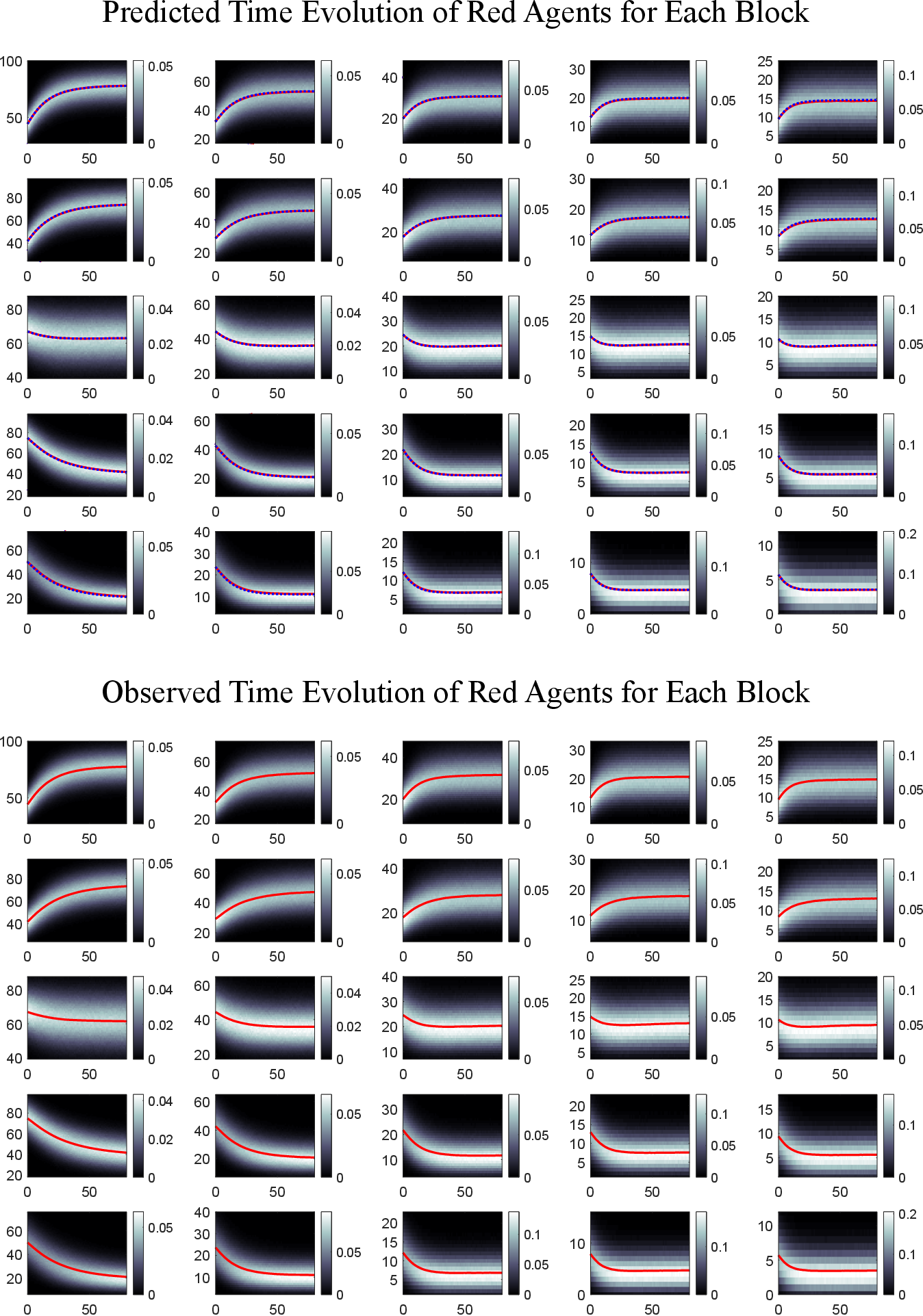}
	\caption{\textbf{Predicted versus observed time evolution of red agents for all blocks} Fig. 4b and d of the main text showed such a comparison for block SE, and here we show this comparison for all blocks. Results are presented for each individual block, arranged to reflect their location in the city: top-right panel is for block NE, bottom-left plot is for block SW, etc. Refer to the main text figure for axes labels and legends for each plot. We observe good agreement between TD-DFFT model/MVE and observations for all blocks.}\label{ABt}
\end{figure}

\begin{figure}[H]
	\centering
	\includegraphics[width=0.79\linewidth]{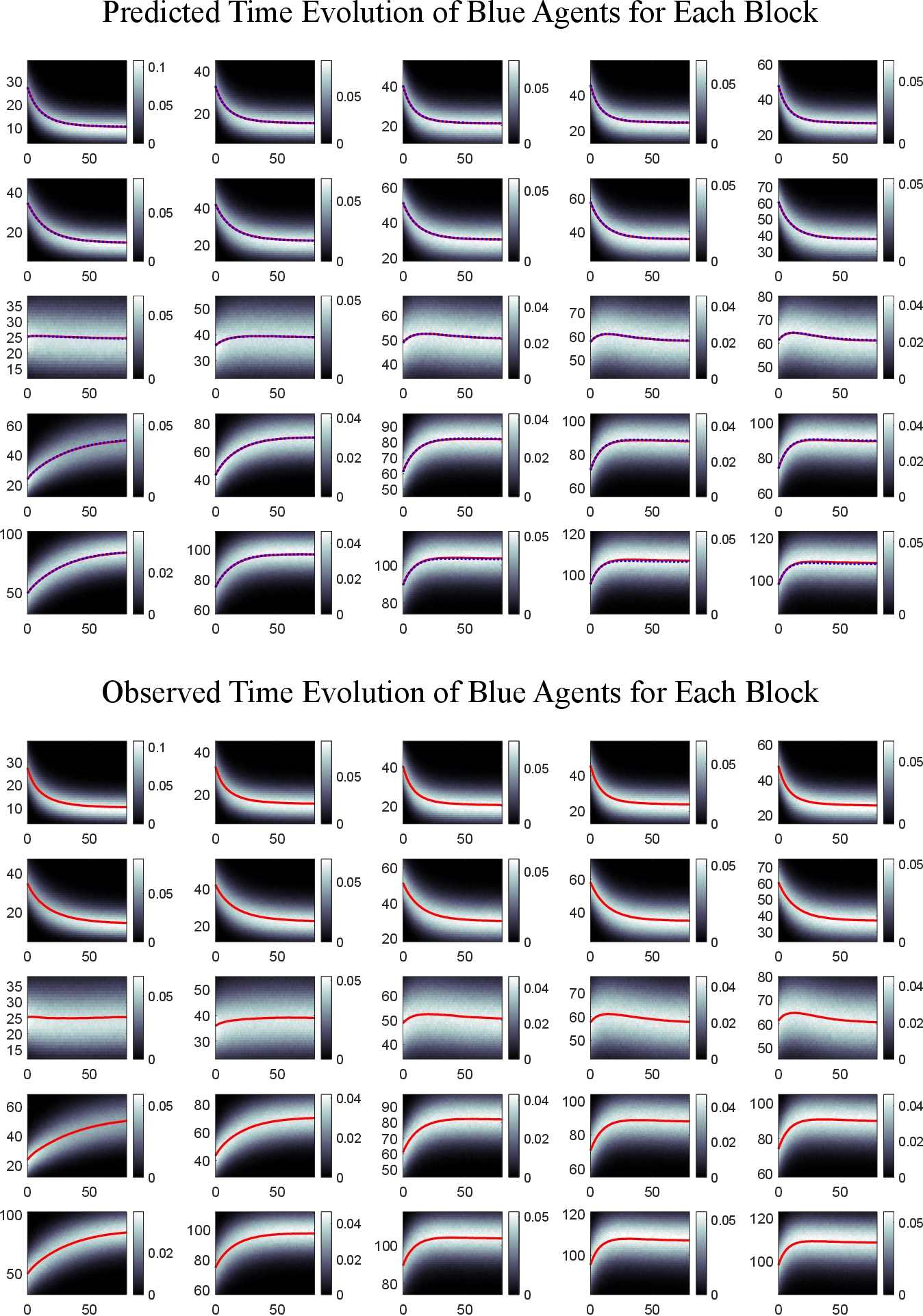}
	\caption{\textbf{Predicted versus observed time evolution of blue agents for all blocks} Fig. 4b and d of the main text showed such a comparison for block SE, and here we show this comparison for all blocks. Results are presented for each individual block, arranged to reflect their location in the city: top-right panel is for block NE, bottom-left plot is for block SW, etc. Refer to the main text figure for axes labels and legends for each plot. We observe good agreement between TD-DFFT model/MVE and observations for all blocks.}\label{ABtb}
\end{figure}

\begin{figure}[H]
	\centering
	\includegraphics[width=0.79\linewidth]{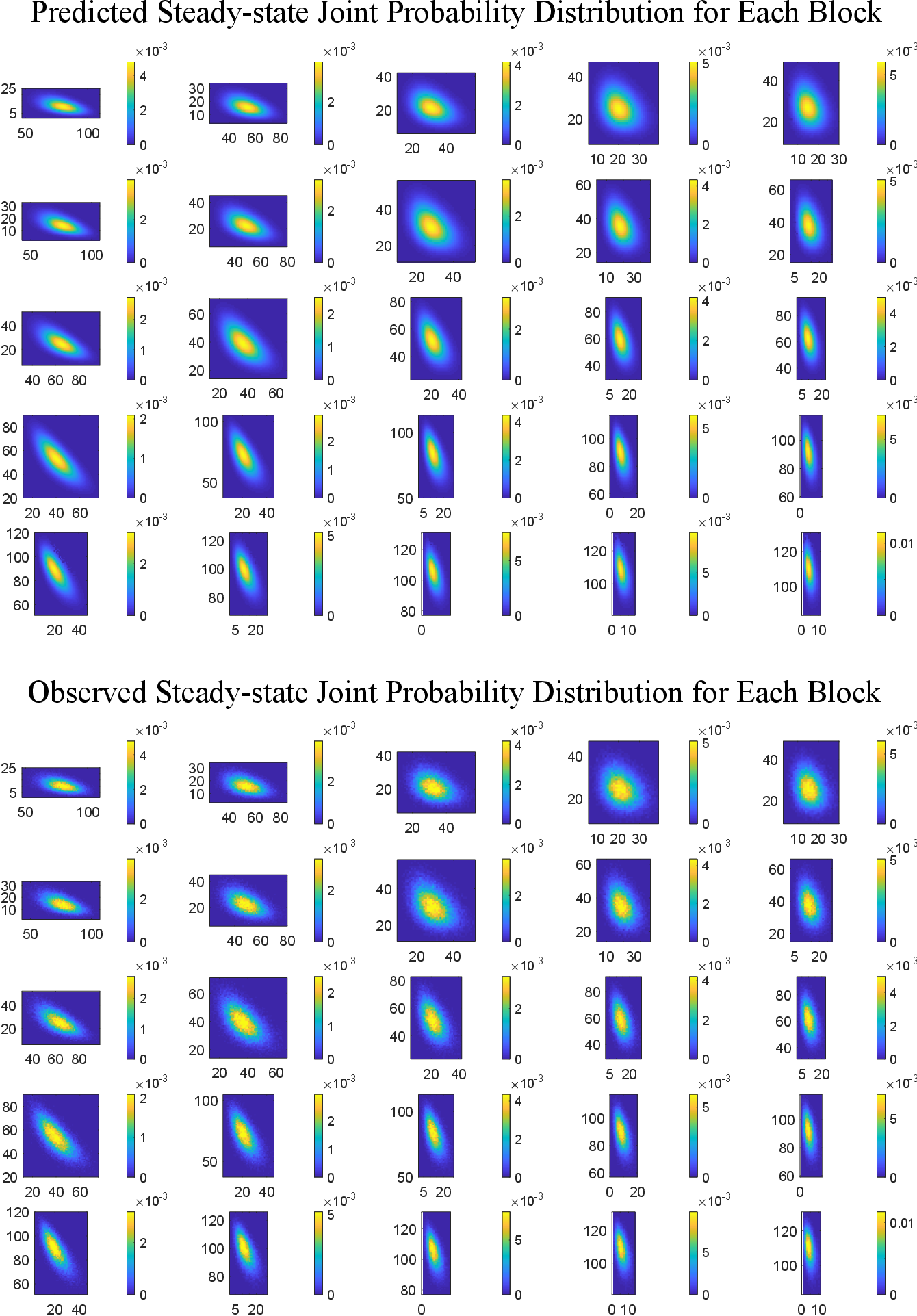}
	\caption{\textbf{Predicted versus observed new steady-state joint probability distributions for all blocks} Fig. 5a of the main text showed this comparison for block SE, and here we show this comparison for all blocks. Results are presented for each individual block, arranged to reflect their location in the city: top-right panel is for block NE, bottom-left plot is for block SW, etc. Refer to the main text figure for axes labels and legends for each plot. W e observe good agreement between DFFT analytical predictions and observations for all blocks.}\label{ABP}
\end{figure}

The main text presented predictions of the time evolution and final steady states for a particular block (SE) in our model city. Figs.~\ref{ABt} --\ref{ABP} of this section presents the results for all blocks.

\section{Discrepancies with Extreme Boundary Interactions}

The coarse graining of the underlying migration model into a sum of \emph{independent} contributions from each block (Equation~\eqref{eq:CoarseH}) 
in section \ref{sec:Under} implicitly assumes that the utility function for agents does not depend significantly on agents in neighboring blocks. When the block dimensions are large compared to the interaction range of agents, this represents a reasonable approximation because each agent in the Schelling simulation interacts directly with only its 8 direct neighbors and thus only agents on the very edges and corners of each block can be affected by agents in neighboring blocks. For example, for the 12-by-12 blocks employed in the main text that contain 144 sites total, there are only 44 edge and corner sites, so that the majority of sites have no interaction with neighboring blocks.  Moreover, of the edge and corner sites, 40 are sites along edges and, for these, the majority of potential interaction sites (five out of eight) remain within the block.

\begin{figure}[H]
	\centering
%	\begin{center}
	\includegraphics[width=1\linewidth]{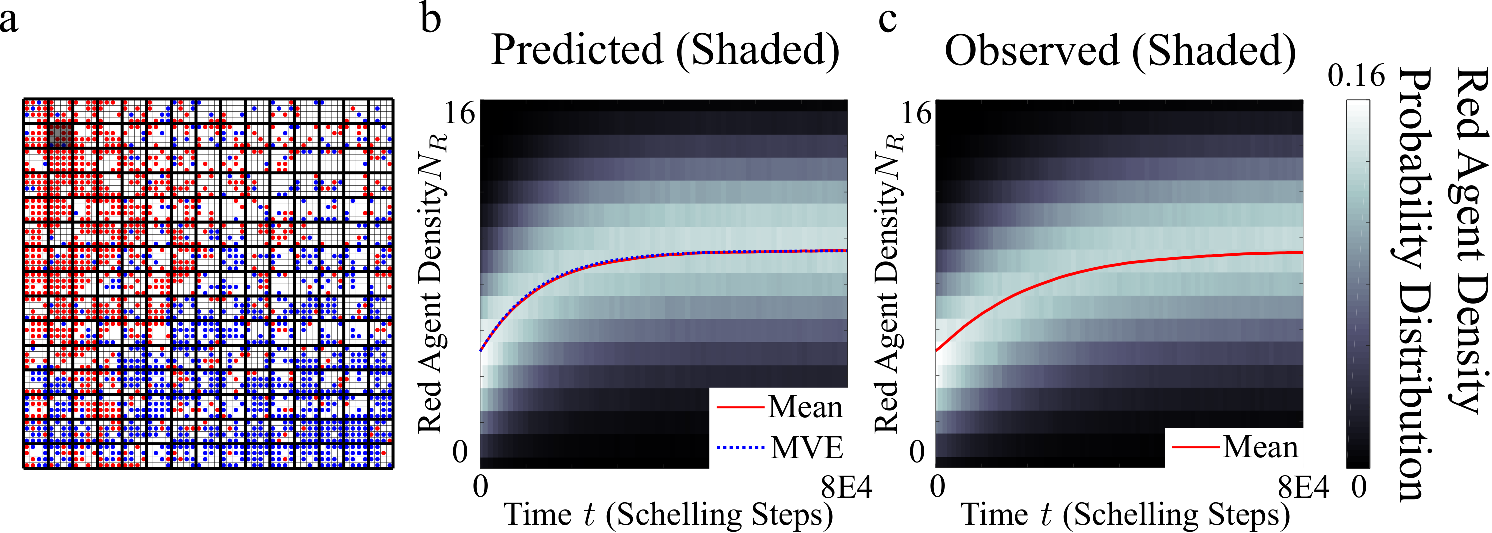}
%	\end{center}
	\caption{\textbf{Discrepancy in DFFT predictions due to boundary interactions for small blocks} \textbf{(a.)} Analysis corresponding to Fig.~2f of the main text but now with small, 4-by-4 blocks.
	\textbf{(b.)} Predicted time evolution of red agent number with TD-DFFT model and MVE for the shaded block (analogous to Fig. 4b of main text). \textbf{(c.)} Observed time evolution of red agent number in the shaded block (analogous to Fig.~4d of main text). Using small blocks for the analysis, we find predictions that evolve about 60\% faster than the observation. This deviation is caused by the strong boundary interactions of agents in the shaded block with neighboring blocks.}\label{Boundary}
\end{figure}

We do expect, however, this approximation to begin to break down for smaller blocks. For example, when using 4-by-4 blocks (Fig.~\ref{Boundary}a), a significant number of the sites (twelve of sixteen) are along the boundaries and edges. Indeed, when using such blocks to analyze the simulations described in the main text, we find that blocks whose neighboring blocks show large deviations away from the initial steady state experience noticeable errors for DFFT predictions.
\footnote{Intuitively, the extracted frustration and vexations of a block capture some of the boundary interactions during the initial steady state. When boundary interactions of the block change due to deviations of neighboring blocks away from the initial steady state, DFFT would fail to capture this change, leading to larger prediction errors for the block.}
For example, Figs.~\ref{Boundary}b,c compare the predicted and observed time evolution of the number of red agents of the shaded block (Fig.~\ref{Boundary}a), showing a 60\% distortion in the time scale to approach equilibrium when analyzing the regional-scale demographic change (Fig. 4a of the main text). 

To verify that the discrepancy in Fig.~\ref{Boundary} is due to boundary interactions, we also performed a test using a Schelling-like simulation with no inter-block interactions. To accomplish this, we replaced all out-of-block interactions of edge and corner sites for each 4-by-4 block with interactions with corresponding sites on the opposite side of the same block (periodic boundary conditions for each 4-by-4 block). In this case, we found that the DFFT predictions are once again accurate, confirming that the discrepancy seen in Fig.~\ref{Boundary} is due indeed to interactions with neighboring blocks.

\section{Analysis of the Schelling Simulation with Agent Rule (2)}\label{sec:altR}
This section explores the impact on DFFT and TD-DFFT results when the underlying agent rule is unknown and replaced with a somewhat different rule. Specifically, we repeat the analysis of our Schelling simulation from the main text but replace the multiplicity factors from Section~\ref{sec:uniform} with those from \ref{sec:grav} in the extraction of vexation and frustration functions and replace the agent rule from Section~\ref{sec:uniform} with that from \ref{sec:grav} when performing the TD-DFFT simulation. Such replacements may be necessary in the analysis of real data when, for example the maximum occupancy $s$ of each block is not known \emph{a priori}, making it impossible to evaluate the multiplicity factors from Equation~\eqref{eq:omega1}, and making it impossible to ensure that the move-in rate for each block is proportional to the number of available vacant locations.

As evident in the comparison between Fig.~\ref{newpred1}f and the corresponding panel in the main text (Fig.~4f), although the version of section \ref{sec:grav} performs well in predicting final states and the path followed toward those states, this less realistic model does slightly distort time scales at different rates across the system, as seen in the noticeably larger, but still small, discrepancies between the red and black arrows in Fig.~\ref{newpred1}f as opposed to Fig.~4f. These results underscore the importance of including realistic scalings in the agent-rules used in the TD-DFFT calculation (such as how movement probabilities scale with the number of available options) in order to obtain the most accurate results. On the other hand, these results also demonstrate that adjusting the TD-DFFT time scale by the ratio $M^{\mathrm{TDDFFT}}/M^{\mathrm{obs}}$ as recommended in Section~\ref{sec:timescale} results in predictions that are relatively insensitive to the choice of agent rule used for the TD-DFFT calculation.

\section{DFFT and TD-DFFT Analyses for More Complex Utility Functions}
In this section, we present the same DFFT analysis as the main text for a class of more complicated social utility functions that has been studied in the literature \parencite{zhang2004residential,grauwin2012dynamic}, specifically
\begin{equation}
U_R^\text{so}(N_R^\text{ne},N_B^\text{ne})=\begin{dcases}
0.5\cdot N_R^\text{ne} & \text{if $N_R^\text{ne}\le 4$}\\
-0.25\cdot N_R^\text{ne}+3 & \text{otherwise}
\end{dcases},\label{eq:compSUR}
\end{equation}
and
\begin{equation}
U_B^\text{so}(N_R^\text{ne},N_B^\text{ne})=\begin{dcases}
0.5\cdot N_B^\text{ne} & \text{if $N_B^\text{ne}\le 4$}\\
-0.25\cdot N_B^\text{ne}+3 & \text{otherwise}
\end{dcases},\label{eq:compSUB}
\end{equation}
as shown in Fig.~\ref{compSU}.

Our DFFT approach succeeds extracting reliable frustration and vexation functions even for Schelling systerms with these more complex utility functions (Fig.~\ref{newextract2}). We also find very good agreement between DFFT predictions and observations, as the detailed comparisons in Figs.~\ref{newpred2_1}~and~\ref{newpred2_2} show.

\begin{figure}[H]
	\centering
	\includegraphics[width=1\linewidth]{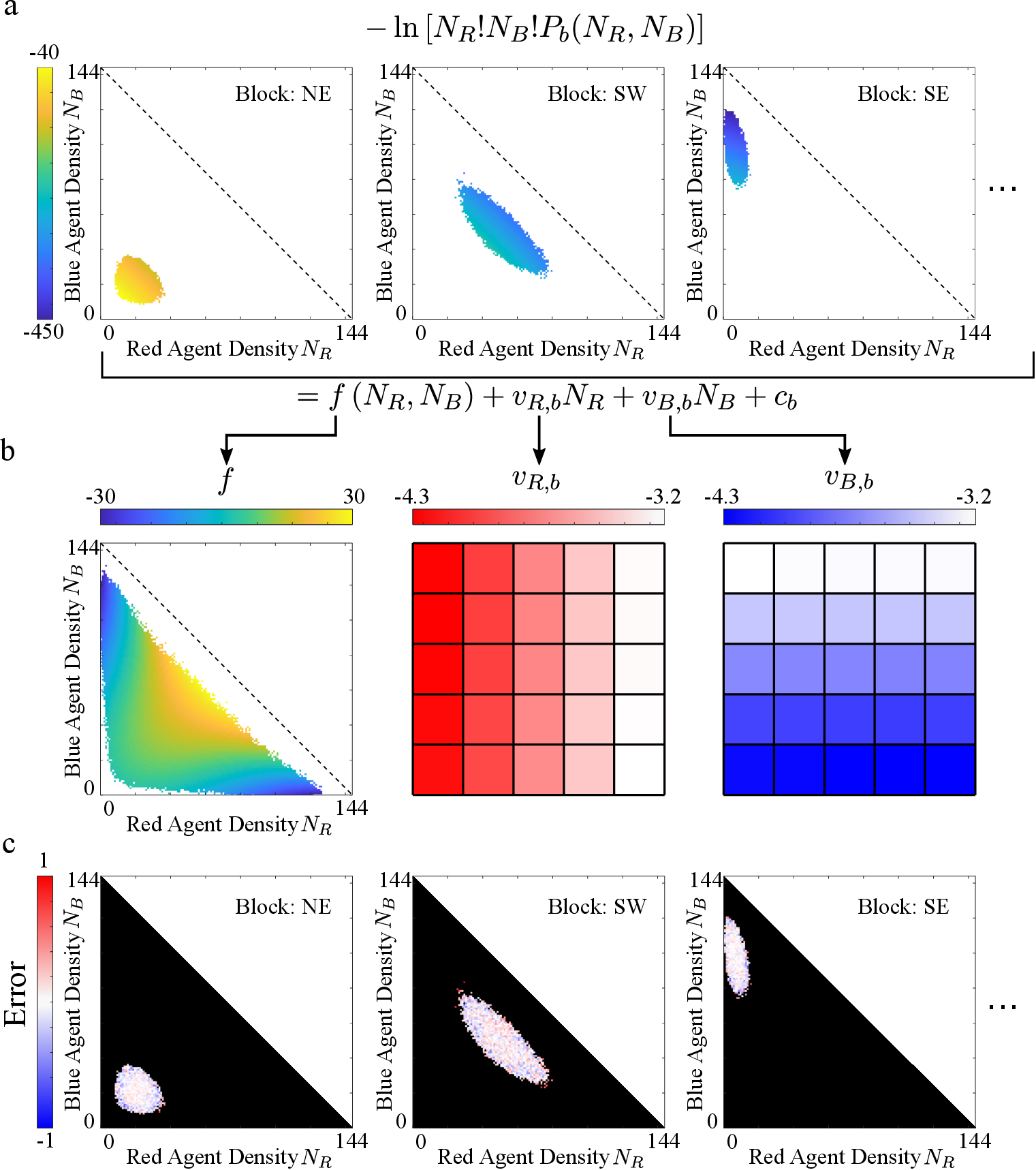}
	\caption{\textbf{Extraction of effective social and spatial preferences using multiplicity factors from Section~\ref{sec:grav}}. (Compare to Fig. 3, main text.) One key difference in these results is that the term $\ln[(s-N_R-N_B)!]$ is now absorbed into the frustration. Indeed, we observe that the extracted frustration here is different from its counterpart in the main text. The vexations (up to constant shifts according to Eq.~\eqref{eq:gauge}) and errors remain roughly the same (depending on the implementation of the DFFT function extraction method).}\label{newextract}
\end{figure}

\begin{figure}[H]
	\centering
	\includegraphics[width=1\linewidth]{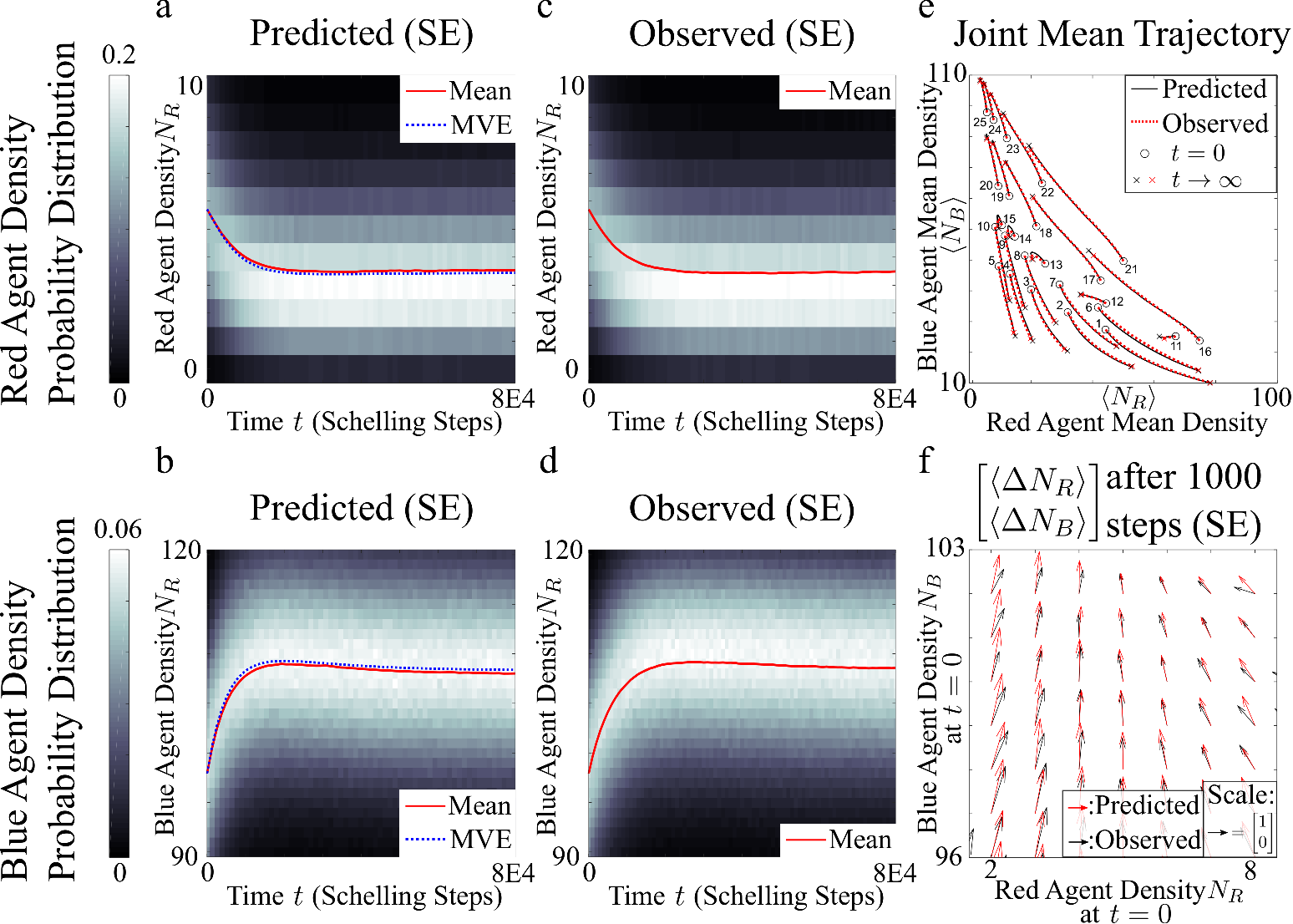}
	\caption{\textbf{Prediction of time evolution using multiplicity factors from Section~\ref{sec:grav}} (Compare to Fig.~4, main text.)  We find that for block SE, 
	the predictions with this different multiplicity factor lead to noticeably larger, but still small, discrepancies when compared to the observations. Such differences underscore the importance of using that finding the appropriate rule for agents to propose moves will improve the prediction to some extent.}\label{newpred1}
\end{figure}

\begin{figure}[H]
	\centering
	\includegraphics[width=0.79\linewidth]{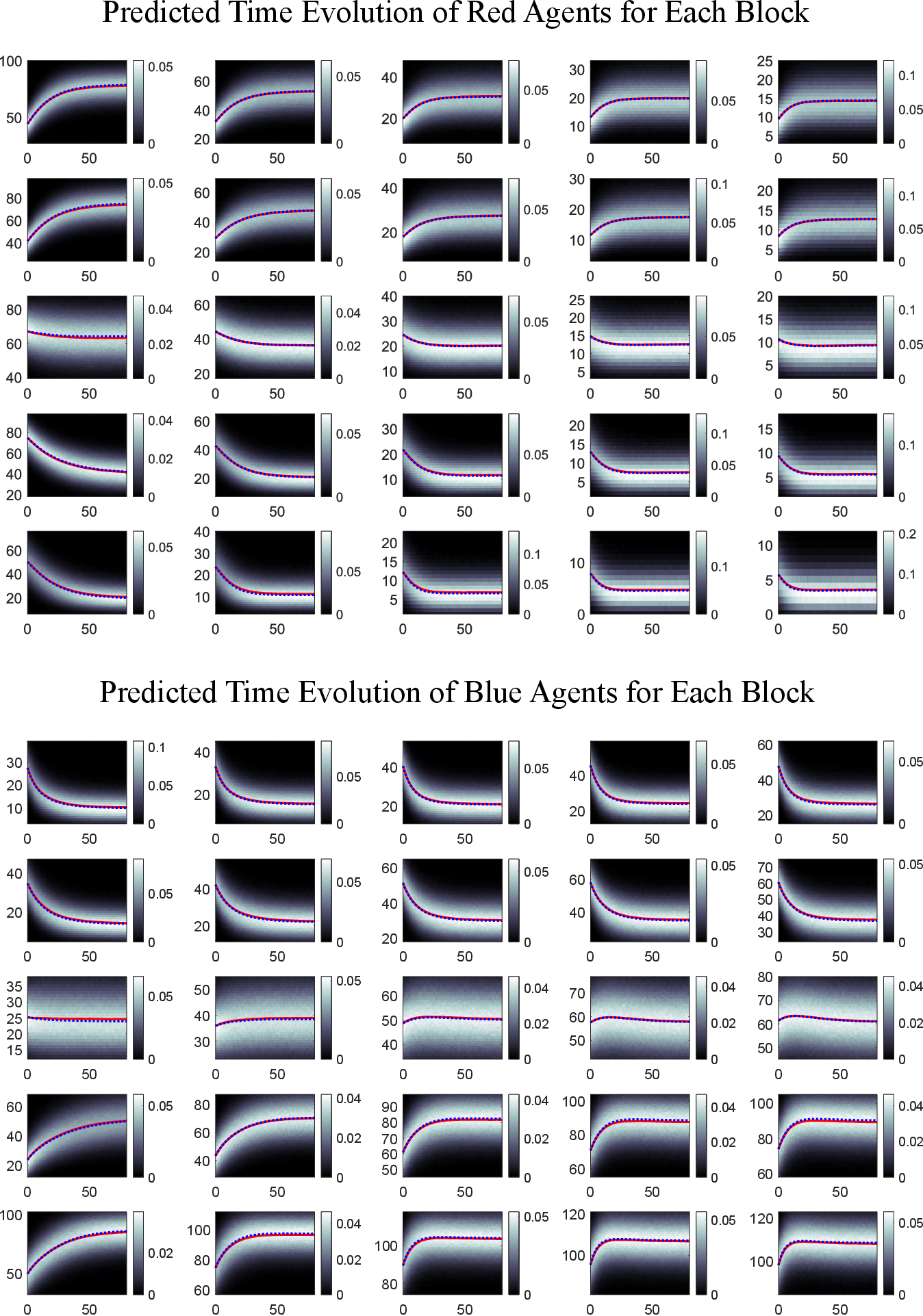}
	\caption{\textbf{Prediction of time evolution of red and blue agents for all blocks} (Compare to Figs.~\ref{ABt}~and~\ref{ABtb}.) We find that, for most blocks, even though the errors are larger, the time evolution is still predicted relatively well even when not using the precise multiplicity factors and agent rule.}\label{oldABt}
\end{figure}

\begin{figure}[H]
	\centering
	\includegraphics[width=1\linewidth]{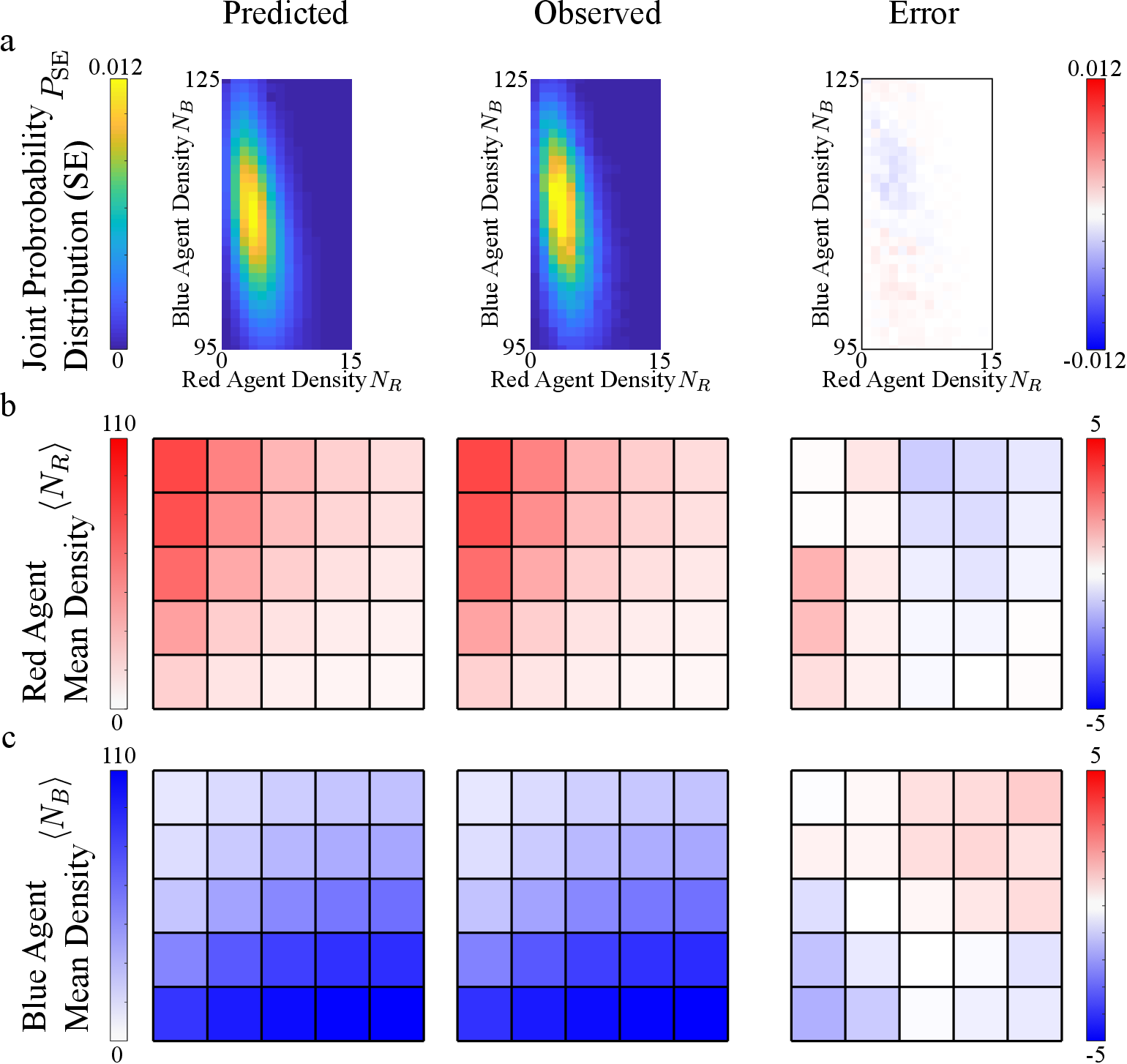}
	\caption{\textbf{Analytic prediction of new steady state} (Compare to Fig. 5, main text). As expected, the steady-state predictions remain unchanged because the differences in multiplicity factors are folded into the new, extracted frustrations.}\label{newpred2}
\end{figure}

\begin{figure}[H]
	\centering
	\includegraphics[width=0.66\linewidth]{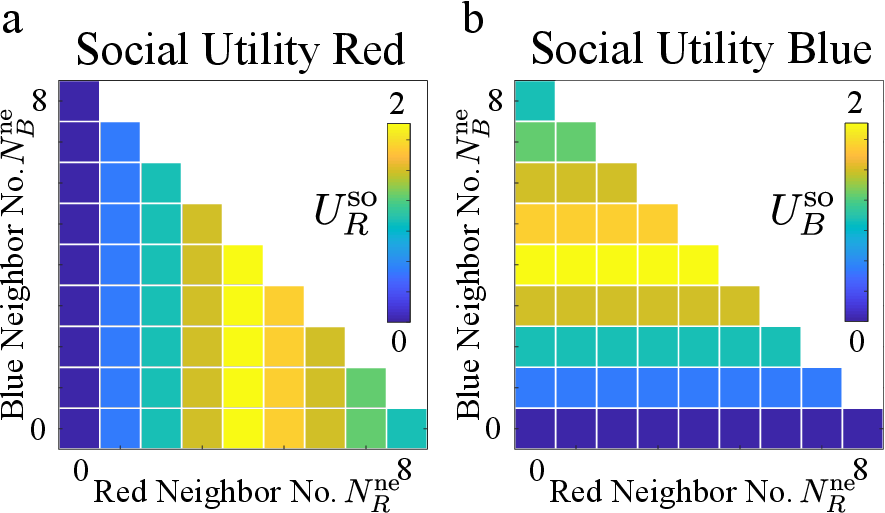}
	\caption{\textbf{More Complex Social Utilities} \textbf{(a.)} Social utility for red agents, as defined in Eq.~\eqref{eq:compSUR}, replacing the linear social utility in the main text (Fig. 2b). This utility exhibits an asymmetric peak at $N_{R}^\text{ne}=4$. \textbf{(b.)} Social utility for blue agents, as defined in Eq.~\eqref{eq:compSUB}, replacing the linear social utility in the main text (Fig. 2c). This utility exhibits an asymmetric peak at $N_{B}^\text{ne}=4$. }\label{compSU}
\end{figure}

\begin{figure}[H]
	\centering
	\includegraphics[width=1\linewidth]{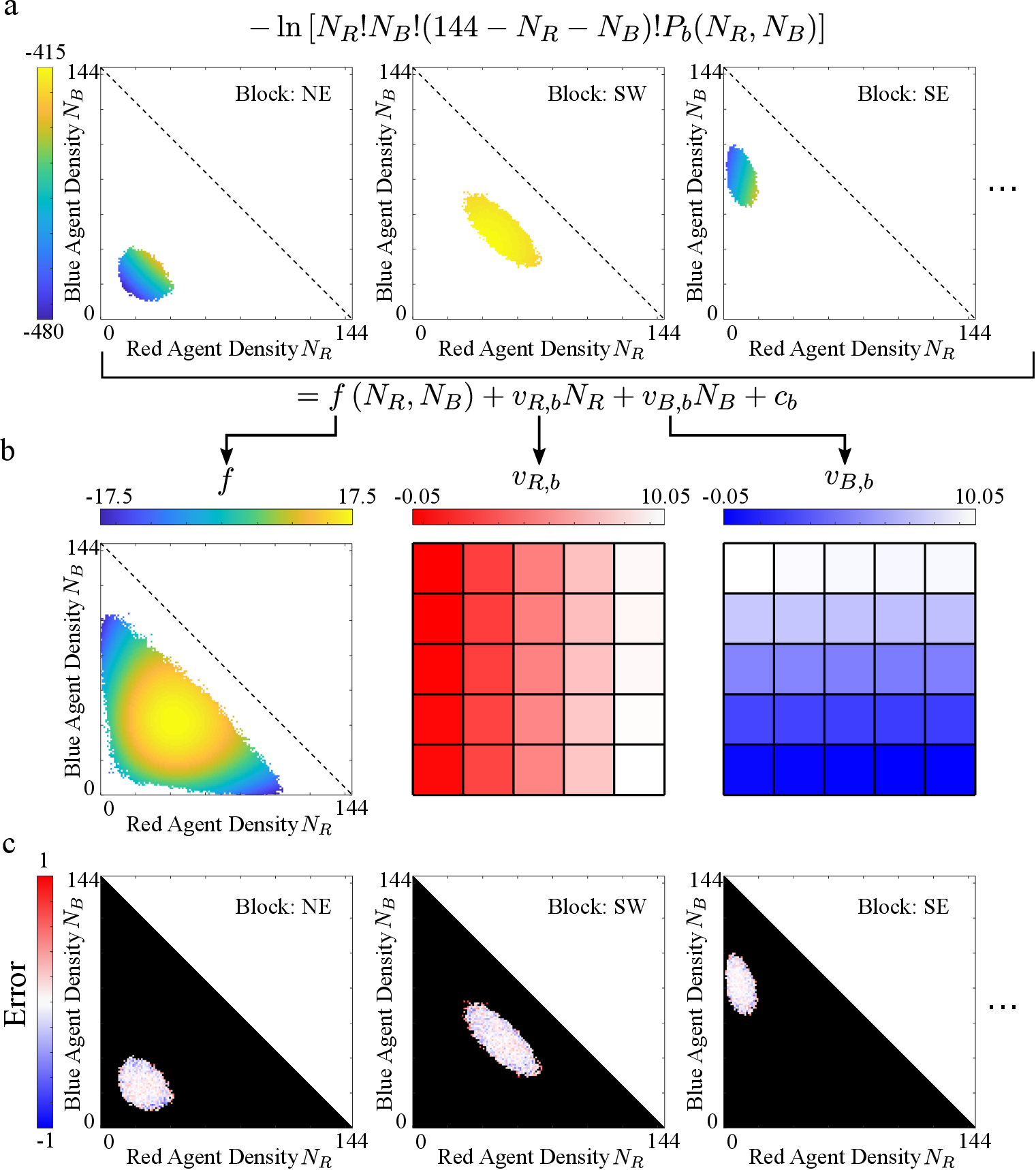}
	\caption{\textbf{Extraction of effective social and spatial preferences} (Compare to Fig.~3, main text.) The extracted frustration here is different from its counterpart in the main text to capture the more complex utility function, while the vexations (up to constant shifts according to Eq.~\eqref{eq:gauge}) remain roughly the same. The extracted DFFT functions fit the data well with small errors.}\label{newextract2}
\end{figure}

\begin{figure}[H]
	\centering
	\includegraphics[width=1\linewidth]{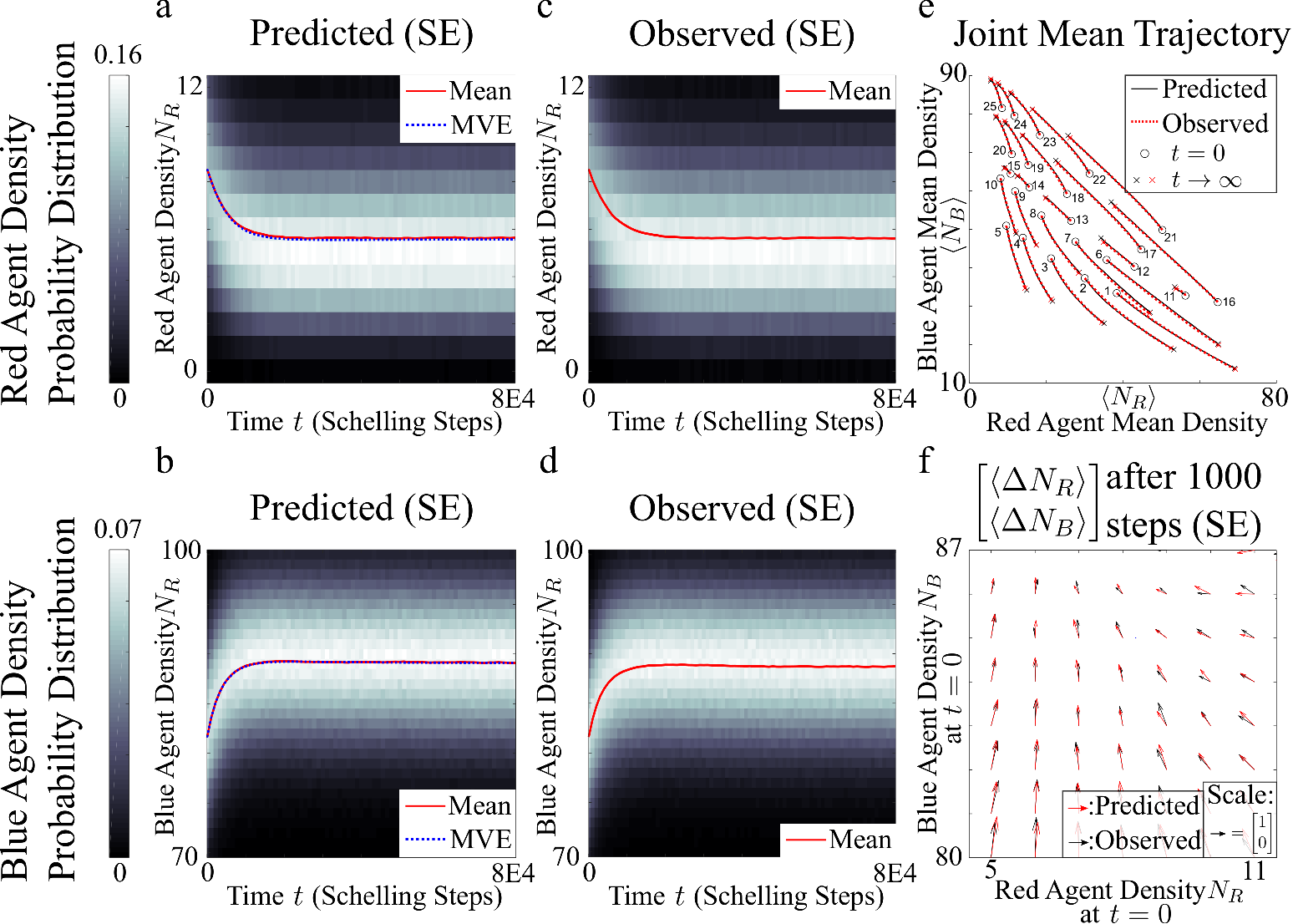}
	\caption{\textbf{Prediction of time evolution} (Compare to Fig.~4, main text.) The time evolution predictions agree well with observations. Interestingly, the more complicated social utilities do not generate more complicated time evolutions.}\label{newpred2_1}
\end{figure}

\begin{figure}[H]
	\centering
	\includegraphics[width=1\linewidth]{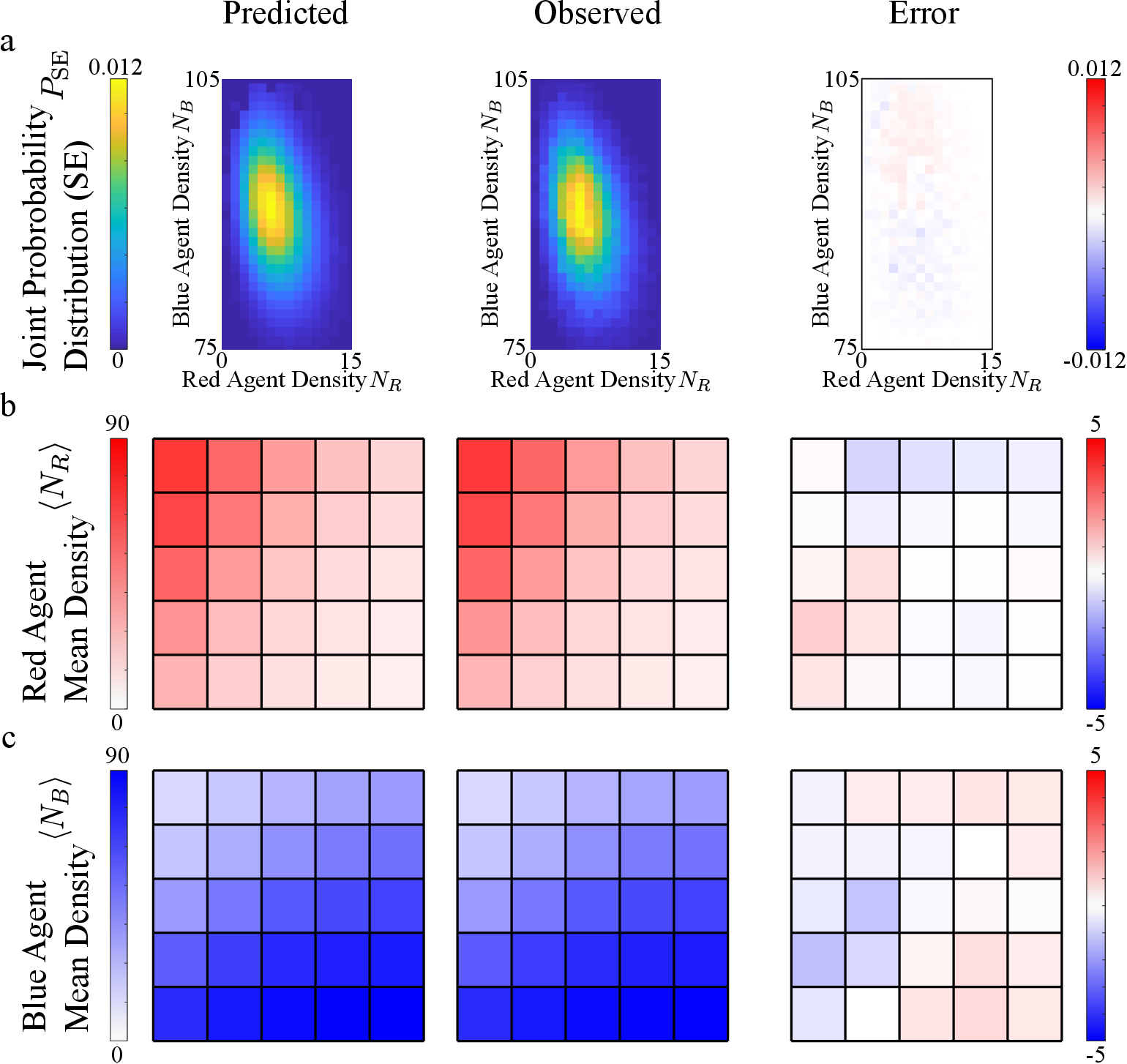}
	\caption{\textbf{Analytic prediction of new steady state} (Compare to Fig. 5, main text.) Predictions agree well with observations, even for more complex utility functions.}\label{newpred2_2}
\end{figure}

\newpage

\singlespace

\begin{xltabular}{\textwidth}{|l|p{10cm}|l|}
\caption{List of Important Variables in SI} 
\\
\hline
Variable & Meaning & Usage \\ [0.5ex]
\hline 
$P_\text{Schelling}$ & Probability of moving for each Schelling step&\\ [0.5ex]
$U^\text{so}_R$ & Social Utility for a red agent& \multirow{9}{4em}{Schelling}  \\  [0.5ex]
$U^\text{so}_B$ & Social Utility for a blue agent&\\ [0.5ex]
$\beta_R$ & Slope of linear $U^\text{so}_R$ &\\ [0.5ex]
$\beta_B$ & Slope of linear $U^\text{so}_B$ &\\ [0.5ex]
$U^\text{sp}_R$ & Spatial Utility for a red agent&\\ [0.5ex]
$U^\text{sp}_B$ & Spatial Utility for a blue agent&\\ [0.5ex]
$N_R^\text{ne}$ & Number of red agents in the 8-connected neighborhood &\\ [0.5ex]
$N_B^\text{ne}$ & Number of blue agents in the 8-connected neighborhood &\\ [0.5ex]
\hline 

$b$ & Block index & \multirow{15}{4em}{Schelling \& DFFT}\\ [0.5ex]
$b_\text{tot}$ & Total number of blocks in a city &\\ [0.5ex]
$x$ & Location in the city &\\ [0.5ex]
$t$ & Time &\\ [0.5ex]
$s$ & Maximum agent occupancy in a block (number of cells in a Schelling block)& \\ [0.5ex]
$s_\text{tot}$ & Maximum agent occupancy in a city (total number of cells in a Schelling city) &\\ [0.5ex]
$N_\text{tot}$ & Total number of agents in a city &\\ [0.5ex]
$N_R^\text{tot}$ & Total number of red agents in a city &\\ [0.5ex]
$N_B^\text{tot}$ & Total number of blue agents in a city &\\ [0.5ex]
$P_b$ & Marginal Block Distribution: probability distribution of agents in block $b$      &\\ [0.5ex]
$N_{R,b}$ & Number of red agents in block $b$ (also used interchangeably as density of red agents $n_R$ when $A_b=1$)    & \\ [0.5ex]
$N_{R}$ &  Abbreviated $N_{R,b}$ when there is no ambiguity   &\\ [0.5ex]
$N_{B,b}$ & Number of blue agents in block $b$ (also used interchangeably as density of red agents $n_B$ when $A_b=1$)     &\\ [0.5ex]
$N_{B}$ & Abbreviated $N_{B,b}$ when there is no ambiguity &\\ [0.5ex]
\hline 

$z_b$ & Normalization constant for $P_b$     &\\ [0.5ex]
$h$ & Dissatisfaction function &\\ [0.5ex]
$f_R$ & Frustration function for red agents  &\\ [0.5ex]
$f_B$ & Frustration function for blue agents  &\\ [0.5ex]
$V_R$ & Vexation function for red agents  &\\ [0.5ex]
$V_B$ & Vexation function for blue agents  &\\ [0.5ex]
$f$ & Global frustration function & \\ [0.5ex]
$n_R$ & Density of red agents ($n_R\equiv N_R/A_b$ in the coarse-grained case)  & \\ [0.5ex]
$n_B$ & Density of blue agents ($n_B\equiv N_B/A_b$ in the coarse-grained case)  & \\ [0.5ex]
$v_{R,b}$ & Average vexation for red agents in block $b$    & \multirow{1}{4em}{DFFT}\\ [0.5ex]
$v_{B,b}$ &  Average vexation for blue agents in block $b$    & \\ [0.5ex]
$H$ & Global headache functional/function &\\ [0.5ex]
$P$ & Probability distribution of states of the entire city &\\ [0.5ex]
$Z$ & Normalization constant for $P$ &\\ [0.5ex]
$\Omega$ & Effective multiplicity of a state of the entire city &\\ [0.5ex]
$\omega_b$ & Multiplicity factor for block $b$ such that $\Omega=\prod_b\omega_b$&\\ [0.5ex]
$v_{i,j}^\text{M}$ & Strength of migratory interaction between blocks $i$ and $j$ &\\ [0.5ex]
$A_b$ & Area of block $b$ &\\ [0.5ex]

$H_b$ & Headache function for block $b$& \\ [0.5ex]
$P_{b\to b'} $ & Probability of an agent accepting a transition from block $b$ to $b'$ (dependent on the type of agent and the state of the system)& \\ [0.5ex]

$\nu_{R,b\to b'}$ & Overall rate of transition for red agents from block $b$ to $b'$ (dependent on the state of the system) &\\ [0.5ex]
$\nu_{B,b\to b'}$ & Overall rate of transition for blue agents from block $b$ to $b'$ (dependent on the state of the system)&\\ [0.5ex]
$\mu_R$ & Red agent potential in a city &\\ [0.5ex]
$\mu_B$ & Blue agent potential in a city &\\ [0.5ex]
$v_{R,b}^\text{eff}$ & Effective potential (vexation) for red agents in block $b$ &\\ [0.5ex]
$v_{B,b}^\text{eff}$ & Effective potential (vexation) for blue agents in block $b$ &\\ [0.5ex]
\hline
\end{xltabular}

\pagebreak
\theendnotes
\pagebreak

\printbibliography[title={References}]

\pagebreak

\end{document}